\documentclass[%
 pre,%
 amsmath,amssymb,
 reprint,%
]{revtex4-2}

\usepackage{epsdice}
\usepackage{skak}
\usepackage[T1]{fontenc}
\usepackage[utf8]{inputenc}
\usepackage{amsmath}
\usepackage{amssymb}
\usepackage{graphicx}
\usepackage{yfonts}
\usepackage{mathtools}

\DeclarePairedDelimiter\floor{\lfloor}{\rfloor}
\usepackage[
  citecolor=blue,
  colorlinks,
  linkcolor=blue,
  urlcolor=blue,
]{hyperref}
\usepackage[dvipsnames]{xcolor}
\usepackage{dcolumn}
\usepackage{bm}
\newcommand{\bs}[1]{\boldsymbol{#1}}
\newcommand{\mc}[1]{\mathcal{#1}}
\newcommand{\mb}[1]{\mathbb{#1}}
\begin{document}

\preprint{AIP/123-QED}

\title{{Eclectic Notes on Uncertainty, Information, and Classical Dynamics}}
\author{Sagar Chakraborty}
\email{sagarc@iitk.ac.in}
\affiliation{
  Department of Physics,
  Indian Institute of Technology Kanpur,
  Uttar Pradesh 208016, India
}
\date{\today}
  
\begin{abstract}
%
%
Information is everywhere in nature which is very uncertain and unpredictable. But information, in itself, is a very ambiguous term. In this cursory write-up, we attempt to understand the formal meaning of information by quantifying uncertainty and discuss how it naturally appears in two core topics of classical physics---classical mechanics and statistical mechanics. In the process, we witness how the concepts of the information theory render a unique viewpoint in physics. Except for the presentation of the material, the author claims no originality; however, the responsibility of every inadvertent error lies entirely with the author. 
\end{abstract}
\maketitle
\tableofcontents
\section{Quantifying Uncertainty: Probability}
\hskip 2.2 cm``\emph{Probability does not exist.}''---de Finetti
\\
\\
\indent {\LARGE \textgoth{S}}urvival in this world is an adventure because we are so ignorant and everything around us seems so uncertain. In fact, it is hard to think of any realistic phenomenon that is completely free from one of the three kinds of uncertainty---deterministic (chaotic), stochastic, and quantum; more often than not, all of the three are present in various degrees. Thus, in order to gain any knowledge about any realistic system, one must understand how to characterize, quantify, and interpret these uncertainties. 

\subsection{Interpreting Probability}
\label{sec:I}

If we can define the state of a system in some appropriate unambiguous manner, then the implication of the aforementioned uncertainties is that we can only have some \emph{degree of belief} or \emph{credence} about the state at any given time. Assuming we are rational, our credences better be \emph{coherent}: In other words, we are adopting the philosophical view of \emph{probabilism} that requires a rational being's credences to form a probability distribution.

Any signal we receive in relation to the state of the system may convey some information. What information does is that it changes the (credence) probability distribution over time. One could ask: What is the quantity and the content of the information received? It is apparent that an answer to this question is intricately tied to the concept of the probability distribution described by a probability function as discussed below.

A probabilistic system is mathematically described by a triple $(\mathcal{S},\mathcal{E},P)$: $\mathcal{S}$ is the \emph{sample space}---set of all possible outcomes, $\mathcal{E}$ is the \emph{event space}---set of all possible (measurable) sets of outcomes, and $P:\mathcal{E}\rightarrow[0,1]$ is the \emph{probability measure or probability function} that obeys \emph{Kolmogorov's axioms}:
\begin{enumerate}
\item \emph{Normalization:} $P(\mathcal{S})=1$.
\item \emph{Non-negativity:} $P(E)\ge0$ $\forall E\in\mathcal{E}$.
\item \emph{Finite additivity:} $P(\cup_{i=1}^nE_i)=\sum_{i=1}^nP(E_i)$ for any finite sequence, $\{E_i\}_{i=1}^n$, of mutually disjoint events.
\end{enumerate}
For an infinite sample space, a fourth axiom---\emph{Continuity at zero}---needs to be added: If $E_1\supseteq E_2\supseteq E_3\supseteq\cdots$ is a decreasing sequence of events that tends to empty set, then $\lim_{i\rightarrow\infty}P(E_i)=0$. Equivalently, rather than adding the fourth axiom, the third axiom may be modified for an infinite sample space; it becomes: (\emph{Countable additivity}) $P(\cup_{i=1}^\infty E_i)=\sum_{i=1}^\infty P(E_i)$ for any infinite sequence, $\{E_i\}_{i=1}^\infty$, of mutually disjoint events. By taking almost all events as empty set, the countable additivity can be seen to imply finite additivity; and the `continuity at zero' axiom can be obtained from the countable additivity by choosing events, $E_i=F_{i-1}\backslash F_i$ where $F_i\equiv \cup_{k=1}^{\infty}E_k\backslash \cup_{k=1}^{i}E_k$ ($\{F_i\}$ is a decreasing sequence of events approaching empty set). 

Observe that if $\mathcal{S}$ is countably infinite (or finite) and $\mathcal{E}$ is taken to be the power set $2^\mathcal{S}$, then $\mathcal{E}$ satisfies following properties:
\begin{enumerate}
\item  $\mathcal{S}\in\mathcal{E}$.
\item  If $E\in\mathcal{E}$, then $E^c\in\mathcal{E}$.
\item If $E_i\in\mathcal{E}\,\, \forall i\in\{1,2,\cdots,\infty\}$, then $\cup_{i=1}^\infty E_i\in\mathcal{E}$ (or, for finite $\mathcal{S}$: if $E_i\in\mathcal{E}\,\, \forall i\in\{1,2,\cdots,n\}$, then $\cup_{i=1}^nE_i\in\mathcal{E}$). 
\end{enumerate}
Note that it follows that the impossible event $\phi\in\mathcal{E}$ (for which $P(\phi)=0$) and $\cap_{i=1}^\infty E_i\in\mathcal{E}$ (or, for finite $\mathcal{S}$, $\cap_{i=1}^nE_i\in\mathcal{E}$). In technical jargon, $2^\mathcal{S}$ is a \emph{sigma algebra} (or \emph{algebra} if $\mathcal{S}$ is finite; all sigma algebras are algebras). In simpler words, a sigma algebra is just a set of sets that is closed under complementation, countable union, and countable intersection.

Interestingly, there may exist smaller sigma algebra $\mathcal{E}'\subset 2^\mathcal{S}$ for countable sample space: e.g., for the sample space $\mathcal{S}=\{1,2,3,4\}$, $\mathcal{E}'=\{\phi,\{1,3\},\{2,4\},\mathcal{S}\}$ is a sigma algebra. However, it doesn't make sense to work with such smaller non-power sets in the probability theory as they do not exhaust all possible events.

Problem arises when one deals with uncountably infinite sample space: There are events with which one cannot associate probability measure. A classic example is that of the Vitali set. How should probability space be defined then? In this case, smaller non-power sets come to rescue. If the event space $\mathcal{E}$ has to be taken as the set of measurable sets of outcomes, we should take the one that is a sigma algebra even though it may not be the power set. For example, Borel sigma algebra on the sample space $[0,1]\subset\mathbb{R}$ is a set of all the open intervals $(a,b)$ with $0\le a\le b\le 1$, all their complements, all their finite or countably infinite unions, and all complements of these unions.

The probabilistic system has to be further supplemented with the notion of \emph{conditional probability}---$P(E_i|E_j)$ for any two events in the event space---that describes the probability of event $E_i$ given $E_j$ has occurred; there need not be any causal connection between the two events. It can either be introduced as a definition in the restricted sample space:
\begin{equation}
P(E_i|E_j)=\frac{P(E_i\cap E_j)}{P(E_j)}
\end{equation}
(termed \emph{ratio formula}) with $P(E_j)\ne0$ or as an independent axiom for the conjunction:
\begin{equation}\label{eq:indt}
{P(E_i\cap E_j)}=P(E_i|E_j){P(E_j)},
\end{equation}
where $E_i$ and $E_j$ need not occur at the same time.

Naturally, $E_i$ and $E_j$ can be said to be \emph{independent} if 
\begin{equation}
P(E_i|E_j)={P(E_i)};
\end{equation}
i.e., in the light of Eq.~(\ref{eq:indt}), if
\begin{equation}
{P(E_i\cap E_j)}=P(E_i){P(E_j)}.
\end{equation}
A related concept is that of \emph{conditional independence}: Two events, say $E_i$ and $E_j$, are known as conditionally independent given an event $E_k$ if 
\begin{equation}
P(E_i\cap E_j|E_k)=P(E_i|E_k)P(E_j|E_k),
\end{equation}
or equivalently, if
\begin{equation}
P(E_i|E_j,E_k)={P(E_i|E_k)}.
\end{equation}
Here, $P(E_i|E_j,E_k)$ reads as probability of $E_i$ given both $E_j$ and $E_k$. It is neither necessary nor sufficient condition for independence between two events $E_i$ and $E_j$. 

Next important concept is that of \emph{conditionalization}. The principle of conditionalization governs how one's credences in a proposition changes upon receiving new evidence/data concerning the proposition. In particular, under the simple principle of conditionalization, if $H$ is the proposition and $D$ is the data acquired between times $t_i$ to $t_j>t_i$, then
\begin{equation}
P_{t_j}(H)=P_{t_i}(H|D)
\end{equation}
gives a rule to go from old \emph{prior} $P_{t_i}(H)$  to new prior $P_{t_j}(H)$ through \emph{posterior} $P_{t_i}(H|D)$; the subscripts denote the corresponding times. In this context, it helps to realize that a relation known as \emph{Bayes' theorem} trivially follows from the definition of the conditional probability:
\begin{eqnarray}
&&P(E_i|E_j)=\frac{P(E_j|E_i){P(E_i)}}{P(E_j)}\nonumber\\
\implies&&P(E_i|E_j)=\frac{P(E_j|E_i){P(E_i)}}{P(E_j|E_i){P(E_i)}+P(E_j|E^c_i){P(E^c_i)}}.\quad
\end{eqnarray}
Note that the expanded form of denominator as been in the last expression is an application of what is known as the \emph{law of total probability}: If $\{E_1,E_2,\cdots,E_n\}$ is a partition of a sample space, then any event $E$ satisfies the relation, $P(E)=\sum_{i=1}^nP(E\cap E_i)$. Here \emph{partition} means a set of non-overlapping, non-empty subsets of the sample space such that their union is the sample space itself.

If one uses the ratio formula and Bayes' theorem, then the aforementioned conditionalization is essentially what is known as the \emph{Bayesian updating}. The five core \emph{normative} rules---the three axioms of Kolmogorov, the ratio formula, and conditionalization---form the \emph{Bayesian rules} that are necessary (but not sufficient) condition for the credences to be rational. It may be remarked that there exist other proposed rules of conditionalization.

But what is meant by the word probability is not unanimously agreed upon! Broadly, there are three formulations of probability: \emph{formalist}, \emph{objectivist}, and \emph{subjectivist}. Although all of them use the three axioms of Kolmogorov and the ratio formula, the probability has distinctly different interpretations.

From the formalist viewpoint, probability is nothing more than a notation in used in an axiomatic structure. One can use the consistent and correct syntax to pose and prove various propositions unambiguously staying within the paradigm of the axiomatic structure. It has existence even without anything to do with the real world---living or non-living. Beyond the mathematical curiosity of generating artful consistent structures, it may serve as an ideal model to be put to use in practice to model certain phenomena.

The objectivist viewpoint considers repeatable random events that show limiting frequency distribution for infinite number of independent repetitions. It is supposed that to the question of probability of an event, only a unique answer is acceptable. Note, however, in such a formulation, probability itself is not an observable. An objectivist wants to estimate probability of an event given its past record of occurrences. 

The subjectivist viewpoint is very personal in nature. There is no need for the repeatable nature of the events; with every event any individual can associate some probability that can be observed directly by some individual who can communicate it numerically as well. Of course, for the exact same event the probability can have different values for different individuals depending on her knowledge about the state of the corresponding system. Most importantly, it entails that randomness is not the property of the event but a property of the knowledge of the observers. A subjectivist with uncertain knowledge is interested in logical coherent inferencing. 

Along the line of the aforementioned objective and subjective viewpoints, many different interpretations of probability are possible. Broadly, philosophers are most interested in the following five:
\begin{itemize}
\item \emph{Classical}: This objective \emph{conjectural} interpretation does not call for any experiments. Calculations can be done a priori. When there is no evidence or there is symmetrically balanced evidences, one can resort to it to calculate probability of an outcome as the ratio of number of chances with which the outcome can occur to the number of chances with which it can or cannot occur. One objection is that while treating all the outcomes of same kind to occur as `equally likely', one is getting trapped in a circular argument as to define `equally likely' one already needs the concept of probability! In order to deal with it, one raises this aspect to the level of a principle---the \emph{principle of insufficient reason} or the \emph{principle of indifference}: When there is no reason to believe in any bias, then one should treat all the outcomes as equally likely.  However, this principle can give rise to incompatible results when applied carelessly.  For example, consider \emph{Bertrand's paradox}: A factory produces metal cube magnets. The length of the sides can be between $0$ to $1$ unit. Of course, the classical probability of picking up a cube with side-length up to $1/3$ is $1/3$. Suppose this situation was presented as a factory produces metal cube magnets with face-area between $0$ to $1$ unit. Then the classical probability of picking a cube with face-area up to $1/9$ (i.e., side-length up to $1/3$) is $1/9$! In such cases, an \emph{invariant} assignment of probability should be done by reformulating the cases properly so that they are based on same equivalent knowledge. Note that, while pointing out the paradox, we have tacitly applied the concept of classical probability analogously and straightforwardly to mutually exclusive and exhaustive events (sets of outcomes) rather than outcomes.
\item \emph{Logical}: It is also implemented a priori and is \emph{conjectural}. It essentially extends the classical formulation by introducing unequal weights with events and tackling asymmetrically balanced evidences. There is a bit of debate whether it is an all out objective interpretation as it deals with the evidences available to someone whose degree of beliefs are under consideration---hence, this interpretation is also called \emph{evidential} sometimes. Perhaps, it is appropriate to interpret it as objective Bayesian (see below).
\item \emph{Frequentist}: This objective formulation is \emph{empirical}: the probabilities are defined in terms of actual real events. It calculates the relative frequency of an event's occurrence and interprets that as probability. Of course, the probability will fluctuate depending on the number of trials. However, a stronger form of the view, \emph{hypothetical frequentism}, defines the probability as the limiting frequency which has an additional serious problem that it may violate countable additivity owing to the fact that the domain of the limit frequency's definition may not be a field. Moreover, limiting relative frequency may depend on the order of trials, and hence, technically may not even exist. Last but not the least, it appears that the frequentism suffers from \emph{reference class problem} in a sense and gives only conditional probabilities; it means that the same event can have more than one probability associated with it: E.g., `what is the probability that a cow will die at the age of 15' depends on which class of cows it is being compared with.
\item \emph{Propensity}: This metaphysical and objective interpretation treats the probabilities as real properties of events: Probability is seen as the property that defines the tendency of the corresponding system to yield the event. One of the problems (\emph{Humphrey's paradox}) this interpretation has is that the probability axioms imply Bayes' theorem which goes both ways for two given events, but the propensity argument seems to be conceptually at odds with it as the propensity interpretation has an inbuilt bias towards causality.
\item \emph{Bayesian}: Also called personalist or subjective, this formulation associates probability of an event with the degree of belief in the event; and as new knowledge is gained, the degree of belief is updated using Bayes' theorem and conditionalization. The main problem (formally known as a \emph{reference class problem}) of this interpretation is how to choose \emph{prior} probabilities. This issue brings us to the following categorization: 
\subitem$\bullet$ \emph{Objective Bayesian}: An objective Bayesian assumes that probability assertions possess truth-conditions independent of who is the individual. In fact, she follows \emph{uniqueness thesis}: The probabilities are uniquely constrained by the evidences present.
\subitem$\bullet$ \emph{Subjective Bayesian}: The probabilities are only constrained by the aforementioned five normative Bayesian rules.
\end{itemize}
It should and can be proved that each of the aforementioned five interpretations of probability actually satisfies the Kolmogorov axioms so as to justify its usage as a bonafide probability measure.

In the case of subjective view---where one can associate credences with the propositions associated with the events that need not be repeatable---for the credence to be treated as a probability in line with Kolmogorov axioms, we have to restate the axioms more contextually. To this end, let us assume that an \emph{agent} assigns some credence to the propositions in some formal language. The language contains some \emph{atomic} propositions (assertions that must be either true or false) that can be combined to make more propositions using logical connectives like \emph{not} or \emph{negation} ($\neg$), \emph{or} or \emph{disjunction} ($\vee$), and \emph{and} or \emph{conjunction} ($\wedge$). In order to formalize what we are after, it helps to consider the concept of set of all \emph{possible worlds} that is a maximally specified structure: For any possible world, any event either occurs or does not occur in that world; and any logically possible combination of the events occurs in some possible world. A proposition that is true in every possible world is called a \emph{tautology}. Now consider that sets of possible worlds corresponding to the propositions form a sigma algebra since the propositions are closed under negation, disjunction, and conjunction. In this formalism, the credences---assignment of real numbers to the propositions---are probabilities (and hence the usage of $P$ below) if they satisfy the following three axioms of Kolmogorov:
\begin{enumerate}
\item \emph{Normalization:} For any tautology $T$, $P(T)=1$.
\item \emph{Non-negativity:} $P(H)\ge0$ for any proposition $H$.
\item \emph{Finite additivity:} $P(H_1\vee H_2)=P(H_1)+P(H_2)$ for any two mutually exclusive propositions $H_1$ and $H_2$.
\end{enumerate}
As before, the finite additivity can be extended to the countable additivity. Note that while associating probabilities to the propositions, one is equivalently associating probabilities to some sets of possible worlds; e.g., the first axiom equivalently means that probability associated with the set of all possible worlds is unity.

It is intriguing to point out that objective Bayesian can derive the Kolmogorov axioms from the following three desiderata:
\begin{enumerate}
\item\emph{Divisibility and comparability:} Depending on the information we have related to a proposition, the plausibility of the proposition is a real number.
\item\emph{Common sense:} The plausibilities should be in qualitative correspondence with common sense.
\item\emph{Consistency:} If the plausibility of a proposition can be computed in many ways, then all the ways should lead to the same result.
\end{enumerate}
Under this view, the probability may be interpreted as describing \emph{degree of plausibility} of  events and the plausible reasoning may be seen as an extension of Aristotelian logic system such that when one is fully certain about events, the Aristotelian logic system is recovered. By the way, some arguments why probability should be a real number are (i) the some definite physical process must be going on in brain during plausibility calculations and we denote measure of real quantities with real numbers; also, (ii) demanding completeness (any two propositions can be compared) among a set of plausibility propositions (proposition $A$ is more/less/as plausible than/than/as proposition $B$) and transitivity in degrees of plausibility among them, automatically mean that real numbers (in fact, rational numbers!) should be associated with probability. 

On the other hand, the subjective Bayesians, who insist that `probability does not exit', can get to the standard probability axioms using the notion of \emph{coherence}: To various events one can assign very personal probabilities and she can never be made to lose for certain in a betting based on the events (i.e., no \emph{Dutch book} can be made against her). While it may feel uncomfortable to define probabilities in terms of betting preferences---that too without explicitly invoking any requirement of consistency---for certain problems, e.g., in psychology, it is quite insightful. (It may be pointed out that consistency requirements automatically incorporate  the property of coherence into the plausibility reasoning rules.)
\subsection{Elaborating Bayes' Theorem}
Bayes' theorem is not exclusive to adherent of any particular interpretation of probability. Bayesian inference---a branch of statistical inference---that is closely related to the Bayesian updating, is merely an application of the theorem. The theorem has much simpler applications in simple problems of elementary probability irrespective of the latter's intriguing interpretations; all formalists, objectivists, and subjectivists can use Bayes' theorem as per their needs to deal with the appropriate questions in hand. Let us clarify this through some examples.

As the first example, consider that 99\% of non-vegetarians (defined as meat-eaters), who constitutes 2\% of the population of a predominantly vegetarian state, eat eggs. The over-enthusiastic secret service of the state wants to apprehend the non-vegetarians and after a lot of hard work manages to catch a person red-handed who is seen consuming a small hard-boiled egg---a hard evidence acceptable in the court for conviction. However the problem is that the secret service knows, while covertly and illegally monitoring egg consumption in the state, that 6\% of the vegetarians also secretly eat eggs. So what is the probability that the person caught is actually a non-vegetarian as per the objectivist secret service? Using obvious notations, we are essentially asking what $P(\operatorname{{\rm non-vegetarian}}|\operatorname{{\rm egg-eater}})$ is. To find the answer to the question, we note that $P(\operatorname{{\rm non-vegetarian}})=0.02$, $P(\operatorname{{\rm egg-eater}}|\operatorname{{\rm non-vegetarian}})=0.99$, and $P(\operatorname{{\rm egg-eater}}|{\rm vegetarian})=0.06$. Also, we find $P(\operatorname{{\rm vegetarian}})=1-P(\operatorname{{\rm non-vegetarian}})=0.98$. Straightforward use of Bayes' theorem yields that the chance that the person caught is non-vegetarian is approximately 25\%. It is interesting to note that when the demographic data was being gathered, non-vegetarian individuals' frequency was found by noting down the eating preferences of individuals one after the other---thus, in a way, creating a frequentist viewpoint of the probability (of the non-vegeterian population).

Suppose a rational Bayesian human rights officer from abroad arrives in a cafe in the above mentioned dystopian state which has kept the its demography hidden from the human rights officer. She has a prior degree of belief, $P(\operatorname{{\rm non-vegetarian}})$,  for a person sitting in the cafe to be a non-vegetarian. She then notices that the person orders an omelette secretly in the cafe. She now has to update her belief for the person's being non-vegetarian given the evidence/data. She must use the same Bayes' formula to find the posterior:
\begin{eqnarray}
P(N|E)&&=\frac{P(E|N)P(N)}{P(E)},\\
&&=\frac{P(E|N)P(N)}{P(E|N)P(N)+P(E|V)P(V)}.
\end{eqnarray}
where $E$ means egg-eater, $N$ means non-vegetarian, and $V$ means vegetarian. She may choose $P(N)=0.02$ either coincidentally or through the knowledge of all the relevant data pertaining to the state; most likely she would realistically choose some value different from $0.02$. Next the \emph{likelihood} $P(E|N)$ is hard to find and essentially depends on how she models the eating habit of the population. Similarly, $P(E)$, the \emph{marginal likelihood} or the \emph{evidence}, is also hard to find. (If one is interested only in ratio of posteriors, the marginal likelihood term disappears from calculations.) In principle, the objectivist secret service who knows all these from the survey data, can provide her with the values of the likelihood and the evidence; and she can update her prior to $P(N|E)\approx0.25$ about the person. Otherwise, she has to do her own survey of eating habits of the citizens and slowly accumulate data while sequentially updating her (different kinds of) beliefs; e.g., if she is interested in the degree of belief about the person in cafe to be non-vegetarian and see keeps seeing him ordering omelette everyday, then sequential belief-updating should lead her to the value $0.25$; if she is interested in her degree of belief about an arbitrary person's being non-vegetarian, then belief-updating should finally lead her to $0.02$ in the limit of infinite sequence of data. 

Next consider a coin toss experiment. Suppose a coin is tossed twice independently and shows head ($H$) on both occasions. One is interested in knowing whether the coin is unbiased ($U$) or completely biased ($B$); we are assuming possibility of only two types of coins. Let us see how a Bayesian would handle this problem. First of all, she has to associate a probability to the prior $P(U)$ (or equivalently, $P(B)$). If she is an objective Bayesian, then she might argue that due to the principle of insufficient reasons, one must assign $0.5$ to $P(U)$ and that should be done by any other Bayesian; but if she is a subjective Bayesian, she would claim that any value between $0$ to $1$ is allowed. (We remark here that some subjective Bayesians would additionally impose the \emph{regularity principle}---any logically contingent proposition cannot have zero unconditional probability---to rule out $0$ as possible value for the prior.) In any case, after observing two heads, she wants to update her prior about the coin's being fair; thus, essentially, following calculation for the posterior is in order:
\begin{eqnarray}
P(U|HH)=\frac{P(HH|U)P(U)}{P(HH)},
\end{eqnarray}
where the terms in the R.H.S. are all known: $P(U)=0.5$ (say), $P(HH|U)=0.25$, and $P(HH)=P(HH|U)P(U)+P(HH|B)P(B)=0.625$. Therefore, P(U|HH)=0.2. It is interesting to note that $P(HH|U)$ and $P(HH|B)$ (which is unity) have been calculated using the classical interpretation of probability.

Before we discuss this problem from another angle, let us try to understand the mathematical space where the probabilities are being defined. The easiest is to consider the algebra formed by the sets of possible worlds using the atomic propositions---``coin is unbiased'', ``first toss yields $H$'' and ``second toss yields $H$''. If we treat both the type of the coin and what it shows on toss as random events defined respectively through the sample spaces $\mathcal{S}_{c}$ and $\mathcal{S}_t$, then in our problem we are implicitly working in a sample space $\mathcal{S}=\mathcal{S}_c\times\mathcal{S}_t$ such that we can consider as an outcome in $\mathcal{S}$ any ordered pair of `type of coin' and `what it shows'.

Coming back to the example, we note that the conditionalization adopted in the Bayesian updating is \emph{commutative} and \emph{cumulative}. Rather than seeing both the tosses' result together, let us say the agent first sees the first toss's result and updates her prior, and then does the same after subsequent toss; actually, the order of tosses doesn't matter---the updating is commutative. If the conditionalization is cumulative, the sequence of updates $P(U)\rightarrow P(U|H)\rightarrow P(U|HH)$ should yield $0.2$ for $P(U|HH)$. So, after first toss
\begin{eqnarray}
P(U|H)=\frac{P(H|U)P(U)}{P(H)},
\end{eqnarray}
where the terms in the R.H.S. are all known: $P(U)=0.5$ (say), $P(H|U)=0.5$, and $P(H)=P(H|U)P(U)+P(H|B)P(B)=0.75$. Therefore, P(U|H)=1/3. $P(U|H)$ is the new prior that now needs to be updated:
\begin{eqnarray}
P(U|H,&&H)=\frac{P(H|U,H)P(U|H)}{P(H|H)},\\
=&&\frac{P(H|U,H)P(U|H)}{P(H|U,H)P(U|H)+P(H|B,H)P(B|H)},\quad\\
=&&\frac{P(H|U)P(U|H)}{P(H|U)P(U|H)+P(H|B)P(B|H)},\quad\\
=&&\frac{0.5\times \frac{1}{3}}{0.5\times \frac{1}{3}+1\times (1-\frac{1}{3})}=0.2,
\end{eqnarray}
as expected. Here we explicitly use comma to separate the newer data. It is expected that if the true nature of the coin can be divulged by infinite such updatings starting from any arbitrary non-zero prior; e.g., if the coin is actually biased, then the observation data should all be heads only and the sequence $\{P(U|HH\cdots n\,{\rm times})\}_{n\ge1}$ is monotonically decreasing sequence with $\lim_{n\rightarrow\infty}P(U|HH\cdots n\,{\rm times})=0$. In this problem, the true nature of the coin can be divulged by finite updatings as well if a single tail appears; any appearance of the tail immediately makes the posterior unity---implying belief that the coin is unbiased with certainty---irrespective of the value of the non-zero prior.

In the context of the coin toss, we take a digression to answer an interesting question: With what probability any arbitrary pattern consisting of heads and tails appears in an infinite sequence of the (possibly biased but not completely) coin toss experiment? This can be answered using the \emph{Borel--Cantelli lemma}: In a sequence of infinite events $\{E_i\}$ with $\sum_{i=1}^\infty P(E_i)<\infty$, only finitely many of the events $\{E_i\}$ occur with probability one; and in case events $\{E_i\}$ are independent with $\sum_{i=1}^\infty P(E_i)=\infty$, infinitely many of the events $\{E_i\}$ occur with probability one. Consider an event that is repetition of $HT$ (say, $n$ times): $HTHT\cdots HT$. Let $A_{i}$ ($i\in\mathbb{N}$) is the event that $H$ and $T$ shows at $(2i-1)$-th and $(2i)$-th tosses respectively. Consider the event $E_i$ that $HTHT\cdots HT$ appears as a sequence between $i$th and $(i+2n-1)$-th tosses, i.e., $E_i=A_i\cap A_{i+1}\cap\cdots\cap A_{i+n-1}$ with $i\in\mathbb{N}$ and note that $\{E_{kn+1}\}_{k=0}^{\infty}$ is a set of independent events such that $\sum_{k=0}^\infty P(E_{kn+1})=\sum_{k=0}^\infty p^{n}(1-p)^n=\infty$; $p\in(0,1)$ is the probability of getting $H$. Thus, owing to the Borel--Cantelli lemma, the pattern $HTHT\cdots HT$---repetition of $HT$ $n$ times---will repeat infinitely often. This is obviously true for any other pattern one can come up with.

Last but not the least, let us look at a problem that shows how use of Bayes' theorem can be used to rectify the incorrect results that careless implementation of the frequentist's interpretation of probability may lead to. Following is essentially a variant of the coin toss problem but masked under a catchy story. Consider Ameera ($A$) and Bheem ($B$) are playing a game in which the first person to get $5N+1$ ($N$ is a natural number) points wins the game. The way each point is decided is as follows: Their friend Carol ($C$) has a carrom board that $A$ and $B$ can't see. Before the game begins, $C$ rolls an initial ball onto the table, which comes to rest at a completely random position, which $C$ marks. Then, each point is decided by $C$ rolling another ball onto the table randomly. If it comes to rest to the left of the initial mark, $A$ wins the point; otherwise $B$ wins the point. $C$ reveals nothing to $A$ and $B$ except who won each point.  
Imagine $A$ is already winning $5N$ points to $3N$. We want to find out what is the expected probability that a player goes on to win the game. Clearly, the probability that $A$ wins a point is the fraction of the table to the left of the mark---call this probability $p$; and $B$'s probability of winning a point is $1-p$. Because C rolled the initial ball to a random position, before any points were decided every value of $p$ was equally probable. The mark is only set once per game, so $p$ is the same for every point. So, specifically we ask: what is the probability that $B$ will win the game in terms of $p$? 

Let us discuss the case for $N=1$. A frequentist would find $p$ as $5/8$ and hence claim that the probability that $B$ wins the game is $P(B{\rm\,wins})=(1-5/8)^3=27/512$ (because $B$ must win three consecutive rounds to win the game). Simulations show that this is incorrect result. Of course, the problem is in the estimation of $p$ with the limited data available; in principle, any value of $p$ is possible. We must take the bull by the horns and realize that $B$'s winning probability is actually given by: 
\begin{equation}
P(B{\rm\,wins})=\int_0^1(1-p)^3P(p|A=5,\,B=3)dp,
\end{equation}
where $P(p|A=5,\,B=3)$ is the probability that a particular value of $p$ is true given the data that $A$ has 5 points and $B$ has 3 points. Now we use Bayes' theorem:
\begin{equation}
P(p|A=5,\,B=3)=\frac{P(A=5,\,B=3|p)P(p)}{\int_0^1P(A=5,\,B=3|p)P(p)dp}.
\end{equation}
Choosing a (uninformative) uniform prior (by invoking the principle of insufficient reason) and binomial distribution for the likelihood, we arrive at
\begin{equation}
P(B{\rm\,wins})=\frac{\int_0^1p^5(1-p)^6dp}{\int_0^1p^5(1-p)^3dp}=\frac{1}{11}
\end{equation}
---the correct result. If one redoes the problem with $N\rightarrow\infty$, both the approaches give the correct result, now more easily gettable from the (hypothetical) frequentist's view: $P({\rm B\,wins})=(1-1/2)^3=1/8$. One could solve an analogous problem where the first person to get $N+5+1$ points wins the game when $A$ is already winning $N+5$ points to $N+3$; both approaches again converge on to each other as $N\rightarrow\infty$.

Three important characteristics of the Bayesian approach has been brought to the fore by the above example: (i) integration over uncertain parameter potentially leading to computational problems, (ii) choosing a seemingly ad hoc prior leading to objections by frequentists, and (iii) interpreting parameter (or hypothesis) as a random variable to associate a probability with this one-time, non-repeatable event. 
\subsection{Basic Ideas on Random Variables}
A single-valued function $X(s)$ is called a \emph{random variable} (actually a `function' in traditional sense!) if it maps each outcome $s$ of sample space $\mathcal{S}$ to a real number; two distinct outcomes may be mapped to same real number but not vice versa. One can then use the random variable to define events. 

\emph{Cumulative distribution function} (cdf) $F_X(x)$ of $X$ is defined as
\begin{eqnarray}
F_X(x)=P(X\le x),\, x\in(-\infty,+\infty).
\end{eqnarray}
If a cdf changes values only in a countable number of jumps and is otherwise constant between two subsequent jumps, the corresponding $X$ is a \emph{discrete random variable}. Essentially, a discrete random variable is the one for which $\{X(s): s\in\mathcal{S}\}$ is finite or countably infinite. On the other hand, the discrete random variable is \emph{continuous} if $\{X(s): s\in\mathcal{S}\}$ is either a single interval or a set of disjoint intervals of real line. Note that for a continuous random variable, $F_X(x)$ is continuous and its piecewise continuous derivative exists everywhere except maybe at a finite number of points. One can construct a \emph{mixed} random variable that has a cdf with properties of both the discrete and the continuous random variables. The notion of \emph{percentile} is defined via cdf: The $u\,(\in[0,1])$ percentile of $X$ is the smallest real number $x_u$ that satisfies $u=F_X(x_u)$.

For the discrete random variable, one can define \emph{probability mass function} (pmf):
\begin{eqnarray}
p_X(x)=P(X=x),
\end{eqnarray}
where
\begin{eqnarray}
P(X=x_i)=F_X(x_i)-F_X(x_{i-1}).
\end{eqnarray}
Here subscript $i$ denotes the points of jumps for the cdf.

Equivalently for the continuous random variable, one can define \emph{probability density function} (pdf):
\begin{eqnarray}
p_X(x)=\frac{d}{dx}F_X(x).
\end{eqnarray}
One can invert this to get,
\begin{eqnarray}
F_X(x)=P(X\le x)=\int_{-\infty}^{x}p_X(x)dx.
\end{eqnarray}
Although we are using the same symbol for pmf and pdf, it should not be source of a confusion as the context with make it clear.

It is straightforward to define \emph{conditional cdf} of an event given some event $E$:
\begin{eqnarray}
F(x|E)=P(X\le x|E)=\frac{P(\{s:X(s)\le x\}\cap E)}{P(E)}.\quad
\end{eqnarray}
Using this definition, the corresponding definitions for \emph{conditional pmf} or \emph{conditional pdf} follow in analogy with what was done before for the unconditional cases.

An all important concept in the theory of random variable is that of the functions of random variables. Consider a function $f(x)$. Notationally, 
\begin{equation}
Y=f(X)
\end{equation}
defines a new random variable such that for any subset $\mathcal{I}$ with $f(x)\le y$ and $\mathcal{I}\subseteq({\rm range~of~}X)\subseteq\mathbb{R}$ ,
\begin{equation}
\{s:Y=f(X(s))\le y\}=\{s:X(s)\in \mathcal{I}\}.
\end{equation}
Naturally, for continuous random variable,
\begin{eqnarray}
F_Y(y)=P(Y\le y)=P(X\in \mathcal{I})=\int_\mathcal{I}p_X(x)dx\,.\quad
\end{eqnarray}
The pdf of $Y$ can then be obtained through,
\begin{equation}\label{eq:conversionp}
p_Y(y)=\sum_i\frac{p_X(x_i)}{|f'(x_i)|}\,,
\end{equation}
where $x_i$'s are the real roots of $y=f(x)$.
In passing, we note that in case of the discrete random variable the pmf $p_X(x)$ under transformation $y=f(x)$ rather trivially implies
\begin{equation}
p_Y(y)=\sum_{x\in f^{-1}(y)}p_X(x).
\end{equation}

The \emph{expectation} of a function of a random variable is given by
\begin{equation}
E[Y]=E[f(X)]=\sum_if(x_i)p_X(x_i)\,{\rm and}\,\int_{-\infty}^{+\infty}f(x)p_X(x)dx,
\label{eq:expectation}
\end{equation}
for discrete and continuous cases respectively. Expectations of the following three particular functions are useful:
\begin{itemize}
\item \emph{Moment generating function}:
\begin{equation}
M_X(t)\equiv E[e^{tX}],\,t\in\mathbb{R}.
\end{equation}
$M_X(t)$ need not exist as the integral or the sum may not converge for all $t$. The beauty of $M_X(t)$ is that $n$th \emph{moment} $m_n$ ($n\in\mathbb{N}$)---defined as $E(X^n)$---can be obtained from it through
\begin{equation}
m_n\equiv E[X^n]=\left.\frac{d^n}{dt^n}M_X(t)\right|_{t=0}.
\end{equation}
An important result is that if two random variables have same moment generating functions, then they necessarily have same probability distribution. Sometimes it is convenient to work with \emph{central moments} $\mu_n$ ($n=2,3,\cdots$) defined as
\begin{equation}
\mu_n\equiv E[(X-m_1)^n],
\end{equation}
that measures deviation of the random variables from the \emph{mean} $m_1$ (mostly commonly denoted by $\mu$ with no subscript; hence no confusion with the central moments). Mean measures the \emph{central tendency} of a probability distribution, i.e., it is a measure that locates where the data are concentrated. Two other widely used measures of the central tendency are: \emph{median} and \emph{mode}. Median is value of the random variable such that the probability of getting smaller and greater values are equal; and mode is the most probable value of the probability distribution. Note that median is $0.5$ percentile of $X$.
\item \emph{Characteristic function}:
\begin{equation}
\phi_X(t)\equiv E[e^{\sqrt{-1} tX}],\,t\in\mathbb{R}.
\end{equation}
Unlike $M_X(t)$, $\phi_X(t)$ always exists. One can prove that if two random variables have same caracteristic functions, then they necessarily have same probability distribution.
\item \emph{Cumulant generating function}: The cumulant generating function
\begin{equation}
K_X(t)\equiv\ln E[e^{tX}]=\sum_{n=1}^\infty\kappa_n\frac{t^n}{n!},\,t\in\mathbb{R},
\end{equation}
yields \emph{cumulants} $\kappa_n$ through the following relation:
\begin{equation}
\kappa_n=\left.\frac{d^n}{dt^n}K_X(t)\right|_{t=0}.
\end{equation}
The cumulants are related to moments and central moments. In fact, the first cumulant $\kappa_1$ is the mean ($\mu$); and the next two, $\kappa_2$ and $\kappa_3$, are respectively equal to the central moments $\mu_2$ (usually denoted by $\sigma^2$, commonly known as the \emph{variance}) and $\mu_3$ (whose scaled form---$\mu_3/\sigma^3$---is called \emph{skewness}). Square root of the variance is called \emph{standard deviation} and is denoted as $\sigma$, or $\sigma_X$, if the corresponding random variable needs to be emphasized. Positive skewness means the probability distribution is skewed towards right and negative means the distribution is skewed towards left. The fourth cumulant $\kappa_4$ is not equal to any central moment but is related to $\mu_4$ as follows: $\kappa_4=\mu_4-3\sigma^4$. The scaled version of $\kappa_4$ is known as \emph{kurtosis} $\kappa\equiv\kappa_4/\sigma^4$. A positive kurtosis makes the distribution \emph{leptokurtic} meaning that the distribution's tail is heavier than in a \emph{Gaussian} (also called \emph{normal}) distribution $\mathcal{N}(\mu,\sigma^2)$ specified by
\begin{equation}
p_X(x) = \frac{1}{\sqrt{2\pi\sigma^2}}e^{-\frac{(x-\mu)^2}{2\sigma^2}}.\label{eq:gdnd}
\end{equation}
 Whereas a negative value of kurtosis indicates that distribution is \emph{platykurtic} meaning that the tail is lighter than in a Gaussian distribution. (Near zero value for kurtosis means that the distribution is \emph{mesokurtic}. Interestingly, the ``lightness'' or ``heaviness'' in these Greek names do not refer to the tails; rather they refer to the region around the peak: Because of the normalization of probability distribution, while a heavier tail implies a lighter weight in the peak region, a lighter tail implies more weight around the peak region.)
\end{itemize}

Let us mention three extremely important practical results concerning sequences of \emph{iid} (independent, identically distributed) random variables. While the mathematically precise meaning of the term iid is defined later but the idea should be clear from the following intention that we have: Our intention is to say something concrete about a sample that contains a sequence ${X_1,X_2,\cdots,X_N}$ (called \emph{sample vector} or \emph{random sample};  to be denoted as $\{X_i\}_{i=1}^{i=N}$ or simply $\{X_i\}$) of $N$ observations for a particular experimental setup or phenomenon. Because we are interested in a particular fixed experimental setup, it is natural to assume that all the observations are modelled by identical random variables. Furthermore, assume that each observation is randomly generated independent of the other observations. One can define the \emph{sample mean}, $\bar{X}_N\equiv(\sum_i^NX_i)/N$, for the sample. If many such samples are collected, what is the statistical behaviour of the sample mean?
\begin{itemize}
\item \emph{(Strong) Law of large numbers}: If each of the iid random variables have a finite mean $\mu$, then for any positive $\epsilon$,
\begin{equation}
P\left(\lim_{N\rightarrow\infty}|\bar{X}_N-\mu|>\epsilon\right)=0.
\end{equation}
It essentially says that the sequence of sample means converges to the mean of random variable $X_i$ with full certainty. The weak law---$\lim_{N\rightarrow\infty}P\left(|\bar{X}_N-\mu|>\epsilon\right)=0$---can be inferred from the strong law. If $E[X_i]=\infty$, the strong law does not hold but using the idea of conditionally convergent series, sometimes a finite expectation value may be defined and the weak law may be seen to be obeyed with convergence towards the redefined expectation value.
\item \emph{Central limit theorem} (CLT): If each of the iid random variables have a finite mean $\mu$ and a finite variance $\sigma^2$, then 
\begin{equation}
\lim_{N\rightarrow\infty}\left(\frac{\bar{X}_N-\mu}{\sigma/\sqrt{N}}\right)\xrightarrow[]{\text{distribution}}\mathcal{N}(0,1),
\end{equation}
i.e.,  asymptotic distribution of the random variable, $\bar{X}_N$, is Gaussian with mean $\mu$ and variance $\sigma^2/N$.

The CLT provides more detailed description about the sample mean than the laws of large numbers. It allows to approximately find a probability of $P(\bar{X}_N>x)$ because we know that we just have to use $\mathcal{N}(0,1)$. However, such an approximation by the CLT will not be very accurate in the tails---where $x$ is quite distant from $\mu$---if $N$ is not large enough. Moreover, the CLT does not furnish any detail about the convergence of the tail probabilities with the sample size. This is where a theorem from the theory of large deviations comes to the rescue.
\item \emph{Cram\'er's theorem}: If each of the iid random variables have a finite mean $\mu$ and a cumulant generating function $K_{X_i}(t)=\ln E[e^{tX_i}]$, then $\bar{X}_N$ satisfies the \emph{large deviation principle}, i.e.,
\begin{equation}
\lim_{N\rightarrow\infty}-\frac{1}{N}\ln P(\bar{X}_N\ge x)=I(x)\,\,\forall x>\mu,
\end{equation}
where $I(x)$---the Legendre--Fenchel transform of $K(t)$, i.e., $I(x)\equiv\sup_t(tx-K_{X_i}(t))$---is called \emph{rate function} or \emph{Cram\'er's function}. Sometimes, this is less rigorously presented as 
\begin{equation}
P(\bar{X}_N\ge x)\approx e^{-NI(x)},
\end{equation}
where `$\approx$' sign means that only the dominant exponential term has been kept and any sub-exponential terms have been ignored (hence the limit in the rigorous definition).
\end{itemize}

Readers must have already realized a technical point: $\bar{X}_N$ is defined on combined sample space---$\mathcal{S}^N\equiv\mathcal{S}_1\times \mathcal{S}_2\times\cdots\times\mathcal{S}_N$---where $\mathcal{S}_i$ is the sample space corresponding to the random variable $X_i$. Thus, by independent random variable we mean that $X_i(s_1,s_2,\cdots,s_N)=X_i(s_i)$, where $s_i\in\mathcal{S}_i$. Moreover, if $\mathcal{S}_i=\mathcal{S}_j=\mathcal{S}$ $\forall i,j$ and $X_i(s_i)=X(s_i)$ $\forall i$ (where $X$ is some random variable defined on $\mathcal{S}$), then $X_i$'s can be said to be identically distributed. This way of looking at multiple random variables begs the question what if two or more random variables are not independent or not identical; more generally, how to generalize the concepts in single random variable theory to multiple random variables. While all the preceding ideas can be generalized to the case of multiple random variables, some newer concepts emerge. Let us list them. For convenience, we shall only work with the two random variables unless specified otherwise; further generalization is straightforward but notationally cumbersome.

First of all, we must realise that, for consistency of the concepts below, we should define a \emph{bivariate} or \emph{two-dimensional} random variable $(X,Y)$ on the same sample space. ($Y$ should not be confused as the function of $X$ as used earlier in this section.) The \emph{joint cumulative distribution function} can be defined as
\begin{eqnarray}
F_{XY}(x,y)=P(X\le x,Y\le y);\,x,y\in(-\infty,+\infty).\qquad
\end{eqnarray}
The \emph{joint probability mass function} is
\begin{eqnarray}
p_{XY}(x,y)=P(X=x,Y=y),
\end{eqnarray}
and \emph{joint probability density function} is
\begin{eqnarray}
p_{XY}(x,y)=\frac{\partial^2}{\partial x\partial y}F_{XY}(x,y).
\end{eqnarray}
One defines \emph{marginal cumulative distribution function} as
\begin{eqnarray}
F_X(x)=\lim_{y\rightarrow\infty}F_{XY}(x,y)=P(X\le x,Y\le \infty).
\end{eqnarray}
Furthermore, \emph{conditional} pmf or pdf can now be defined as
\begin{equation}
p_{X|Y}(x|y)\equiv\frac{p_{XY}(x,y)}{p_Y(y)},\,p_Y(y)\ne0.
\end{equation}
Symmetry of $p_{XY}(x,y)$ in the arguments $x$ and $y$, allows to write \emph{Bayes' rule} for the random variables:
\begin{equation}
p_{X|Y}(x|y)\equiv\frac{p_{Y|X}(y|x)p_X(x)}{p_Y(y)},\,p_Y(y)\ne0.
\end{equation}

In the above notations, the random variables $X$ and $Y$ are defined to be independent if
\begin{eqnarray}
F_{XY}(x,y)=F_X(x)F_Y(y),
\end{eqnarray}
and then, for such variable, we find that 
\begin{eqnarray}
&&p_{XY}(x,y)=p_X(x)p_Y(y), \\
&&p_{X|Y}(x|y)=p_X(x)\,{\rm and}\,p_{Y|X}(y|x)=p_Y(y).
\end{eqnarray}
Here, $p_X(x)$ and $p_Y(y)$ are the marginal versions obtained after summing (in the case of pmf) or integrating (in the case of pdf) over all possible values of $y$ and $x$ respectively.

One can furthermore define $(i,j)$th moment as
\begin{equation}
m_{ij}\equiv E[X^iY^j].
\end{equation}
Similarly, one can define bivariate central moments and cumulants; and also different generating functions. Two concepts are of particular interest: \emph{orthogonal} $X$ and $Y$, and \emph{uncorrelated} $X$ and $Y$. $X$ and $Y$ are said to be orthogonal if $m_{11}=0$; and they are said to be uncorrelated (otherwise, they are correlated) if their \emph{covariance} ${\rm Cov}(X,Y)$ (or $\sigma_{XY}$), given by
\begin{equation}
\sigma_{XY}\equiv E[(X-m_{10})(Y-m_{01})],
\end{equation}
is zero. Note that since
\begin{equation}
\sigma_{XY}= E[XY]-E[X]E[Y],
\end{equation}
independent random variables are uncorrelated but the converse need not be true. It may be pointed out that (Pearson's) correlation coefficient,
\begin{equation}
 {\rm Corr}(X,Y)\equiv\frac{\sigma_{XY}}{\sigma_X\sigma_Y}\,,
\end{equation}
which ranges between $-1$ to $+1$, quantifies the degree to which the two random variables are linearly correlated:  ${\rm Corr}(X,Y)=1$ means completely linear increasing relationship, while ${\rm Corr}(X,Y)=-1$ means completely linear decreasing relationship.

We end this section by introducing a rather useful special multivariate random distribution: \emph{$n$-variate Gaussian (normal) distribution}. Denoting $\boldsymbol{X}$ as a column vector $[X_1\,X_2\,\cdots\,X_n]^T$ and its particular realization as $\boldsymbol{x}\equiv [x_1\,x_2\,\cdots\,x_n]^T$, we define $\{X_i\}_{i=1}^{i=n}$ an $n$-variate Gaussian (normal) random variable if its joint pdf is
\begin{equation}
p_{\boldsymbol{X}}(\boldsymbol{x})=\mathcal{N}_n(\boldsymbol{\mu},{\sf C})\equiv\frac{1}{\sqrt{(2\pi)^n|\textsf{C}|}}e^{-\frac{1}{2}(\boldsymbol{x}-\boldsymbol{\mu})^T{\sf C}^{-1}(\boldsymbol{x}-\boldsymbol{\mu})},
\end{equation}
where $\boldsymbol{\mu}$ is a column vector $[E[X_1]\,E[X_2]\,\cdots\,E[X_n]]^T$ and ${\sf C}_{ij}\equiv\sigma_{X_iX_j}$ is the $(i,j)$-th element of the $n\times n$ matrix ${\sf C}$---the \emph{covariance matrix}---whose determinant has been denoted by $|{\sf C}|$ in the pdf. The set of such $n$ variables are equivalently termed \emph{jointly normal}, an apt term because the sum $\sum_{i=1}^{n}k_ix_i$ can be shown to be a normal random variable for any set of numbers, $k_i$'s. We remark that if additionally the variables are uncorrelated, then they are independent as well. Moreover, the central limit theorem extended to the case of multivariate random variables says: $\lim_{N\rightarrow\infty}\sqrt{N}\left({\bar{\boldsymbol{X}}_N-\boldsymbol{\mu}}\right)\xrightarrow[]{\text{distribution}}\mathcal{N}_n(\boldsymbol{0},{\sf C})$.
\subsection{Bare Basics of Markov Chain}
A (real) \emph{stochastic or random process} is a single valued function $X(s,t)$ that maps each outcome $s$ of an experiment, with a sample space $\mathcal{S}$, at each parameter (usually, time) value $t$ of a parameter set $\mathcal{T}\subseteq\mathbb{R}$ to a real number: $X:\mathcal{S}\times\mathcal{T}\rightarrow\mathbb{R}$. If $\mathcal{T}$ is a countable set, then the process is \emph{discrete-parameter process}; if $\mathcal{T}$ is an interval of $\mathbb{R}$, then the process is \emph{continuous-parameter process}. A discrete-parameter chain is more conspicuously denoted by $X_n(s)$  ($n\in\mathbb{Z}$) rather than $X(s,t)$. It is customary to omit the reference to the outcome $s$ in the random process and denote it simply as $X(t)$ (or $X_n$). This is rather a relief as  specifying the outcome a priori is explicitly neither easy nor practically useful: An $s\in\mathcal{S}$ for practical purpose merely symbolizes a sequence of experimental realization (outcome),  say, $(X_0(s),X_1(s),X_2(s),X_3(s))$.  The elements or outcomes of sample space $\mathcal{S}$, thus, are able to specify abstractly every possible experimental realization or outcome. E.g., in an experiment of tossing a unbiased coin two times, $\mathcal{S}$ is not $\{H,T\}$ but set of all possible sequences of heads and tail---$\{(H,H), (H,T),(T,H),(T,T)\}$; of course, any of the two steps of the experiment is modelled by a random variable with sample space $\{H,T\}$.

At a fixed $t$, $X(t)$ is a random variable that can take all possible allowed values to which at least one of the experimental outcomes (not to be confused with $s$ mentioned above) at that $t$ is associated; every such value is called a \emph{state}. The set of all possible states across $\mathcal{T}$ is called \emph{state space} $\mathcal{X}$ of the process. If the state space is countable, then the process is \emph{discrete-state process} (commonly referred as \emph{chain}); otherwise, the process is \emph{continuous-state process}. For a fixed $s$, $X(t)$ is a \emph{sample/sample-path/sample-function/realization} of the process. Finally, with neither $s$ nor $t$ held fixed, $X(s,t)$ can be seen as specifying an \emph{ensemble} of time functions depending on outcome: Ensemble is the collection of all possible sample paths.

For the probabilistic description of a stochastic process, \emph{$n$th-order distribution}---defined as the joint distribution cdf
\begin{equation}
F(\{x_i\}_{i=1}^{i=n};\{t_i\}_{i=1}^{i=n})\equiv P(\{X(t_i)\le x_i\}_{i=1}^{i=n})
\end{equation}
---is a useful theoretical quantity. In principle, for the complete knowledge of $X(t)$ is equivalent to knowing all the $n$th-order distributions formed using all possible $t_i$'s. In place of such an impractical specification, following few quantities are commonly used to (partically) specify $X(t)$:
\begin{itemize}
\item \emph{Ensemble average} or \emph{mean}: $\mu_X(t)\equiv E[X(t)]$.
\item \emph{Autocorrelation}: $R_X(t,t')\equiv E[X(t)X(t')]$.
\item \emph{Autocovariance}: $C_X(t,t')\equiv\sigma_{X(t)X(t')}=R_X(t,t')-\mu_{X}(t)\mu_{X}(t')$.
\end{itemize}
An ideal but widely used important process is that of the \emph{white noise}: It is a stochastic process such that its autocovariance is of the form $C(t,t')=q(t)\delta(t-t')$ (or $C(n,n')=q(n)\delta_{nn'}$ for discrete-time process in obvious notations); usually, the mean of the process is taken as identically zero.

Fortunately, for some random processes of interest, their complete specification requires one to know a much smaller set of distributions than all possible $n$-th order distributions. Some of the important ones are as follows:
\begin{itemize}
\item \emph{Independent process}: If $X(t_i)$ and $X(t_j)$ are independent random variables $\forall i\ne j$, then $X(t)$ is an independent process. In this case, obviously any $n$th-order distribution can be written as a product of the first-order distributions. If  an independent process is white noise, then it is also called \emph{strictly white noise}.
\item \emph{$k$th order stationary process}: A stochastic process $X(t)$ is $k$th order stationary if $\forall i\le k$ and for every possible values of the set of parameters $\{t_i\}_{i=1}^{i=k}$, $F(\{x_i\}_{i=1}^{i=k};\{t_i\}_{i=1}^{i=k})=F(\{x_i\}_{i=1}^{i=k};\{t_i+\tau\}_{i=1}^{i=k})$ $\forall\tau\in\mathbb{R}$. 
\item \emph{Strict-sense stationary process}: Sometimes simple called the \emph{stationary process}, it satisfies the condition for the $k$th-order stationarity for all possible values of $k$. Evidently, since the $1$st-order distribution must be time-independent, the ensemble average, the variance $C_X(t,t)$  and similar quantities derived from the distribution must be constants. The quantities, like autocorrelation, calculated using the $2$nd-order distribution must depend on the time-difference only, i.e., $R_X(t,t')=R_X(|t-t'|)$. For stationary processes, sometimes \emph{mean-ergodicity} is assumed, i.e., it is assumed that the time-average of a sample (as time goes to infinity) is equal to its corresponding ensemble average; if the assumption of equality is true for any measurable function for the random variable, then we can say that \emph{ergodicity} has been assumed or that the ergodicity with respect to the probability distribution function has been assumed---the most strict notion of ergodicity. It is possible to have non-stationary processes that are mean-ergodic (and even additionally \emph{autocorrelation-ergodic}).
\item\emph{Weak-} or \emph{wide-sense stationary process}: Also aptly known as \emph{covariance stationarity}, it is defined through the collective requirements of finite variance (at all times), time independent mean, and time-difference-dependent autocorrelation. All strict-sense stationary processes are wide-sense stationary processes (but not vice-versa).
\item\emph{Gaussian} or \emph{normal process}: A stochastic process $X(t)$ is a Gaussian process, if for any $n$ and any $\{t_i\}_{i=1}^{i=n}\subseteq \mathcal{T}$, the sequence of random variables $\{X(t_i)\}_{i=1}^{i=n}$ is jointly normal. It is completely specified by the $2$nd-order distributions. A wide-sense stationary Gaussian process is strict-sense stationary as well.
\item\emph{Markov process}: Another important process that is completely specified by the $2$nd-order distributions is Markov process defined as the stochastic process such that $P(X(t_{n+1})\le x_{n+1}|\{X(t_i)=x_i\}_{i=1}^{i=n})=P(X(t_{n+1})\le x_{n+1})|X(t_n)=x_n)$ $\forall n$ and $\forall x_i$'s. It is also called \emph{memoryless process}. In the light of the commonly found definition of \emph{$p$th-order Markov process} that obeys $P(X(t_{n+1})\le x_{n+1})|\{X(t_i)=x_i\}_{i=1}^{i=n})=P(X(t_{n+1})\le x_{n+1})|\{X(t_i)=x_i\}_{i=n-p+1}^{i=n})$ $\forall n$ and $\forall x_i$'s, the memoryless process is also called $1$st-order Markov process and the $0$th-order Markov process is actually independent process.

\item\emph{Process with stationary independent increments}: It is a special class of \emph{non-stationary} Markov process $X(t)$ such that it has stationary increments---i.e., $X(t)-X(t')$ and $X(t+\epsilon)-X(t'+\epsilon)$ have same distribution $\forall t,t'\in\mathcal{T}$ with $t>t'$ and $\forall \epsilon\ge0$---and independent increments---i.e., $\forall t_i\in\mathcal{T}(=\mathbb{R}_{\ge0}$,~say, for convenience) with $0<t_1<t_2<\cdots<t_n$, $X(0),\,X(t_1)-X(0),\,X(t_2)-X(t_1),\cdots,\,X(t_n)-X(t_{n-1})$ are independent.
\end{itemize}

One specific continuous-time random process---\emph{Wiener process} or \emph{Brownian motion process} $B(t)$---is of immense practical interest. Mathematically, $B(t)$ with $t\in\mathcal{T}=\mathbb{R}_{\ge0}$ is the Wiener process if (i) $B(0)=0$, (ii) $E[B(t)]=0$, and (iii) $B(t)$ has stationary independent increments that are normally distributed. One can show that $B(t)$ is \emph{continuous in probability}, i.e., $\forall\epsilon>0$ and $\forall t\in\mathcal{T}$, $P(|B(t+h)-B(t)|>\epsilon)=0$ when $h\to0$. It is easy to see that $B(t)$ is a Gaussian process with variance proportional to $t$; when the proportionality constant is unity, $B(t)$ is called \emph{standard} Wiener/Brownian process. For standard $B(t)$, the covariance is $\min(t,t')$ $\forall t,t'\in\mathcal{T}$---a signature of non-stationarity. In fact, the Wiener process is a Markov process as well. This mathematical construction of $B(t)$ and its properties appear rather intuitive when $B(t)$ is seen as a limit of a random walk which we now proceed to learn.

The \emph{(simple) random walk}, $X_n$ with $n\in\mathcal{T}=\mathbb{Z}_{\ge0}$, is a discrete-state, discrete-time stochastic process defined as $X_0\equiv0$ and $X_n\equiv\sum_{i=1}^nY_i$ where $\{Y_i\}$ is a sequence of iid random variables with $P(Y_i=+1)=p$ and $P(Y_i=-1)=1-p$ $\forall i\in\mathbb{N}$. Note it is a Markov process since
\begin{eqnarray}
&&P(X_{n+1}=x_{n+1}|X_0=0,\{X_i=x_i\}_{i=1}^{i=n})\nonumber\\
=&&P(Y_{n+1}+x_n=x_{n+1}|X_0=0,\{X_i=x_i\}_{i=1}^{i=n})\nonumber\\
=&&P(Y_{n+1}=x_{n+1}-x_n)\nonumber\\
=&&P(X_{n+1}=x_{n+1}|X_n=x_n).
\end{eqnarray}
Thus, random walk is a discrete-time \emph{Markov chain}. 

A \emph{symmetric random walk} is the random walk with $p=0.5$. It can be shown to converge to the standard Wiener process in the limit of very large $n$ and infinitesimal step size as follows: Consider a stochastic jump process, $Z_{\delta,\varepsilon}(t)\equiv\{\varepsilon X_{\floor*{t/\delta}}:t\in\mathbb{R}_{\ge0}\}$ ($\delta,\varepsilon\in\mathbb{R}^+$), with jumps of size $\pm\varepsilon$ at every $k\delta$ ($k\in\mathbb{Z}_{\ge0}$).  It is obvious that $E[Z_{\delta,\varepsilon}(t)]=0$ and $E[Z_{\delta,\varepsilon}(t)^2]=\varepsilon^2\floor*{t/\delta}$ because $E[X_n]=0$ and $E[X_n^2]=n$. In the light of the CLT, the choice of $\varepsilon=\sqrt\delta$ ensures that $Z_{\delta,\sqrt\delta}(t)\rightarrow\mathcal{N}(0,t)$ as $\delta\rightarrow0$. It follows from the nature of the normal distribution that the increments of  $Z_{\delta,\sqrt\delta}(t)|_{\delta\rightarrow0}$ should be normally distributed as well. We also hope (without rigorous proof) that the property of the stationary independent increments that $X_n$ is endowed with, owing to being the partial sum of iid random variables, carries over to the continuous time process in the limiting process discussed above. Furthermore, by construction, $Z_{\delta,\sqrt\delta}(0)|_{\delta\rightarrow0}$ is obviously zero. In conclusion, $Z_{\delta,\sqrt\delta}(t)|_{\delta\rightarrow0}$ is the standard Wiener process.

The random walk has its own huge literature and practical applications; so has the topic of Markov chain whose example is the random walk. Let us elaborate on the concepts related to discrete-time Markov chain which for brevity will be called Markov chain henceforth. Without any loss of generality and for the simplification of notations, the state space $\mathcal{X}$ of a Markov chain is taken as $\mathbb{Z}$ or $\mathbb{Z}_{\ge0}$ or $\mathbb{N}$ (or a finite subset of these, in which case the chain is termed \emph{finite}) and the specific $N$ possible states are denoted by $i$, $j$, $k$, $l$ etc.; consequently, $P(X_n=i)$ is to be read as the probability of the system to be in state $i$ at time-step $n$. Thus, the fact that in a Markov chain the probability of the system to be in a future state, conditioned on the present state, is independent of the past states is encapsulated in the fact that the \emph{one-step transition probabilities} $p_{ij}^{(1)}(n)=p_{ij}(n)\equiv P(X_{n+1}=j|X_n=i)=P(X_{n+1}=j|X_n=i,X_{n-1}=k,\cdots,X_0=l)$ $\forall n,i,j,k,\cdots,l$. $p_{ij}$ can be thought of as the $(i,j)$-th element of an $N$-dimensional (possibly infinite) square matrix ${\sf P}(n)$, called the \emph{transition probability matrix} (also \emph{Markov matrix, stochastic matrix, transition matrix, and probability matrix}). Obviously, $\sum_{j\in\mathcal{X}}p_{ij}=1\,\forall i\in\mathcal{X}$. One furthermore defines \emph{state probability (row) vector} at time $n$ as ${\boldsymbol{\pi}(n)}=[\pi_0(n)\,\pi_1(n)\,\cdots\,\pi_{N-1}(n)]\equiv[P(X_n=0)\,P(X_n=1)\,\cdots\,P(X_n=N-1)]$  that specifies the probability distribution of the chain at time $n$. 

If ${\sf P}(n)$ is independent of $n$, i.e., \emph{stationary} (not to be confused with stationarity of a chain), then the Markov chain is \emph{time-homogeneous}; the chain is \emph{nonhomogeneous} otherwise. However, this confusing dual meaning of the term---`stationary'---is somewhat unified on noting that applying Bayes' rule on $n$th order joint pmf, any stationary Markov chain can easily be shown to be time-homogeneous (but not vice versa).  Henceforth, we consider only (time-)homogeneous Markov chain unless specified otherwise. We note that
\begin{eqnarray}
&&P(X_{n+1}=j)=\sum_{i\in\mathcal{X}} P(X_{n+1}=j|X_n=i)P(X_n=i),\,\,\quad\\
&&\implies\pi_j(n+1)=\sum_{i\in\mathcal{X}} p_{ij}\pi_i(n),\\
&&\implies\boldsymbol{\pi}(n+1)=\boldsymbol{\pi}(n){\sf P},\\
&&\implies\boldsymbol{\pi}(n)=\boldsymbol{\pi}(0){\sf P}^n,
\end{eqnarray}
meaning that any time-homogeneous Markov chain is completely specified by initial probability vector $\boldsymbol{\pi}(0)$ and the transition matrix ${\sf P}$, i.e., equivalently, by the 2nd order distributions. For example, the $3$rd order joint pmf $P(X_2=i,X_1=j,X_0=k)=P(X_2=i|X_1=j,X_0=k)P(X_1=j,X_0=k)=P(X_2=i|X_1=j)P(X_1=j|X_0=k)P(X_0)=\pi_k(0)p_{kj}p_{ji}$; and so on for any higher order joint pmf.

Similarly, the \emph{$n$-step transition probabilities}, $p_{ij}^{(n)}\equiv P(X_n=j|X_0=i)$, are merely the elements of ${\sf P}^n$ as can be seen from the following. Suppose the system reaches state $k$ from state $i$ after $n$ steps and then reaches state $j$ after $n'$ more steps, then because of the Markov property, $p_{ij}^{(n+n')}$ can be obtained by summing over all possible intermediate states $k$; therefore,
\begin{eqnarray}
p_{ij}^{(n+n')}=\sum_{k\in\mathcal{X}}p_{ik}^{(n)}p_{kj}^{(n')},
\end{eqnarray}
known as the \emph{Chapman--Kolmogorov equation}. For $n=n'=1$, we note $p_{ij}^{(2)}$'s are merely the elements of ${\sf P}^2$. Extending the argument inductively, it is obvious that $p_{ij}^{(n)}$ are the elements of ${\sf P}^n$ and the Chapman--Kolmogorov equation is essentially the matrix identity, ${\sf P}^{n+n'}={\sf P}^{n}{\sf P}^{n'}$ $\forall n,n'\in\mathbb{Z}_{\ge0}$ (${\sf P}^0$ is taken to be identity matrix ${\sf I}$). 

One of the most interesting question in the theory of Markov chain is that of the \emph{limiting distribution}, $\boldsymbol{\pi}^\infty$ that is such a distribution that $\lim_{n\rightarrow\infty}\boldsymbol{\pi}(n)=\boldsymbol{\pi}^\infty$ $\forall \boldsymbol{\pi}(0)$. By definition, $\boldsymbol{\pi}^\infty$ is insensitive to $\boldsymbol{\pi}(0)$ and hence it can also be equivalently defined in component form as $\pi^\infty_j=\lim_{n\rightarrow\infty}p^{(n)}_{ij}$ $\forall i,j\in\mathcal{X}$. It means that the rows of ${\sf P}^n$ are more and more similar to the limiting state probability vector as $n$ becomes larger and larger; and consequently $\lim_{n\rightarrow\infty}\boldsymbol{\pi}(0){\sf P}^n\rightarrow\pi^\infty$ $\forall \boldsymbol{\pi}(0)$. The limit distribution is usually found by looking for \emph{stationary distribution} (or \emph{steady-state distribution}) $\boldsymbol{\pi}^*$ that is defined as the one that satisfies the following relation: $\boldsymbol{\pi}^*=\boldsymbol{\pi}^*{\sf P}$. Note that a time-homogeneous Markov chain is stationary iff the distribution $\boldsymbol{\pi}(0)$ is a stationary distribution of the Markov chain. $\boldsymbol{\pi}^\infty$ is a  $\boldsymbol{\pi}^*$ (but the converse need not be true) since
\begin{eqnarray}
&&\boldsymbol{\pi}^\infty=\lim_{n+1\rightarrow\infty}\boldsymbol{\pi}(0){\sf P}^{n+1},\\
\implies&&\boldsymbol{\pi}^\infty=\left[\lim_{n\rightarrow\infty}\boldsymbol{\pi}(0){\sf P}^n\right]{\sf P},\\
\implies&&\boldsymbol{\pi}^\infty=\boldsymbol{\pi}^\infty{\sf P}.
\end{eqnarray}
We naturally intuit that the answer to the question of the existence of the limiting distribution in a Markov chain should depend on the nature of the states in the Markov chain. 

Some important definitions classifying the states are as follows:
\begin{itemize}
\item \emph{Accessible state}: A state $j$ is accessible from a state $i$ (denoted by $i\rightarrow j$) if $p^{(n)}_{ij}>0$ for some $n\in\mathbb{Z}_{\ge0}$. It follows that any state is accessible from itself because $p^{(0)}_{ii}=1$. If $i\rightarrow j$ and $j\rightarrow i$, i.e., notationally $i\leftrightarrow j$, the two states are said to \emph{communicate}. 
\item \emph{Recurrent} or \emph{persistent state}: State $i$ is recurrent if the probability $f_{ii}$, that the system starting from state $i$ comes back to $i$, is unity. General $f_{ij}$ is defined as follows: Let random variable, \emph{first passage time}, $\tau_j>0$ be the number of steps (starting from zero time) required to visit state $j$ for the first time. ($\tau_j=\infty$ means that state $j$ is never visited.) Define \emph{first passage probability in $n$ steps}, $f^{(n)}_{ij}\equiv P(\tau_j=n|X_0=i)$. Consequently, the probability of visiting state $j$ eventually starting from state $i$ is the \emph{first passage probability} $f_{ij}=\sum_{n=0}^{\infty}f_{ij}^{(n)}=P(\tau_j<\infty|X_0=i)$. For $j=i$, the sequence $\{f^{(n)}_{ii}\}$ is aptly said to give distribution of \emph{recurrence times}. A recurrent state is further classified as follows: If  the \emph{mean recurrence time} for state $i$, $E[\tau_i|X_0=i]<\infty$, then the recurrent state $i$ is \emph{positive} or \emph{nonnull recurrent}; otherwise the recurrent state is \emph{null recurrent}. This finer classification of the recurrent states makes sense only for infinite Markov chains; a null recurrent state in a finite Markov chain is obviously an impossibility.
\item \emph{Transient state}: State $i$ is transient if $f_{ii}<1$; i.e., there is a finite probability that it will not be visited ever again.
\item \emph{Periodic state}: State $i$ is said to be periodic if the period of the state, $d(i)\equiv{\rm gcd}\{n\ge 1: p_{ii}^{(n)}>0\}>1$. Thus, it is a recurrent state such that only certain recurrence times (with $d(i)$ as their greatest common divisor) are allowed. $d(i)$ itself may not be a recurrence time.
\item \emph{Aperiodic state}: State $i$ is said to be aperiodic if $d(i)=1$. Thus, it is a recurrent state such that all recurrence times are allowed. It is easily argued that any state $i$ with $p_{ii}\ne0$ is aperiodic.
\item \emph{Absorbing state}: State $i$ is termed absorbing if $p_{ii}=1$; the state is \emph{nonabsorbing} otherwise. By definition, an absorbing state must be recurrent state (not vice versa) and aperiodic state (not vice versa).
\item \emph{Ergodic state}: State $i$ is called ergodic if it is aperiodic and positive recurrent. 
\end{itemize}

We realize that every state in a Markov chain must either be transient or recurrent. Think about a recurrent state: It is such that with full certainty it will be revisited; by Markov property, once revisited, it will be re-revisited with probability one, and so on. This means that the state will be visited infinitely many times with unit probability and hence, the mean of number of returns to the state must be infinity. Now think about a transient state: It is such a state that with a finite probability $1-f_{ii}$ it will not be revisited. If we refer `not returning' to the state as `success', then sequence of $k$ `failures' followed by a success is equivalent to a sequence independent Bernoulli trials with probability $f_{ii}^{k}(1-f_{ii})$.Therefore, the mean number of failures or returns is $\sum_{k=0}^{\infty}kf_{ii}^{k}(1-f_{ii})=f_{ii}/(1-f_{ii})$---a finite number. From a different angle, mean number of returns to any arbitrary state $i$ is nothing but $\sum_{n=0}^{\infty}p^{(n)}_{ii}$. Therefore, recurrent state and transient state are, respectively, equivalent to the states satisfying $\sum_{n=0}^{\infty}p^{(n)}_{ii}=\infty$ and $\sum_{n=0}^{\infty}p^{(n)}_{ii}<\infty$. Along similar line of arguments, it is obvious that given $j\rightarrow i$, the mean number of times one reaches $i$, starting from $j$ should be given by $\sum_{n=0}^{\infty}p^{(n)}_{ji}$ and this should be finite if $i$ is transient. But for finiteness (i.e., convergence) of this sum, it must be that ${\lim_{n\rightarrow\infty}}p^{(n)}_{ji}=0$. Note that trivially $p^{(n)}_{ji}=0$ $\forall n$ if $j\nrightarrow i$. Summing up, if $i$ is transient, then ${\lim_{n\rightarrow\infty}}p^{(n)}_{ji}=0$ $\forall j\in\mathcal{X}$ necessarily. We mention, without proof, that this limit is also satisfied by any null recurrent state which is, thus, such a state that it is returned infinitely often with probability one but the probability that it will be occupied as $n\rightarrow\infty$ is zero.

Having formed some intuition about the recurrent and the transient states, we can now present a rather simple proof of an important proposition that \emph{any finite Markov chain has at least one positive recurrent state}: Observe that $\sum_{j\in\mathcal{X}}p^{(n)}_{ij}=1$ $\forall i,n$ because starting from a state, one must be somewhere in the state space at any $n$. Taking the limit $n\rightarrow\infty$ on both sides and letting the limit enter the summation (possible under the assumption of finite Markov chain), we get $\sum_{j\in\mathcal{X}}\lim_{n\rightarrow\infty}p^{(n)}_{ij}=1$. If every $j$ is either transient or null recurrent state, then the L.H.S. must be zero which is an impossibility. Hence, at least one $j$ must be a positive recurrent state. 

The relation `$\leftrightarrow$' (being reflexive, symmetric, and transitive) is an equivalence relation and hence all the states of the Markov chain can be partitioned into \emph{communicating classes}. Note that if two states are in the same class, then either both of them are recurrent or both of them are transient. If two recurrent states are in the same class, then either both of them are null or both of them are nonnull. On a similar note, all the states of the same class have same period. Thus, it makes sense to talk about null recurrent class, nonnull/positive recurrent class, transient class, periodic class of a particular period, and aperiodic class; the definitions are obvious. 

Recurrent (null or nonnull) class is special in the sense that it is \emph{closed}, i.e., in case $i$ is recurrent then $p_{ij}=0$ where $j$ is any state not the equivalent class of $i$. This is because if it is so that $p_{ij}\ne0$ and ergo $i\rightarrow j$, then we must have $j\rightarrow i$ since $i$ is recurrent; but this is not possible since, by assumption, $j$ is not in the equivalent class of $i$. Summarizing, $p_{ij}\ne0$ is an impossibility. Thus, if there are more than one equivalent recurrent classes, one could just practically assume that the corresponding Markov chain has only that equivalent class where the chain starts initially. 

A Markov chain with one and only one communicating class is known as \emph{irreducible} or \emph{indecomposable}. An irreducible Markov chain consisting of only ergodic states is called \emph{ergodic Markov chain}. Thus, irreducible, aperiodic, and positive recurrent Markov chain is ergodic Markov chain. Note again that we can use the adjectives---aperiodic, nonnull recurrent, etc.---without any ambiguity, for the chain itself that is irreducible as the adjectives are class properties and we have a single class.

Now coming to the issue of existence and uniqueness of stationary distribution, it is known that \emph{in an irreducible Markov chain, a stationary distribution exists iff all the states are positive recurrent; furthermore, the $\boldsymbol{\pi}^*$ is unique with ${\pi}^*_i=(E[\tau_i|X_0=i])^{-1}$}. (For comparison, recall that for null recurrent states, $E[\tau_i|X_0=i]=\infty$.) It should be noted that the condition of irreducibility is crucial for the uniqueness. For example, recall that otherwise one could have the possibility of many positive recurrent classes; any such class (being closed) could lead to a stationary distribution formed by finding the appropriate values of the probability vector's components corresponding to the class and associating zeros to the other components' values for the states outside the class.  It is furthermore interesting to observe that there is no condition on the periodicity of the chain. Note that for finite chain, irreducible condition is enough to guarantee the existence and the uniqueness of the stationary distribution. This is so because, in the light of the fact that any finite Markov chain has at least one positive recurrent state, a finite irreducible Markov chain must have exclusively positive recurrent states. 

Actually, any (time-homogenous) finite Markov chain has at least one stationary state. This may be argued as follows: The row transition matrix has a trivial right-eigenvector---one with unity as all components---with right-eigenvalue equal to one. Since the right and the left eigenvalues of any square matrix are same, there must be a left-eigenvector, corresponding to the unity eigenvalue, that serves as the stationary distribution. It is in line with Brouwer's fixed-point theorem (stating that for any continuous function mapping a nonempty compact convex set to itself, there is a point in the set that is mapped to itself) applied to compact convex set of all probability distributions of the finite set, $\{1,2,\cdots, n\}$ ($n$ being the order of the transition matrix).  As an aside, it is of interest to note that the spectral radius of any transition matrix is one; and hence, in accordance with the Gershgorin circle theorem on the spectrum of complex square matrices, no eigenvalue has absolute value greater one.

Whether $\boldsymbol{\pi}^*$ is reached asymptotically, i.e., whether $\boldsymbol{\pi}^\infty=\boldsymbol{\pi}^*$, is a question that requires constraint on the periodicity of the Markov chain for an affirmative answer. An intuitive idea behind this constraint can be seen as follows: Recall that $\pi^\infty_j=\lim_{n\rightarrow\infty}p^{(n)}_{ij}$ $\forall i,j$. It may be argued that a non-zero value of $p^{(n)}_{ii}$ in the limit of large $n$ (and hence $\pi^\infty_i$) does not make much sense if $i$ is a periodic state with period $d(i)$, since $p_{ii}^{(n)}=0$ for any $n$ not a multiple of $d(i)$. What can be proven is the following: For an aperiodic irreducible Markov chain with exclusively positive recurrent states, a unique $\boldsymbol{\pi}^*$, which is $\boldsymbol{\pi}^\infty$ as well, exists. In other words, \emph{a unique limiting invariant probability distribution exists for any ergodic Markov chain}. As an aside, we remark that the existence of a limiting distribution does not mean that the corresponding Markov chain must be ergodic, because limiting distribution can exist for reducible Markov chains: e.g., consider a Markov chain with an upper triangular transition matrix.

We now finish off the theoretical discussion on the Markov chain with the important concept of (time-)reversibility. Any general stochastic process $X(t)$ is called \emph{reversible} if any sequence $\{X_{t_n}\}$ has same probability distribution as its \emph{reversed process},  $\{X_{\tau-t_n}\}$ $\forall n\in\mathbb{N}$, $\forall \tau\in\mathbb{R}$ and for all $t_i$'s ($t_1<t_2<\cdots<t_n$). From this definition, we get that for reversible process both $\{X_{t_n}\}$ and $\{X_{\tau'+t_{n}}\}$ have same distribution as $\{X_{-t_n}\}$ on choosing $\tau=0$ and $\tau=\tau'$ respectively; and hence, we conclude that a reversible process is stationary (not necessarily vice versa). So while discussing reversible Markov chain, one must only consider stationary Markov chain. It is easy to see that a \emph{reversed chain} $\{X_{\nu-n}\}$ (some arbitrary $\nu\in\mathbb{Z}$) corresponding to a Markov chain $\{X_n\}$ must be Markov because independence of events is a symmetry property. Note that the reversed chain can always be constructed irrespective of whether the process is reversible or not. Now, for a reversible Markov chain, on choosing $\nu=n+n'$, we must have
\begin{eqnarray}
&&P(X_{n}=i,X_{n'}=j)=P(X_{n'}=i,X_n=j),\quad\\
\implies&&\pi_ip_{ij}=\pi_jp_{ji},
\end{eqnarray}
$\forall i,j\in\mathcal{X}$. This is called the \emph{local balance} or the \emph{detailed balance} condition. (We have omitted the argument $n$ while specifying the probability vector because it necessarily has to be a stationary distribution.) Since the above steps can be reversed, we have the following fact: Detailed balance implies and is implied by reversibility in a stationary Markov chain. The stationary distribution in a reversible Markov chain is also called \emph{equilibrium distribution}. 
Summing the detailed balance condition over states $i$, we get $\boldsymbol{\pi}=\boldsymbol{\pi}{\sf P}$---the stationarity condition---that can now be interpreted as the \emph{global balance} condition: The rate of flow $\sum_{i\in\mathcal{X}}\pi_jp_{ji}$ (=$\pi_j$) out of state $j$ is equal to the rate of flow $\sum_{i\in\mathcal{X}}\pi_ip_{ij}$ into the state $j$. Of course, local balance implies global balance but not vice versa. By definition, global balance is required for stationarity of the Markov chain. So, there is no reason that a stationary Markov chain must satisfy local balance condition and hence be reversible. Therefore, even a ergodic Markov chain need not be reversible and hence, the corresponding limiting distribution need not be an equilibrium distribution. In passing, we mention that \emph{Kolmogorov's criterion} for ergodic Markov chain states that the necessary and sufficient condition for the chain to be reversible is that starting from any state, the probability of coming back to the state along any sequence of states should be unchanged for the reversed sequence.

Let us illustrate some of the aforementioned concepts in some variants of random walks. Firstly, consider simple random walk. Its state space is $\mathcal{X}=\mathbb{Z}$. The transition matrix ${\sf P}$ is given by
\[
p_{ij}=\left\{
\begin{array}{lc}
 p & {\rm if}\, j=i+1,   \\
1-p  & {\rm if}\, j=i-1,    \\
 0 & {\rm otherwise}.   
\end{array}\right.
\]
We note $\forall i,j$, $i\leftrightarrow j$; i.e., the chain is irreducible. For $p\ne0.5$, every state is transient; while for $p=0.5$, every state is null recurrent. This can be shown through the explicit calculation of $\sum_{n=0}^{\infty}p^{(n)}_{ii}$ and finding that it is finite in the former case, while it is infinite in the latter case. Furthermore, the chain is periodic with period $2$. Naturally, there cannot be any stationary or limiting or equilibrium distribution.

Secondly, consider random walk with one end unbounded and the other with self-transition $p_{00}\in(0,1)$, i.e., its state space is $\mathcal{X}=\mathbb{Z}_{\ge0}$. The corresponding transition matrix ${\sf P}$ is
\[
p_{ij}=\left\{
\begin{array}{lc}
1-p  & {\rm if}\, j=i=0, \\
 p & {\rm if}\, j=i+1,   \\
1-p  & \phantom{xxxxxii}{\rm if}\,i>0\,\&\, j=i-1,    \\
 0 & {\rm otherwise}.   
\end{array}\right.
\]
First we observe that the chain is irreducible. State $0$ is aperiodic, and hence so is every other state in the chain. Rather than trying to find whether the chain is recurrent or transient using their definitions, let us use the concept that if a stationary distribution exists in an irreducible Markov chain, then the chain is positive recurrent. To this end, we assume that $\boldsymbol{\pi}^*$ exists. Consequently, it is easy to see that $\pi^*_i=r\pi^*_{i-1}$ for $i\in\mathbb{N}$ and $r\equiv p/(1-p)$; and using this in the normalization condition $\sum_{i\in\mathcal{X}}\pi^*_i=1$, we get $\pi^*_0=1-r$ if $r<1$, i.e., if $0<p<1/2$. Concluding, for $p<1/2$, a stationary distribution [$\pi_i^*=r(1-r)$] exists, and hence the chain is positive recurrent. Due to aperiodicity, we can also claim that the distribution is limiting distribution as well. Furthermore, we observe that the condition of detailed balance is not satisfied $\boldsymbol{\pi}^*$ and so the limiting distribution is not an equilibrium distribution. Similar arguments for the case $p\ge1/2$ tells us that no nonzero value of $\pi^*_0$ satisfies the normalization condition. Therefore, in this case, no stationary or limiting or equilibrium distribution exists. Ergo, the chain is either transient or null recurrent. 

Thirdly, consider random walk with absorbing boundaries. Its state space can be chosen as $\mathcal{X}=\{1,2,3,\cdots,N\}$. The transition matrix ${\sf P}$ is given by
\[
p_{ij}=\left\{
\begin{array}{ll}
 p & {\rm if}\, i>1\,\&\, j=i+1,   \\
1-p  & {\rm if}\, i<N\,\&\, j=i-1,    \\
1  & {\rm if}\, i=j=1\,{\rm or}\,N,   \\
 0 & {\rm otherwise}.   
\end{array}\right.
\]
Since $1\nleftrightarrow N$, it is not irreducible. In fact there are three equivalent classes: $\{1\}$, $\{N\}$, and $\{2,3,\cdots,N-1\}$---respectively absorbing (positive recurrent and aperiodic), absorbing (positive recurrent and aperiodic), and transient. Note that the absorbing states are trivially ergodic states but the chain is not ergodic. Furthermore, note that there are non-unique stationary distributions: e.g., $\boldsymbol{\pi}^*=[1\,0\,\cdots\,0\,0]$ and $\boldsymbol{\pi}^*=[0\,0\,\cdots\,0\,1]$ (in fact, any convex combination of them); one doesn't reach them from anywhere unless possibly one is starting in the corresponding equivalent class (of one state, in this case). We are witnessing a trivial illustration of a recurrent class's being closed.

Finally, consider random walk on a ring. Its state space can be chosen as $\mathcal{X}=\{1,2,3,\cdots,N\}$. The transition matrix ${\sf P}$ is given by
\[
p_{ij}=\left\{
\begin{array}{ll}
 p & {\rm if}\, j=i+1,   \\
  p & {\rm if}\, i=N\,\&\,j=1,   \\
1-p  & {\rm if}\, j=i-1,    \\
1-p  & {\rm if}\, i=1\,\&\,j=N,   \\
 0 & {\rm otherwise}.   
\end{array}\right.
\]
We note $\forall i,j$, $i\leftrightarrow j$; consequently, the chain is irreducible. For even $N$, every state is positive recurrent (the chain is finite) and periodic with period $2$. For odd $N$, every state is positive recurrent and aperiodic, and hence the chain is ergodic. Hence, unique stationary distribution, $\pi^*_i=1/N$ $\forall i$ exists for any $N$, but it is the limiting distribution only when $N$ is odd. For the stationary case, the detailed balance is satisfied only when $p=1/2$ because only in that case, $p_{ij}/N=p_{ji}/N$. Hence the ergodic random walk on a ring is reversible only if $p=1/2$ and the equilibrium distribution exists in this case only.
\section{Quantifying Uncertainty: Entropy}
\hskip 2.5cm``\emph{Only entropy comes easy.}''---Chekhov
\\
\\
\indent {\LARGE\textgoth{W}}e would perhaps prefer \emph{uncertainty} to \emph{ignorance}: If we think of a situation where we are ignorant---i.e., we cannot even associate a probability to a state of a system,  then uncertainty does not appear so bad an alternative to rather have; at least, we could do some predictions about the corresponding uncertain system by characterizing it using probabilities. Among any two events, the one with relatively higher probability is, by definition, less uncertain; on observing that random event obviously relatively less amount of information is gained. We can associate this information gained to be contained in the event itself---hence, sometimes it is called the \emph{self-information} or \emph{information content}. However, we shall refrain from using these terms as they are also used in other different contexts. We shall rather use the term \emph{Shannon information} to denote it.

\subsection{Shannon Entropy}
The Shannon information, $I(E)$, of an event $E$ is formally defined (within a constant multiplicative factor) as $I(E)=-\log_b{P(E)}$ ($b$ is the base of the logarithm) by demanding that (i) it should be inversely related to the corresponding probability, (ii) a certain event yields zero information, and (iii) total information of two independent events should be additive. The unit of the Shannon information is \emph{shannon} (\emph{Sh} or \emph{bit}), \emph{hartley} (\emph{Hart} or \emph{dit} or \emph{ban}), and \emph{nat} when $b$ is chosen as $2$, $10$, and $e$, respectively. Similarly, for any discrete random variable $X$ with probability mass function $p_X(x)$, the Shannon information can be defined as $I_X(x)=-\log_bp_X(x)$. We note that while $P(E)\in[0,1]$, $I(E)\in[0,\infty]$, i.e., the Shannon information is nothing but the probability itself mapped bijectively to a larger interval on the extended real line. Since the Shannon information is effectively the \emph{degree of surprise} (contrast this with degree of belief used for probability), sometimes it is called \emph{surprisal} as well.

Suppose now in place of asking about the information gained by observing an event (i.e., Shannon information of an event), we ask about the average information gained by observing one sample outcome from the sample space of the underlying probabilistic system. Of course, the answer depends on our knowledge about the details of the sample space $\mathcal{S}$; the states of the knowledge can be represented as the partitioning $\mathcal{P}$---sometimes called the \emph{information partition}---of the sample space. The elements $\mathcal{I}_n$ $(n=1,2,\cdots,N$ where $N$ is cardinality of $\mathcal{P}$) of $\mathcal{P}$ may aptly be called \emph{information sets} as it represents the state of knowledge of the observer when she is informed of occurrence of $\mathcal{I}_n$. For example, our average degree of surprise or the average information gained on seeing an outcome should intuitively be different for different partitions: $\mathcal{P}_1=\{\mathcal{S}\}$ or $\mathcal{P}_2=\{\{1\},\{2\},\{3\},\{4\},\{5\},\{6\}\}$, or $\mathcal{P}_3=\{\{1\},\{2\},\{3,4\},\{5,6\}\}$ or $\mathcal{P}_4=\{\{1,3,5\},\{2\},\{4,6\}\}$, so on. If we denote the average degree of surprise as a functional $H(\mathcal{P})$, then intuitively $H(\mathcal{P}_1)=0$ and $H(\mathcal{P}_2)>H(\mathcal{P}_3)>H(\mathcal{P}_4)$. Here, we are most simply interpreting $H(\mathcal{P})$ as average of the Shannon information over the information sets of the information partition $\mathcal{P}$ specifying the states of knowledge about the probabilistic system.

The functional $H(\mathcal{P})$, termed \emph{Shannon entropy} for historical reasons, can be formally defined via the requirement of the four intuitive \emph{Shannon--Khinchin axioms} for a measure of average information. Suppose $\{p_1,p_2,\cdots,p_N\}$ be the probability mass distribution over the information sets $\mathcal{I}_n\in\mathcal{P}$ ($n=1,2,\cdots,N$). Then the desiderata assumed to be obeyed by $H(\mathcal{P})\equiv H(\{p_i\})=H(p_1,p_2,\cdots,p_N)$---a multivariable function---are as follows:
\begin{enumerate}
\item \emph{Continuity}: $H(\{p_i\})$ is continuous with respect to all its variables.
\item \emph{Maximality}: $H(\{p_i\})$ is the largest when $p_i=1/N$ $\forall i\in\{1,2,\cdots,N\}$.
\item \emph{Expansibility}: $H(\{p_i\})=H(0,\{p_i\})=H(p_1,p_2,\cdots,p_i,0,p_{i+1},\cdots,p_N)=H(\{p_i\},0)$ $\forall N$ and $\forall i\in\{1,2,\cdots, N-1\}$.
\item \emph{Strong-additivity}: Considering arbitrary information partition $\mathcal{Q}$ is  of the same sample space with information sets $\{\mathcal{I}'_{n'}\}_{n'=1}^{n'=N'}$, let $p_{ij}$ be the probability distribution over \emph{partition product} $\mathcal{P}\vee\mathcal{Q}$ of $\mathcal{P}$ and $\mathcal{Q}$.  The elements $\mathcal{I}_i\mathcal{I}'_{j}$ of $\mathcal{P}\vee\mathcal{Q}$ are all the intersections of $\mathcal{I}_i$'s with all the $\mathcal{I}'_{j}$'s and so $p_{ij}$ is the joint probability of observing $\mathcal{I}_i$ and $\mathcal{I}'_{j}$. Then, assuming strong-additivity means that $H(\{p_{ij}\})=H(\{p_i\})+\sum_{i=1}^Np_iH(\{p_{ij}/p_i\})$, where $H(\{p_{ij}/p_i\})\equiv H(p_{i1}/p_i, p_{i2}/p_i, \cdots, p_{iN'}/p_i)$ is called the \emph{conditional entropy} of $\mathcal{Q}$ given $\mathcal{I}_i$ has occurred. This axiom is more compactly presented as 
\begin{equation}
H(\mathcal{P\vee Q})=H(\mathcal{P})+H(\mathcal{Q}|\mathcal{P}).
\end{equation} 
$H(\mathcal{P\vee Q})$ is called \emph{joint entropy} and $H(\mathcal{Q}|\mathcal{P})$ is called \emph{conditional entropy}.
Note that if the two partitions are \emph{independent}---i.e., $P(\mathcal{I}_n\mathcal{I}'_{n'})=P(\mathcal{I}_n\cap\mathcal{I}_{n'})=P(\mathcal{I}_n)P(\mathcal{I}_{n'})$ $\forall$ allowed $n,n'$---the condition of strong-additivity reduces to that of \emph{additivity}: $H(\mathcal{P}\vee\mathcal{Q})=H(\mathcal{P})+H(\mathcal{Q})$.
\end{enumerate}
Using these desired axioms, it can be proven rigorously that
\begin{equation}
H(\mathcal{P})=-\sum_{i=1}^{N}p_i\log_bp_i,
\end{equation}
unique up to a positive multiplicative constant which can simply be absorbed in the choice of $b$. We adopt the convention that $0\log_b0=0$ which is easily justified due to the fact that $\lim_{x\rightarrow0}x\log_bx=0$. The definition of the Shannon entropy automatically carries over to a discrete random variable $X$, with pmf $p_X(x)$, defined on a sample space $\mathcal{S}$ and can be written very compactly as follows:
\begin{eqnarray}
H({X})=-\sum_{x\in\mathcal{A}}p_X(x)\log_b{p_X(x)}=E[I_X(x)].\qquad
\end{eqnarray}
The collection of the values $\{X(s):s\in\mathcal{S}\}$ is called \emph{alphabet} $\mathcal{A}$ of $X$; the elements (called \emph{letters}) $x\in\mathcal{A}$  effectively partition the sample space. 

As can be guessed, the value of the Shannon entropy is crucially dependent on the nature of the partition even if the sample space remains the same. Thus, a few comments on the basic concepts of partition is in order. If cardinality of a partition is two, then the partition is called \emph{binary partition}. If the elements of the partition are identically the elements of the sample space, the partition is called \emph{element partition}. A partition $\mathcal{Q}=\{\mc{I}'_j:j=1,2,\cdots,N'\}$ is called a \emph{refinement} of another partition $\mathcal{P}=\{\mc{I}_i:i=1,2,\cdots,N\}$, i.e., $\mathcal{Q}\prec\mathcal{P}$ in symbols, if $\mathcal{I}'_j\subset \mathcal{I}_i$ $\forall j$ and some $i$. In other words, every element of $\mathcal{Q}$ is a subset of an element of  $\mathcal{P}$. Obviously, element partition is refinement of every other partition and $\mathcal{Q}\vee\mathcal{P}$ is a \emph{common refinement} of $\mathcal{Q}$ and $\mathcal{P}$. Owing to the concavity of the $\log$-function, a basic property of  the Shannon entropy is that if $\mathcal{Q}\prec\mathcal{P}$, then $H(\mathcal{Q})\ge H(\mathcal{P})$.

Shannon entropy has a nice interpretation if $b$ is chosen as $2$: $\min(E[N_B])$---denoting the minimum expected number of \emph{binary questions} in obvious notations---required to determine which information set has been observed in single trial lies between $H(\mathcal{P})$ and $H(\mathcal{P})+1$. For a dyadic probability distribution, it interestingly simplifies to $\min(E[N_B])=H(\mathcal{P})$. Specifically, consider a random variable $X$ with $p_X(n)=2^{-n}$ $\forall n\in\mathbb{N}$. Consider the strategy that we ask binary questions of the form `Is $X=n$?' and get a `yes' or a `no' for an answer. Since we are after the minimum number of questions required to specify the single sample observed, we should ask `Is $X=1$?' as $n=1$ has the highest probability of being observed; and it really has been observed the single question does the job of specifying. Otherwise, next question, viz.,  `Is $X=2$?' needs to be asked? This needs to be asked with probability $1/4$ and if $n=2$ has really been observed, aforementioned two questions do the job. Note that half of the time single question ($N_B=1$) terminates---having achieved the goal---the sequence of questions and one-fourth of the time two questions ($N_B=2$) needs to be asked, and so on. $N_B$ is obviously a random variable. Therefore, the expected number of questions $E[N_B]=\sum_{n\in\mathbb{N}}n\cdot 2^{-n}=2=-\sum_{n\in\mathbb{N}}2^{-n}\log_22^{-n}=H(X)$. The strategy followed also intuitively ensures that $H(X)=\min E[N_B]$: Of course, any other sequence of questions would take many more attempts to reach correct conclusion when an outcome with higher probability has actually occurred. We remark that the above conclusion about the dyadic distribution remains unchanged even when the support of the distribution is modified to be a finite subset of $\mathbb{N}$. In other words, we can also say that we require on average $H(X)$ bits of information to describe the random variable $X$.

Before we discuss the Shannon entropy more, let us digress a bit to note that since the ultimate justification for its chosen form is actually its widespread utility, one could relax the axioms if the situation demands. If one thinks about it, the condition of strong-additivity is rather technical, and not very intuitive, required for the uniqueness of $H(\mathcal{P})$. It is, thus, not surprising that there are many proposals of defining entropy by relaxing the strong-additivity condition; all such entropies are called \emph{generalized entropies}. Two of the most studied ones are as follows:
\begin{itemize}
\item \emph{R\'enyi entropy}: Suppose the requirement of strong-additivity is replaced by only additivity. Then it can be argued that rather than the `linear' average of the Shannon information, a more general `average' (Kolmogorov--Nagumo average or quasilinear average)---viz., $\phi^{-1}(\sum_{i=1}^N[p_i\phi(I(\mathcal{I}_i))])$ where $\phi$ is an continuous, strictly monotone, invertible function---should be used. Furthermore, additivity can be demanded that leads to $\phi(x)=cx$ and $\phi=c^{(1-{q})x}$ where $c$ and ${q}$ (${q}\ne 1$) are some positive constants. The former leads to the Shannon entropy and the latter to the R\'enyi entropy given by
\begin{equation}
\qquad R_{q}(\mathcal{P})=\frac{1}{1-{q}}\log_b\left(\sum_{i=1}^{N}p_i^{q}\right)\quad\,({q}>0),
\end{equation}
 The positivity of ${q}$ is required to satisfy the expansibility condition. Here $R_1\equiv R_{{q}\rightarrow1}=H$. We mention that $R_0(\equiv R_{{q}\rightarrow0})$, $R_2$, and $R_{\infty}$ are respectively known as \emph{Hartley entropy} or \emph{max-entropy}, \emph{collision entropy}, and \emph{min-entropy}.
\item \emph{Tsallis entropy}: In case even the additivity condition is not demanded, an important generalized entropy is Tsallis entropy defined by
\begin{equation}
T_{q}(\mathcal{P})=\frac{1}{1-{q}}\left(\sum_{i=1}^{N}p_i^{q}-1\right)\quad\,({q}\in\mathbb{R}),
\end{equation}
where $T_1\equiv T_{{q}\rightarrow1}=H$.
\end{itemize}
In some sense, it is the simplest generalized non-extensive entropy that satisfies \emph{pseudo-additivity}, i.e., $T_{q}(\{p_{ij}\})=T_{q}(\{p_i\})+\sum_{i=1}^Np^{q}_iT_{q}(\{p_{ij}/p_i\})$, in place of additivity.

Coming back to the Shannon entropy, which is both $R_1$ and $T_1$ as well, let us interpret it under various interpretations of probability. At the cost of repetition but for completeness sake, recall that it is merely a functional of partition of the sample space; it yields a non-negative real number for every given partition. This is nothing but the \emph{formalist interpretation}. Also, mentioned earlier is the \emph{subjective interpretation}: The Shannon entropy is the measure of average degree of surprise associated with any single event out of the events of a given partition of the sample space of the underlying probabilistic system.

The \emph{classical interpretation} of the Shannon entropy is as follows. Recall that at the root of the classical interpretation of probability is the principle of insufficient reason due to which a partition $\mathcal{P}$ of cardinality $N$ should have probability $p_i=1/N$ associated with each information set $\mathcal{I}_i$; and hence, the corresponding Shannon entropy, $H(\mathcal{P})$ is $\log_bN$ which is the maximum possible entropy for any collection $\{p_i\}_{i=1}^{i=N}$ subject to the obvious constraint $\sum_{i=1}^{i=N}p_i=1$. This can be also checked by introducing a Lagrange multiplier $\lambda$ and then maximizing the Lagrange function $L$ as follows:
\begin{eqnarray}
&&\frac{\partial L}{\partial p_i}=\frac{\partial} {\partial p_i}\left[-\sum_i p_i\log_bp_i-\lambda\left(\sum_ip_i-1\right)\right],\qquad\\
\implies&&0=-\log_bp_i-1-\lambda,\\
\implies &&p_i=\frac{1}{N},
\end{eqnarray}
where we have used the fact that $b^{-\lambda}=b/N$ (owing to the normalization condition of probability). That $p_i=1/N$ $\forall i$ is maximum (and not minimum or saddle), actually follows from the Shannon--Khinchin axioms. This motivates proposing a principle that is analogous to the principle of insufficient reason: \emph{the principle of maximum entropy}. In line with the Occam's razor, the principle is rooted in the premise that any estimation of the probability distribution of the probabilist system under consideration must be done in a way that one is left with largest remaining uncertainty consistent with the available constraints on the distribution; in other words, one is not invoking any further biases in the estimation beyond what is provided by the constraints. 

In principle, one can extend the principle of maximum entropy beyond the domain of classical interpretation of probability by taking into account any available logical proposition (specifically called \emph{testable information})---other than the obvious normalization condition---about the probability distribution. One could consider the constraints given by testable information as evidences and take the application of the principle beyond the classical interpretation to what may be recognized as \emph{logical interpretaion} of the Shannon entropy formed using the derived probability distribution. (In fact, this principle's usage has been advocated to fix priors for objective Bayesians.) For example, consider that testable information---in the form of some observable expectation values of $m$ random functions $\{f_k(X)\}$---is known; i.e., $m$ number of constraints, $E[f_k(X)]=\sum_{x_i\in\mathcal{A}} p_if_k(x_i)=\left<f_k\right>$ (say), are assumed known. We want to find an underlying pmf $p_i\equiv p(x_i)$ for the random variable with only this much available data such that the observed values of the observables can be obtained from the pmf. The principle of maximum entropy dictates us to use these constraints along with the normalization condition of the probability to construct a Lagrange function $L$ using the Lagrange multipliers $\lambda_k$'s and set
\begin{eqnarray}
&&\frac{\partial L}{\partial p_i}=\frac{\partial} {\partial p_i}\left[-\sum_i p_i\log_bp_i-\lambda_0\left(\sum_ip_i-1\right)\right.\nonumber\\
&&\phantom{\frac{\partial L}{\partial p_i}=}\qquad\,\,\left.-\sum_{k=1}^m\lambda_k\left(\sum_{i}  p_if_k(x_i)-\left<f_k\right>\right)\right]=0.\quad
\end{eqnarray}
This implies the \emph{Gibbs distribution} for the pmf:
\begin{eqnarray}
p(x_i)=p^{(G)}(x_i)\equiv\frac{b^{\sum_{k=1}^m[-\lambda_kf_k(x_i)]}}{Z(\lambda_1,\lambda_2,\cdots,\lambda_m)},
\end{eqnarray}
where $Z=\sum_ib^{\sum_{k=1}^m[-\lambda_kf_k(x_i)]}$ is called \emph{partition function} and is related to the observables via $\left<f_k\right>=-\partial \log_bZ/\partial \lambda_k$. Note that we have technically only extremized the entropy; it remains to be shown that it has actually been maximized. We show it later in the analogous context of continuous state random variable after having developed the concept of relative entropy required for the succinct proof.

Last but not the least, the \emph{frequentist interpretation}, which is empirical in nature, is best understood through the concept of \emph{typical sequence}. A frequentist calls for repeated independent identical trails; in this spirit, consider a (product) sample space $\mathcal{S}^n$ where $\mathcal{S}$ has a partition $\mathcal{P}=\{\mathcal{I}_i:i=1,2,\cdots,N\}$ with $\{p_i\}$ as the pmf. Any sequence of information sets observed has probability in the form $\prod_{i=1}^Np_i^{n_i}$ with $\sum_{i=1}^Nn_i=n$, where $n_i$ is the number of times $\mathcal{I}_i$ is observed in the sequence. If $n\rightarrow\infty$, then $n_i\rightarrow np_i$ with high probability in line with the frequentist interpretation of probability. We define the typical sequences as the ones with $n_i\approx np_i$. More formally, let $X$ be the corresponding random variable associated with $\mathcal{S}$ and let $\{X_i\}$ be the iid sequence of this random variable. We have, by applying law of large numbers on the random variable $\log_b p(X)$, $-\frac{1}{n}\log_bp(\{X_i\})=-\frac{1}{n}\log_b\prod_{i=1}^np(X_i)=-\frac{1}{n}\sum_{i=1}^n\log_bp(X_i)\rightarrow E[-\log_bp(X)]=H(X)$ in probability. That is, for all positive $ \epsilon$,
\begin{equation}
\lim_{n\rightarrow\infty}P\left(\{X_i\}:p(\{X_i\})=2^{-n(H(X)\pm\epsilon)}\right)=1, \label{eq:aep}
\end{equation}
where $b$ has been fixed to $2$. This is known as \emph{asymptotic equipartition property (AEP)}. A typical sequence is, thus, an element of \emph{typical set} $\mathcal{T}_{(n,\epsilon)}$---the set of sequences $\{x_i\}$ such that $p(\{x_i\})\in[2^{-n\{H(X)+\epsilon\}},2^{-n\{H(X)-\epsilon\}}]$. 

We know that the probability of each typical sequence is same as is equal to nearly $2^{-nH(X)}$ and $P(\mathcal{T}_{(n,\epsilon)})\approx1$. Therefore, the total number of such equiprobable sequences is $2^{nH(X)}$ which is much lesser than number of total possible sequences, $N^n(=2^{n\log_2N})$ as $n\rightarrow\infty$ (unless all the events are equally likely in which case $2^{nH(X)}=N^n$). Thus, typical set is a very small set containing most of the probability. Concluding, in the light of Eq.~(\ref{eq:aep}), the frequentist interpretation of the Shannon entropy now can be given as follows: Consider a typical sequence that occurs $m$ times in an experiment $\mathcal{S}^n$ repeated independently $M$ times (note that we are talking about repeated `repeated experiments'!), then the Shannon entropy $H(\mathcal{P})\approx-\frac{1}{n}\log_b\left({m}/{M}\right)$.

Before we end this section, let us highlighting two very important applications of the AEP. Firstly, we note that for any event $E$ in the sample space of $\mathcal{S}^n$:
\begin{eqnarray}
&&P(E)=P(E|\mathcal{T}_{(n,\epsilon)})P(\mathcal{T}_{(n,\epsilon)})+P(E|\mathcal{T}^c_{(n,\epsilon)})P(\mathcal{T}^c_{(n,\epsilon)}),\qquad\\
&&\implies P(E)\approx P(E|\mathcal{T}_{(n,\epsilon)}).
\end{eqnarray}
In plain words, the probability of any event $E$ can be determined by analysing only the small typical set; the result becomes more accurate with increasing $n$. Second important conclusion (without explicit proof) is that any sufficiently long ($n\to\infty$) iid sequence of length $n$ can be represented (encoded) using as few as $nH(X)$ bits on average without losing information. This is the essence of \emph{Shannon's noiseless coding theorem} (or \emph{Shannon's source coding theorem} or \emph{Shannon's first theorem} or the \emph{first fundamental coding theorem}) which deals with the question how much redundancy a message has, so that by neglecting the redundant parts, the signal/message (stream of iid data in this context) can be compressed without losing any information. In other words,  it is impossible to compress the message such that the average number of bits per symbol (i.e., \emph{code rate}) is fewer than the Shannon entropy of the random variable $X$ in the message. It may be mentioned that a \emph{code} can be considered as an algorithm that uniquely represents symbols from some source alphabet, by encoded strings formed using elements of some other target alphabet: E.g., the mapping $C = \{{\rm hot}\mapsto 0,\, {\rm warm}\mapsto 01,\, {\rm cold}\mapsto 11\}$ is a code which represents (encodes) the source alphabet, viz., the set $\{{\rm hot,\,warm,\,cold}\}$ using strings (\emph{codewords}) $\{0,01,11\}$ formed using the target alphabet, viz., the set $\{0,1\}$. 

It should be noted that the stream of iid data cannot be compressed if all the letters of alphabet $\mathcal{X}$ are equiprobable because then every sequence is a typical sequence. Thus, it makes sense to talk about \emph{redundancy}, defined as
\begin{equation}
r\equiv1-\frac{H(X)}{\max_{p(x)}[H(X)]}\times100\%=1-\frac{H(X)}{\log_2N}\times100\%, 
\end{equation}  
which can range from $0\%$ to $100\%$; thus, one would say that the stream of iid data has $r\%$ redundant content that is not needed to faithfully encode the information contained in the data: A uniform pmf over $\mathcal{X}$ imples $0\%$ redundancy, while a pmf with probability equal to one for any one letter means $100\%$ redundancy in the message.

\subsection{Relative Entropy}
We have seen how the job of quantification of uncertainty can be transferred from the probability to the Shannon information and from there to the Shannon entropy. Now let us see how the essence of Bayes' theorem in conditionalization can be seen in information-theoretic perspective. Given a probabilistic system with a sample space $\mathcal{S}$ and its partition $\mathcal{P}=\{\mathcal{I}_i:i=1,2,\cdots,N\}$, let us say there exists a true probability distribution $\{p_i\}$ over $\{\mathcal{I}_i\}$ that describes the corresponding experiment's outcomes. Now before seeing any evidence we have a guess at the $\{p_i\}$; let that guess be $\{q_i\}$. Hence for any event $\mathcal{I}_i$, our Shannon information is $-\log q_i$. (Henceforth, we do not explicitly mention the base $b$ that we choose as $2$ unless specified otherwise.) Thus, the Shannon information averaged over the true distribution is $-\sum_{i=1}^Np_i\log q_i$ which is called \emph{cross entropy}, denoted as $H_\times(p,q)$. It is the measure of average information gained when a single result of the experiment is observed. On observing new data, we update our degree of belief and suppose in the light of enough data, our updated degree of belief coincides with the true distribution $p_i$. Then, intuitively, we hope that the average uncertainty associated with the experiment would decrease post updating. We realize that the final Shannon entropy is $H(\mathcal{P})=H_\times(p,p)$ in the notation of cross entropy. In other words, we expect the quantity $D(p||q)$---\emph{Kullback--Leibler (KL) divergence} (also called \emph{relative entropy})---defined by
\begin{equation}
D(p||q)\equiv H_\times(p,q)-H_\times(p,p)=\sum_{i=1}^Np_i\log \left(\frac{p_i}{q_i}\right),
\end{equation}
to be non-negative, a fact also known as \emph{information inequality}. Furthermore, the fact that $D(p||q)\ne D(q||p)$ in general, is the reflection of the asymmetry possessed by Bayes' updating. Essentially, any degree of belief other than the true one leads to rather inefficient description of the random variable associated with the sample space because it requires extra bits on average equal to the value of the KL-divergence. 

Showing that $D(p||q)$ is indeed non-negative requires invoking \emph{Jensen's inequality} for convex functions: If $f(X)$ is a convex function of random variable $X$, then $E[f(X)]\ge f(E[X])$. Since $-\log(x)$ is a (strictly) convex function, we note that 
\begin{subequations}
\begin{eqnarray}
&&-D(p||q)=\sum_{x\in\mathcal{A}}p(x)\log\left(\frac{q(x)}{p(x)}\right)\\
&&\phantom{-D(p||q)}=\sum_{x\in{\rm supp}(p)}p(x)\log\left(\frac{q(x)}{p(x)}\right)\\
&&\phantom{-D(p||q)}\le \log\left(\sum_{x\in{\rm supp}(p)}p(x)\frac{q(x)}{p(x)}\right)\\
&&\phantom{-D(p||q)}= \log\left(\sum_{x\in{\rm supp}(p)}q(x)\right)\\
&&\phantom{-D(p||q)}\le \log\left(\sum_{x\in\mathcal{A}}q(x)\right)=0.
\end{eqnarray}
\end{subequations}
Here, $\mathcal{A}$ is alphabet of $X$ and ${\rm supp}(p)\equiv\{x:p(x)>0\}$. Hence, we conclude $D(p||q)\le$ with equality holding if $p(x)=q(x)$ $\forall x$. 

Now consider $D(p(x,y)||p(x)p(y))$ defined for two random variables $X$ and $Y$ with alphabets $\mathcal{A}_x$ and $\mathcal{A}_y$, respectively, mapped from the same sample space; $p(x)$ and $p(y)$ are the pmf's of $X$ and $Y$ respectively, and $p(x,y)$ is their joint pmf. This quantity is, thus, the measure of the extra information required to specify $X$ and $Y$ jointly if they are assumed to be independent, whether or not that is truly so. Note,
\begin{eqnarray}
&&D(p(x,y)||p(x)p(y))=\sum_{x\in\mathcal{A}_x}\sum_{y\in\mathcal{A}_y}p(x,y)\log \frac{p(x,y)}{p(x)p(y)},
\qquad\\
&&{\rm or,}\, D(p(x,y)||p(x)p(y))=H(X)+H(Y)-H(X,Y).\quad
\end{eqnarray}
Here, as is notational custom, the joint entropy $H(X,Y)$ uses a comma between $X$ and $Y$. Furthermore, since $H(X,Y)=H(X)+H(Y|X)=H(Y)+H(X|Y)$ (from the Shannon--Khinchin axioms), so using a dedicated notation $I(X,Y)$ for $D(p(x,y)||p(x)p(y))$, we get:
\begin{eqnarray}
I(X;Y)=H(X)-H(X|Y)=H(Y)-H(Y|X).\quad
\end{eqnarray}
 We recall that the conditional entropy $H(X|Y)\equiv\sum_{y\in\mathcal{A}_y}p(y)\sum_{x\in\mathcal{A}_x}p(x|y)\log p(x|y)$. $I(X;Y)$ is the obviously the measure of reduction in uncertainty of $X$ due to knowledge of $Y$, and hence, $I(X;Y)$ is aptly called the \emph{mutual information} as it is the information of one variable contained in the other: If $X$ and $Y$ are independent, then $I(X;Y)=0$; whereas if $X=Y$, $I(X;Y)=H(X)$. Note that, by definition, $I(X;Y)\ge0$ with equality holding only for independent events.

The Shannon entropy, the relative entropy, and the mutual entropy satisfy very useful chain rules written below in obvious notations:
\begin{eqnarray}
&&H(X_1,X_2,\cdots,X_n)=\sum_{i=1}^nH(X_i|X_{i-1},\cdots,X_2,X_1),\nonumber\\
\label{eq:crH}\\
&&I(X_1,X_2,\cdots,X_n;Y)=\sum_{i=1}^nI(X_i;Y|X_{i-1},\cdots,X_2,X_1),\,\,\,\,\nonumber\\
\\
&&D(p(x,y)||q(x,y))=D(p(x)||q(y))+D(p(x|y)||q(x|y)).\nonumber\\ 
\end{eqnarray}
Here the \emph{conditional mutual information} has been defined by $I(X;Y|Z)\equiv H(X|Z)-H(X|Y,Z)$.

The relative entropy can be used to generalize the maximum entropy principle. The motivation comes from the fact that if one considers $D(p||u)$ where $u(x)$ is a uniform pmf over finite alphabet $\mathcal{A}$ of cardinality $N$, then $\min_p[D(p||u)]=\min_p[\log N-H(X)]=\max_p[H(X)]$. Therefore, maximum entropy distribution can be obtained by equivalently minimizing $D(p||u)$. Note that since $D(u||u)=0$ and the relative entropy cannot be negative, a uniform distribution has minimum relative entropy when no testable information (except the normalization of probability) is available. This is compatible with the classical interpretation of the Shannon entropy. Consequently, one can generalize the principle of maximum entropy by proposing the \emph{principle of minimum relative entropy} where one minimizes $D(p||q)$---where $q(x)$ is now an arbitrary distribution---over all possible $p(x)$.

For an illustration, suppose a scenario where the testable information---in the form of some observable expectation values of $m$ random functions $\{f_k(X)\}$---is known; i.e., $m$ number of constraints, $E[f_k(X)]=\sum_{x_i\in\mathcal{A}} p_if_k(x_i)=\left<f_k\right>$ (say), are assumed known. We want to find an underlying pmf $p_i\equiv p(x_i)$ in line with the principle of minimum relative entropy. Consequently, we use these constraints along with the normalization condition of the probability to construct a Lagrange function $L$ using the Lagrange multipliers $\lambda_k$'s and set
\begin{eqnarray}
&&\frac{\partial L}{\partial p_i}=\frac{\partial} {\partial p_i}\left[\sum_i p_i\log_b\left(\frac{p_i}{q_i}\right)-\lambda_0\left(\sum_ip_i-1\right)\right.\nonumber\\
&&\phantom{\frac{\partial L}{\partial p_i}==}\qquad\,\,\left.-\sum_{k=1}^m\lambda_k\left(\sum_{i}  p_if_k(x_i)-\left<f_k\right>\right)\right]=0.\qquad
\end{eqnarray}
This implies following form for the minimum relative entropy pmf:
\begin{eqnarray}
p(x_i)=\frac{q(x_i)b^{\sum_{k=1}^m[\lambda_kf_k(x_i)]}}{Z_q(\lambda_1,\lambda_2,\cdots,\lambda_m)}, \label{eq:mred}
\end{eqnarray}
where $Z_q\equiv\sum_iq(x_i)b^{\sum_{k=1}^m[\lambda_kf_k(x_i)]}$ is a normalization factor. Note that we have technically only extremized the relative entropy; it remains to be shown that it has actually been minimized, which can be shown by contradiction as follows: Assume that there exists another pmf $r(x)\ne p(x)$ such that it satisfies the same constraints and has a smaller value for the relative entropy than what the minimum relative entropy pmf $p(x)$ has. The former condition implies that $\sum_\mathcal{A}p\log(p/q)=\sum_\mathcal{A}r\log(p/q)$ as can be seen by explicitly using Eq.~(\ref{eq:mred}) to replace $p/q$ in summands on both sides of the equality sign. Therefore, $0>D(r||q)-D(p||q)=\sum_\mathcal{A}r\log(r/q)-\sum_\mathcal{A}p\log(p/q)=\sum_\mathcal{A}r\log(r/q)-\sum_\mathcal{A}r\log(p/q)=\sum_\mathcal{A}r\log (r/p)=D(r||p)\ge0$---a contradiction.  We remark that the above proof also serves as the proof for the fact that the Gibbs distribution obtained earlier using the maximum entropy principle is actually the maximum one---one just needs to fix $q=u$. 

A curious point to note is that $\min_p[D(p||q)]$ in the absence of any constraint gives $p(x)=q(x)$. Recall that $q(x)$ may be seen as a guess for $p(x)$ in the definition of the relative entropy. Therefore, the fact that minimizing the relative entropy gives a $p(x)$---compatible with the absence of any constraint---equal to $q(x)$ essentially means that the closest $p(x)$ to the guess is the guess itself. Furthermore, recall that when there is no evidence of any kind, then in line with the principle of insufficient reasons, $u(x)$ is the guess and naturally, $p(x)=u(x)$ appears on minimizing $D(p||u)$. So it is evident that choosing a $q(x)$ is a matter of fixing a guess probability distribution based on evidence at one's disposal. Now evidence may be a very vague entity and sometimes it may not even be clear how to put it in a mathematical form. Thus, the choice of $q(x)$ may be very uncertain; in fact, any possible distribution is allowed from the perspective of subjective Bayesians. However, if the evidence can be put in the form of some mathematical constraints on the probability distribution $p(x)$---the one we want to find out---then one way of finding $q(x)$ (supposing that the constraints completely specify $q(x)$) is to find the maximum entropy distribution for $p(x)$. We could say that $q(x)$ has, thus, been fixed as the maximum entropy distribution; it could be seen as an objective Bayesian viewpoint. If, however, the evidence available has two independent parts---one that be cast in the form of constraints and the other about which one is not sure a priori what to use---then minimizing the relative entropy yields Eq.~(\ref{eq:mred}) where $q(x)$ refers to the latter part.
\subsection{Entropy Rates: Source Entropy}
Let us start with some pertinent ideas of the ergodic theory tuned to our purpose. Consider a \emph{probability space} $(\mathcal{S,E},P)$. 
\begin{itemize}
\item Naturally, $(\mathcal{S,E})$ is a \emph{measurable space}, i.e., a space $\mathcal{S}$ together with a sigma algebra $\mathcal{E}$ of sets; any set $E\in\mathcal{E}$ is a \emph{measurable set}. $P$ is a \emph{finite measure}, i.e., it obeys the three Kolmogorov axioms with countable additivity. Thus, the probability space is a \emph{measure space}. 
\item Given two measurable spaces $(\mathcal{S,E})$ and $(\mathcal{S',E'})$, a function $T:\mathcal{S}\rightarrow\mathcal{S'}$ is called a \emph{measurable function} if $\forall E'\in\mathcal{E}$, $T^{-1}E'=\{s\in\mathcal{S}: Ts\equiv T(s)\in E'\}\in\mathcal{E}$. 
\item A transformation $T:\mathcal{S}\rightarrow\mathcal{S}$ is \emph{measure-preserving} if $T$ is measurable function and $P(T^{-1}E)=P(E)$ $\forall E\in\mathcal{E}$.
\item Equivalently, a measurable transformation $T$ is measure-preserving iff for every integrable function $f:\mathcal{S}\rightarrow\mathbb{R}$, $\int fdP=\int f\circ T dP$.
 \item A set $E\subset\mathcal{S}$ is called \emph{invariant} under a transformation $T$ if $T^{-1}E=E$. 
 \item A measure-preserving transformation $T$ is termed \emph{ergodic} if either $P(E)=0$ or $P(E)=1$ for any $E\in\mathcal{E}$ satisfying $T^{-1}E=E$, i.e., the invariant sets can have either zero or full measure.
 \item A property said to hold \emph{almost everywhere} in a probability space if $\exists E\in\mathcal{E}$ with $P(E)=0$ and the property holds at all $s\in E^c$ (where complementation is with respect to $\mathcal{S}$).
 \item A useful form of \emph{Birkhoff ergodic theorem} is that if $T$ is a measure-preserving transformation in the probability space, then the following statements are equivalent: (i) $T$ is ergodic; (ii) for almost everywhere $s\in\mathcal{S}, $ $\lim_{n\rightarrow \infty}\frac{1}{n}\sum_{i=0}^{n-1}f(T^is)=\int_\mathcal{S} fdP$ $\forall f$ with $\int_{\mathcal{S}}|f|dP<\infty$; and (iii) $\forall E,E'\in\mathcal{E}$, $\lim_{n\rightarrow \infty}\frac{1}{n}\sum_{i=0}^{n-1}P(T^{-i}E\cap E')=P(E)P(E')$.
\end{itemize}

Our interest in this section is to characterize the source of any information. By this we mean any source of series of signals, i.e, technically, we see a source as a generator of a stochastic process with observable values. The source is defined by a set of alphabet $\mathcal{A}$---assumed finite---whose elements, $x_i$'s are called \emph{letters}. Now consider a sequence of letters---$c=\{x_i\}_{i=n}^{i=n'}$ (subscript $i$ denotes the time step) where $n'\ge n$ and $n,n'\in\mathbb{Z}$---as an elementary event. Set of all such elementary events with fixed $n$ and $n'$ is  called a \emph{cylinder}, $C_{nn'}$, and let $\mathcal{C}$ be set of all possible cylinders. Hence, notationally, $c\in C_{nn'}\subset \mathcal{C}$. Thus, using $\mathcal{C}$ as sample space of the infinite probability space, its corresponding event space $\mathcal{E}$, and a probability measure $P$, we can define source as a probabilistic system, $(\mathcal{C,E},P)$, or equivalently and more elementarily as $(\mathcal{A},P)$ that completely characterizes the source's statistical nature. 

Next, we introduce a transformation $T$ such that $Tc\equiv T(c)\equiv\{x_i\}_{i=n+1}^{i=n'+1}$; hence, the $T$ is aptly called the \emph{shift operator}. Of course, $T\mathcal{C}=\mathcal{C}$. The source is called \emph{stationary} if $\forall E\in\mathcal{E}$, $P(TE)=P(E)$. We may note that this is in line with the concept of stationarity of a stochastic process. Furthermore, the source is called \emph{ergodic}, if $T$ is ergodic, i.e., if $T^{-1}E=E$ holds for any $E\in\mathcal{E}$, then either $P(E)=0$ or $P(E)=1$.
 
From information-theoretic context, we are interested in the important question that how much information is conveyed on average by the source as it emits one letter. In other words, we are interested in the rate at which information is emitted by a source. Since we are viewing the source as a generator of a discrete-time stochastic chain $\{X_i\}$ with finite alphabet $\mathcal{A}$ as its state space, following definition is useful: 
\begin{eqnarray}
H_s(\mathcal{A})\equiv\lim_{n\rightarrow\infty}\frac{1}{n}H(X_1,X_2,\cdots,X_n)\label{eq:ose}
\end{eqnarray}
is the \emph{entropy rate} of the stochastic process, assuming that the limit exists. If the stochastic process is associated with a source then the entropy rate is also called \emph{source entropy} because it is obviously a measure of the average information per letter emitted by the source and, thus, an inherent property of the source. Of course, for a source emitting iid random variables, the source entropy is the Shannon entropy of any one of the identical random variables. Note that the unit of source entropy is `bits per symbol' unlike bits for the Shannon entropy.

An important result is that for a stationary source, the source entropy $H_s(\mathcal{A})$ exists (proven below) and 
\begin{eqnarray}
H_s(\mathcal{A})=\lim_{n\rightarrow\infty}H(X_n|X_{n-1},X_{n-2},\cdots,X_1),\label{eq:Hs}
\end{eqnarray}
which also exists because explicitly, we can argue that
\begin{subequations}
\begin{eqnarray}
&&H(X_{n-1}|X_{n-2},X_{n-1},\cdots,X_2,X_1)\nonumber\\
&&\phantom{H(X_n)}=H(X_n|X_{n-1},X_{n-2},\cdots,X_3,X_2)\\
&&\phantom{H(X_n)}\ge H(X_n|X_{n-1},X_{n-2},\cdots,X_2,X_1),\qquad
\end{eqnarray}
\end{subequations}
thereby proving that the sequence $(H(X_n|X_{n-1},X_{n-2},\cdots,X_1))_{n\in\mathbb{N}}$  is a decreasing sequence of non-negative numbers. In the above steps, the equality is consequence of assumed stationarity of the source; while the inequality follows from the fact that conditioning can, on average, only decrease entropy: $H(X|Y)-H(X)=-I(X;Y)\le0$. This result, in conjunction with the general fact that sequences $(a_n)_{n\in\mathbb{N}}$ and $(\sum_{i=1}^na_i)_{n\in\mathbb{N}}$ have same limit point (if it exits), can now be used to show that $H_s(\mathcal{A})$ [defined in Eq.~(\ref{eq:ose})] exists as well and is equal to the R.H.S. of Eq.~(\ref{eq:Hs}): Just observe that by the chain rule given in Eq.~(\ref{eq:crH}), we have $\frac{1}{n}H(X_1,X_2,\cdots,X_n)=\frac{1}{n}\sum_{i=1}^nH(X_i|X_{i-1},\cdots,X_1)$.

 The meaning of the R.H.S. of Eq.~(\ref{eq:Hs}) is obvious when we note that, by the chain rule [see Eq.~(\ref{eq:crH})], $H(X_n|X_{n-1},X_{n-2},\cdots,X_1)$ can be written as difference between two joint entropies, i.e., $H(X_n|X_{n-1},X_{n-2},\cdots,X_1)$=$H(\{X_i\}_{i=1}^{i=n})-H(\{X_i\}_{i=1}^{i=n-1})$. Thus, the source entropy is the extra bits of average information needed to specify the $n$th letter as $n$ tends to infinity. 

Using Eq.~(\ref{eq:Hs}), a stationary Markov chain's source entropy is easily found in terms of the stationary distribution $\boldsymbol{\pi}^*$ and the transition matrix ${\sf P}$ with matrix elements $p_{ij}$: $H_s=\lim_{n\rightarrow\infty}H(X_n|X_{n-1})=H(X_2|X_1)=-\sum_{i,j}\pi^*_ip_{ij}\log p_{ij}$. This formula, with $\boldsymbol{\pi}^*$ replaced by $\boldsymbol{\pi}^\infty$, also works as the source entropy of an ergodic Markov chain starting from any initial distribution. 

In passing, we mention without proof, a formidable result---the \emph{Shannon--McMillan--Breiman theorem}. It is the AEP for the stationary ergodic process (not necessarily Markov): For any finite-valued stationary ergodic chain $\{X_i\}$, 
\begin{equation}
\lim_{n\rightarrow\infty}P\left(\{X_i\}:p(\{X_i\})=2^{-n(H_s(\mathcal{A})\pm\epsilon)}\right)=1,
\end{equation}
for all positive $ \epsilon$. Thus, for a stationary ergodic source, the source entropy, which is an inherent property of the source, can actually be estimated by analysing a single sufficiently long sequence in the typical set; hence we can view the source entropy as a property of any typical sequence. 

In a communication, a source sends a message---string of symbols---through a communication \emph{channel} to a receiver. If $X$ is input and $Y$ is output of a channel, then \emph{channel capacity} is defined as 
\begin{eqnarray}
C\equiv \max_{p(x)}I(X;Y)=\max_{p(x)}[H(X)-H(X|Y)].
\end{eqnarray}
In this context, an interesting interpretation of $H(X|Y)$ is insightful: Consider $Y$ that is input $X$ with an independent additive noise $\eta$  added to it. Hence, $H(Y|X)=H((X+\eta)|X)=H(\eta|X)=H(\eta)$. Thus, $H(Y|X)$ is the entropy of the \emph{channel noise}. It, furthermore, follows that $H(X|Y)=H(X)-I(X;Y)=[H(X)-H(Y)]+H(\eta)$. $H(X|Y)$, also called \emph{equivocation} of $X$ given $Y$, is the average  uncertainty remaining in the input after observing the output; this has an additional contribution from the channel if that is noisy. In a way, it characterizes error in observing $X$ through the lens of the output $Y$. In the absence of noise, if $Y=X$, then $H(X|Y)=0$ and the channel capacity of the noiseless channel is $C=\log N$ (where $N$ is the size of input alphabet). The name channel capacity is apposite in the light of \emph{Shannon's second theorem} (also called \emph{Shannon's noisy channel coding theorem} or the \emph{second fundamental coding theorem}) whose essence is that it is theoretically possible to encode a source so that its information can be transmitted, without almost any error through a noisy channel, at any rate below a limiting rate given by the channel capacity; in other words, if the source entropy exceeds the channel capacity, then the equivocation $H(X|Y)$ cannot be made arbitrarily small using any code.
\subsection{Entropy of Continuous-State Systems}
How would all the concepts---to begin with, the Shannon entropy---presented for a discrete random variable $X$ be extended for the case of a continuous random variable? We address this question by allowing the finite set of $n$ discrete points $\{x_i\}$---the letters of $X$---to become progressively numerous in the interval $[\min\{x_i\},\max\{x_i\}]$. We assume that as $n\rightarrow\infty$, the density of the points approaches a normalized density, $m(x)$: $\lim_{n\rightarrow\infty}\frac{1}{n}({\rm number\, of\, points\, in\,} [a,b])=\int_a^bm(x)dx$, $\forall[a,b]\subset[\min\{x_i\},\max\{x_i\}]$. We have implicitly also assumed that the limiting process is sufficiently well-behaved such that as $x_{i+1}-x_i$  tend to zero with increasing $n$, $n(x_{i+1}-x_i)\rightarrow m(x_i)^{-1}$ $\forall i$. The probability mass function $p_i$, used in defining the Shannon entropy independently goes to some probability density, say, $p(x)$; obviously, $p_i=p(x_i)(x_{i+1}-x_i)$ in the limit. Thus, we have $p_i\rightarrow p(x_i)/[nm(x_i)]$. Plugging this in the definition of the Shannon entropy, $H(X)$, we get
that the Shannon entropy for the discrete $X$ approaches $H^D(X)\equiv-\int p(x)\log[{p(x)}/nm(x)]dx$. Since $\log n$ term diverges, the appropriate finite measure for the entropy for the resultant continuous random variable is then defined as the \emph{limiting density of discrete points}
\begin{eqnarray}
h_{J}(X)\equiv H^D(X)-\log n=-\int p(x)\log\left[\frac{p(x)}{m(x)}\right]dx.\qquad
\end{eqnarray}
We are using subscript $J$ to indicate that Jaynes proposed this definition; he called it \emph{invariant information measure}.

The function $m(x)$ is called \emph{invariant measure}---invariant, because $p(x)$ and $m(x)$ transform in the same way so as to keep $h_J(X)$ invariant under changes in variable; and measure, (in the words of Jaynes) ``only to suggest the appropriate generalization, readily supplied if a practical problem should ever require it''. Note that, by definition, $m(x)$ is a pdf and hence $-h_J(X)=D(p(x)||m(x))$---the KL-divergence, analogously defined for the continuous random function as $\int p(x)\log[p(x)/m(x)] dx$. Consequently, owing to Jensen's inequality, $h_J(X)$ is necessarily always non-positive in stark contrast with the Shannon entropy's range that is always non-negative.

Of course, how to choose $m(x)$ is not very obvious. As a result, a historical predecessor of the limiting density of discrete points---viz., \emph{differential entropy},
\begin{eqnarray}
h(X)\equiv-\int_{{\rm supp}(p)} [p(x)\log{p(x)}]dx,
\end{eqnarray}
(when it exists) is much in vogue even though, in addition to not being non-negative [$h(X)\in(-\infty,+\infty)$], it is not even invariant under the change of variable. ${\rm supp}(p)$---the support of function $p(x)$---is also called the \emph{support set} of the corresponding random variable $X$. Defining $H(X^\Delta)\equiv -\sum_ip_i\log p_i=-\sum_i [p(x_i)\Delta]\log [p(x_i)\Delta]$ for the naively discretized probability $p_i=\int_{x_i}^{x_{i}+\Delta} p(x)dx$, the $h(X)$ may be seen as the limit where $H(X^\Delta)+\log \Delta$ approaches as $\Delta\rightarrow 0$. 

A meaning of the differential entropy can be obtained through the AEP for continuous random variable. One finds that the differential entropy is nothing but the logarithm of the side of the hypercube of the typical set. Consider a (product) sample space $\mathcal{S}^n$ where $\mathcal{S}$ is continuous and the corresponding random variable $X:\mathcal{S}\rightarrow\mathbb{R}$ has $p(x)$ as the pdf. Any sequence of iid random variables $\{X_i\}_{i=1}^{i=n}:\mathcal{S}^n\rightarrow\mathbb{R}^n$ has joint pdf $p(\{x_i\})$=$\prod_{i=1}^np({x_i})$. We have, by applying law of large numbers on the random variable $\log_b p(X)$, $-\frac{1}{n}\log p(\{X_i\})=-\frac{1}{n}\log \prod_{i=1}^np(X_i)=-\frac{1}{n}\sum_{i=1}^n\log p(X_i)\rightarrow E[-\log p(X)]=h(X)$ in probability. That is, for all positive $ \epsilon$,
\begin{equation}
\lim_{n\rightarrow\infty}P\left(\{X_i\}:p(\{X_i\})=2^{-n(h(X)\pm\epsilon)}\right)=1,
\end{equation}
where the base of the logarithm has been fixed to $2$. This is known as \emph{asymptotic equipartition property (AEP)}. One can, thus, define \emph{typical set} $\mathcal{T}_{(n,\epsilon)}$---the set of \emph{typical sequences} $\{x_i\}\in{\rm supp}(p)^n$ defined by $p(\{x_i\})\in[2^{-n\{h(X)+\epsilon\}},2^{-n\{h(X)-\epsilon\}}]$. Obviously $P(\mathcal{T}_{(n,\epsilon)})\approx1$; however, to measure the size of the set, the count of number of typical sequences is not a useful quantity as that is infinity. Therefore, one uses the concept of volume of the set, ${\rm V}(\mathcal{T}_{(n,\epsilon)})\equiv \int_{\mathcal{T}_{(n,\epsilon)}}\prod_{i=1}^ndx_i$. Using the AEP, one can show that $(1-\epsilon)2^{n\{h(X)-\epsilon\}}\le{\rm V}(\mathcal{T}_{(n,\epsilon)})\le2^{n\{h(X)+\epsilon\}}$. Therefore, we can conclude that typical set, $\mathcal{T}_{(n,\epsilon)}|_{\epsilon\rightarrow 0}$, is a small set containing almost all of the probability and occupies a hypercube of side approximately $[2^{nh(X)}]^{1/n}=2^{h(X)}$ in the support set of $\{X_i\}\in\mathbb{R}^n$. An interesting point is that the differential entropy of a discrete random variable is $-\infty$ [as seen by putting $p(x)=\sum_ip_i\delta(x-x_i)$] and this is compatible with the fact that the volume of the typical set, being countable, in such a case should be $2^{-\infty}=0$.

While the differential entropy obviously does not directly relate with the concept of the Shannon information, its practical applicative utility justifies its importance; some derivatives of the differential entropy do keep their information-theoretic meaning as in the discrete random variable case. For example, the KL-divergence using the fine and naive discretization $X^\Delta$ of continuous random variable $X$ is not plagued with the diverging $\log\Delta$ term. In particular, 
\begin{eqnarray}
&&D(p(x)||q(x))=\sum_ip(x_i)\Delta\log\left[\frac{p(x_i)\Delta}{q(x_i)\Delta}\right],\qquad\\
\implies&&D(p(x)||q(x))\xrightarrow[]{\Delta\rightarrow0}\int p(x)\log\left[\frac{p(x)}{q(x)}\right]dx.
\end{eqnarray}
It follows that the same is true for the mutual information, $I(X;Y)$, i.e.,
\begin{eqnarray}
I(X^\Delta;Y^\Delta)\xrightarrow[]{\Delta\rightarrow0}\int p(x,y)\log\left[\frac{p(x,y)}{p(x)p(y)}\right]dx=I(X;Y).\nonumber\\
\end{eqnarray}
Thus, whenever one is interested in the change of entropy---or in over words loss or gain of average information---the differential entropy is a good measure of it.

Defining \emph{joint differential entropy} and \emph{conditional differential entropy} respectively as 
\begin{eqnarray}
&&h(\{X_i\})=-\int p(\{X_i\})\log p(\{X_i\})\prod_idx_i\,\,{\rm and}\qquad\\
&&h(X|Y)=-\int p(x,y)\log p(x|y)dxdy,
\end{eqnarray}
we can show various relations analogous to the discrete case. Some important ones are listed below:
\begin{eqnarray}
&&h(X,Y)=h(X)+h(Y|X)=h(Y)+h(X|Y),\\
&&h(\{X_1,X_2,\cdots,X_n\})=\sum_{i=1}^{n}h(X_i|X_{i-1},\cdots,X_2,X_1),\,\qquad\\
&&I(X;Y)\ge0,\\
&&I(X;Y)=h(X)-h(X|Y)=h(Y)-h(Y|X),\\
&&I(X_1,\cdots,X_n;Y)=\sum_{i=1}^nI(X_i;Y|X_{i-1},\cdots,X_1),\qquad\\
&&D(p(x)||q(x))\ge0,\\
&&D(p(x,y)||q(x,y))=D(p(x)||q(y))+D(p(x|y)||q(x|y)).\,\qquad
\end{eqnarray}

Another reason why the differential entropy is useful is that it can be used to find the continuous probability distributions in accordance with the principle of maximum entropy just as done in the case of the discrete random variables; the diverging $\log\Delta$ term plays no part it its determination because while setting the derivative of the Lagrange function of the problem to zero, the contribution from the diverging term drops out. More often than not, one is interested in the testable information which gives values of expectable values of some functions of the random variable. In such a case, extremizing the Lagrange function of actually yields the distribution with \emph{maximum entropy}; the distribution is aptly called the \emph{maximum entropy distribution}. To show it, let us first specifically state the premise: We want to maximize the differential entropy that is defined using a pdf $p(x)$ that is non-zero only in the support set ${\rm supp}(p)$ of the random variable $X$ under consideration. There are, however, $1+m$ constraints---$\int_{{\rm supp}(p)}p(x)dx=1$ and $\int_{{\rm supp}(p)}p(x)f_k(x)dx=E[f_k(X)]=\left<f_k(x)\right>$ (say) $\forall k=\{1,2,\cdots,m\}$. Therefore, extremizing the corresponding Lagrange functional $L[p]$ with $1+m$ Lagrange multipliers $\lambda_k$ ($k=\{0,1,2,\cdots,m\}$), we get
\begin{eqnarray}
&&\frac{\delta L}{\delta p}=\frac{\delta}{\delta p}\left[-\int p\log_b pdx-\lambda_0\left(\int pdx-1\right)\right.\nonumber\\
&&\phantom{\frac{\partial L}{\partial p}=}\qquad\,\,\left.-\sum_{k=1}^m\lambda_k\left(\int  pf_kdx-\left<f_k\right>\right)\right]=0.\quad
\end{eqnarray}
This implies the \emph{Gibbs distribution} for the pdf:
\begin{eqnarray}
p(x)=p^{(G)}(x)\equiv \frac{b^{\sum_{k=1}^m[-\lambda_kf_k(x)]}}{Z(\lambda_1,\lambda_2,\cdots,\lambda_m)},
\end{eqnarray}
where the partition function, $Z=\int b^{\sum_{k=1}^m[-\lambda_kf_k(x_i)]}dx$. In fact, $p^{(G)}$ is actually a unique maximum because the non-negativity of $D(q||p^{(G)})$, where $q(x)$ is any other pdf that obeys the aforementioned constraints, implies that $-\int q\log qdx\le-\int q\log p^{(G)}dx=-\int p^{(G)}\log p^{(G)}dx$ with equality holding if $q=p^{(G)}$. In passing, we mention that the maximum entropy principle can be generalized to propose the \emph{minimum relative entropy principle} analogous to the case of the discrete random variables.

Analogous to the case of the source with finite alphabet, it is straightforward to define a quantification for a source having continuous alphabet. Considering a source as a generator of a discrete-time stochastic process with a continuous alphabet $\mathcal{A}$ as its state space, we define the \emph{differential source entropy} or \emph{differential entropy rate} as 
\begin{eqnarray}
h_s(\mathcal{A})\equiv\lim_{n\rightarrow\infty}\frac{1}{n}h(X_1,X_2,\cdots,X_n),
\end{eqnarray}
given the limit exists. Note that the differential entropy rate can be defined for a stochastic process whether or not a source is in the picture. It can be shown that for a stationary source, the differential source entropy exists and 
\begin{eqnarray}
h_s(\mathcal{A})=\lim_{n\rightarrow\infty}h(X_n|X_{n-1},X_{n-2},\cdots,X_1)
\end{eqnarray}
holds true.

As an aside, we mention a popular generalization of $h(X)$. Since the differential entropy is actually a functional of probability distribution defined using a (at least) twice differentiable concave function, $-p\log p$, in principle, one could define a generalization of $h(X)$---some generalized entropy functional---by replacing the function $-p\log p$ with some twice differentiable concave function, $\phi(p)$. For the special choice of one parameter family of concave functions, $\phi_\alpha(p)=(p-p^\alpha)/(\alpha-1)$ where $\alpha>0$, one gets the so-called \emph{Havrda--Charv\'at entropy} or the \emph{$\alpha$-order entropy}:
\begin{eqnarray}
h_\alpha(X)\equiv\frac{1}{\alpha-1}\left[1-\int p(x)^\alpha dx\right].
\end{eqnarray}
Obviously, $\alpha\rightarrow1$ yields the usual differential entropy, i.e., $h_1(X)=h(X)$. It is also obvious that the R\'enyi entropy, with its definition tuned to the continuous variable case, is related monotonically to $h_\alpha$ as follows:
\begin{eqnarray}
R_\alpha=\frac{1}{1-\alpha}\log\left[1+(1-\alpha)h_\alpha\right].
\end{eqnarray}
It is easy to note that analogous definition of the Havrda--Charv\'at entropy may be given for the discrete-state systems.

We conclude the discussion of about entropy---a measure quantifying uncertainty---by introducing a necessary condition for establishing causality. The condition uses the concept of mutual information tuned to define a useful quantity called \emph{transfer entropy}. Consider measuring an evolving physical quantity denoted using a random process, without any loss of generality, $Y_t$; subscript $t$ has been used to emphasize that one need not consider discrete time. Also, there is no restriction on whether the corresponding alphabets are discrete or continuous as mutual information is well-defined in either cases. Furthermore, suppose that there is an external noisy influence on $Y_t$---let the resulting additional perturbation be denoted by $X_t$. The transfer entropy is, then, defined as
\begin{eqnarray}
T_{X\to Y}\equiv I(Y_{t_n};Y_{t_{n-1}},X_{t_{n-1}})-I(Y_{t_n};Y_{t_{n-1}}),\quad
\end{eqnarray}
which, by construction, is non-negative. Of course, unlike the mutual information, the transfer entropy is not symmetric, i.e., $T_{X\to Y}\ne T_{Y\to X}$, in general. We note that if $X$ does not affect $Y$ causally at all, then $H(Y_{t_n}|Y_{t_{n-1}},X_{t_{n-1}})=H(Y_{t_n}|Y_{t_{n-1}})$, and hence, $T_{X\to Y}=0$. The degree of causality is, thus, captured by the value of transfer entropy. We might add that while this is not a watertight criterion for causality but it has proven to be a useful one in many studies.
\section{Uncertainty in Deterministic Chaos}
\hskip 0.8 cm``\emph{One must still have chaos in oneself to be able \phantom{xxxxxxxxxxxxx}to give birth to a dancing star.}''---Nietzsche
\\
\\
\indent {\LARGE\textgoth{C}}haos is the phenomenon responsible in making a nonlinear, low dimensional, deterministic dynamical system uncertain. This uncertainty manifests itself as an observer's inability to predict a chaotic system's future. For our purpose, it suffices to consider only autonomous, one-dimensional, bounded, noninvertible maps as simple models to illustrate the ideas behind chaos. Ergo, unless otherwise specified, all the ideas and concepts introduced henceforth are tuned to discuss mostly such systems. 
\subsection{Lyapunov Exponent}
Generally speaking, a $d$-dimensional discrete-time continuous-state deterministic dynamical system is specified by a $d$-dimensional function $\bs{f}$ that maps a $d$-dimensional \emph{phase space} $\mathbb{P}$---e.g., a compact subset of  $\mathbb{R}^d$---onto itself at every discrete time-step $n$ (where $n\in\mathbb{Z}_\ge{0}$ without any loss of generality); the resulting dynamics generated by the mapping is denoted as a \emph{map}: $\bs{x}_{n+1}=\bs{f}(\bs{x}_n)$, where $\bs{x}_n\in\mathbb{P}$ is a \emph{phase point} at time $n$, and the sequence $\{\bs{x}_n\}_{n=0}^{n=\infty}$ (or equivalently, $\{\bs{f}^n(x_0)\}_{n=0}^{n=\infty}$) is the phase space is a \emph{phase trajectory} or \emph{phase orbits}. The map is called \emph{autonomous} if $\bs{f}$ is explicitly independent of $n$. 

The local behaviour of a phase trajectory $\{\bs{x}_n\}$ can be understood by linearizing the map. If $\delta\bs{x}_n$ be the infinitesimal perturbations about $\bs{x}_n$, then its evolution is given by
\begin{eqnarray}
\delta\bs{x}_{n+1}={\bs{\nabla} \bs{f}}(\bs{x_n})\cdot\delta\bs{x}_{n},\,{\rm where}\,{\bs{\nabla} \bs{f}}(\bs{x_n})\equiv\left.\frac{\partial \bs{f}}{\partial{\bs x}}\right|_{\bs{x}=\bs{x}_n}\quad
\end{eqnarray}
is the Jacobian matrix calculated at $\bs{x}_n$. Naturally,
\begin{eqnarray}
\delta\bs{x}_{n}=\prod_{k=0}^{n-1}{\bs{\nabla} \bs{f}}(\bs{x_k})\cdot\delta\bs{x}_{0}.\quad
\end{eqnarray}
In a rather general variety of  maps, \emph{Oseledec's multiplicative ergodic theorem} guarantees existence of the \emph{Lyapunov exponents} (LE), $\lambda^i(x_0)$ ($i=1,2,\cdots,d$)---defined as the logarithm of the eigenvalues of the matrix 
\begin{eqnarray}
{\sf M}(\bs{x}_0)\equiv\lim_{n\rightarrow\infty}\left[\left(\prod_{k=0}^{n-1}{\bs{\nabla} \bs{f}}(\bs{x_k})\right)^T\left(\prod_{k=0}^{n-1}{\bs{\nabla} \bs{f}}(\bs{x_k})\right)\right]^{\frac{1}{2n}}\nonumber\\
\end{eqnarray} 
---for almost every trajectory $\{x_n\}$. Actually, the theorem is essentially a fact about the product of non-commutating square Jacobian matrices in the spirit of the law of large numbers.  

For a one dimensional map denoted by ${x}_{n+1}={f}({x}_n)$, ${\bs{\nabla} \bs{f}}(\bs{x_n})={df(x_n)}/{dx}$ is just a real number parametrized by $n$; we denote the number by $\Delta(n)$. Hence, ${\sf M}({x}_0)$ is a real positive number whose logarithm is the only LE, $\lambda(x_0)$, of the map. We have $\delta{x}_{n}=\prod_{k=0}^{n-1}\Delta(k)\delta x_0$ and hence,
\begin{eqnarray}
&&\lambda(x_0)=\ln{\sf M}({x}_0)=\lim_{n\rightarrow\infty}\frac{1}{n}\sum_{k=0}^{n-1}\ln|\Delta(k)|,\qquad\\
\implies&&\lambda(x_0)=\lim_{n\rightarrow\infty}\lim_{\delta x_0\rightarrow0}\left[\frac{1}{n}\ln\left|\frac{\delta{x}_{n}}{\delta{x}_{0}}\right|\right].
\end{eqnarray}
We observe that we may think about the above relation as if there is `random variable' with an alphabet $\{\ln|\Delta(k)|\}_{k=0}^{k=\infty}$, and whose sample mean converges to the LE in line with the law of large numbers.

An intuition about how LE is related to unpredictability, and hence randomness, may be gained as follows: Note that $n\sim(\ln|{\delta{x}_{n}}|/|{\delta{x}_{0}}|)/\lambda$. Assume $\lambda$ is positive. Consider two initial conditions so close that the value of $\delta x_0$ less than what an observer can resolve; in other words, the two initial conditions appear same to the observer. Now the two initial conditions time-evolve to separate out farther. At the time $n$, when $\delta x_n$ is equal to what the observer can resolve, she suddenly finds that the same two initial conditions have given rise to two different outcomes. Thus, the observer has every right to consider the system to behave unpredictably from time $n$ onwards---practically, the system appears to behave as random to her. In this context, $n$ may be seen a \emph{time horizon of prediction} because till that time the two initial conditions give rise to same unique predictable outcome (within the resolving capability of the observer).

The LE is dependent on the initial condition and so, given that the map is assumed to be smooth (at worst finitely piecewise smooth), each trajectory has a unique LE (if it exists). Moreover, since the limit $n\rightarrow\infty$ is taken in the definition of the LE, the LE of the trajectory should depend on its asymptotic behaviour. In this context, the concept of \emph{forward limit set} or \emph{$\omega$-limit set} of a trajectory $\{{f}^n(x_0)\}$ is useful: $\omega(x_0)\subseteq\mathbb{P}$ is a forward limit set of $\{{f}^n(x_0)\}$ if for any $x\in\omega(x_0)$, $\forall\epsilon\in\mathbb{R}_{>0}$ and $\forall N\in\mathbb{Z}_{\ge0}$, $\exists n>N$ such that $|f^n(x_0)-x|<\epsilon$. Essentially, it is a fixed set of points that find the trajectory infinitely often in their arbitrary small neighbourhoods.

Certain trajectories are their own {$\omega$-limit sets}. For example, consider a \emph{fixed point} $x^*$---solution of $x=f(x)$---and consider the trajectory $\{f^n(x^*)\}$ which essentially is just the fixed point at all times; the {$\omega$-limit set} $\omega(x^*)$ is the singleton set $\{x^*\}$. Next consider any \emph{$m$-period orbit} ($m\in\mathbb{N}$) consisting of $m$ distinct phase points, $x^{(1)},x^{(2)},\cdots x^{(m)}$ such that $x^{(k+1)}=f(x^{(k)})$ $\forall k\in\{1,2,\cdots,m\}$ and $x^{(m+1)}=x^{(1)}$; a fixed point is a $1$-period orbit with $x^{(1)}=x^*$ and \emph{$m$-period orbit} is \emph{$pm$-period orbit} ($p\in\mathbb{N}$) as well unless we demand distinct phase points in the definition of the periodic orbits. We shall refer an $m$-period orbit ($m>1$) as \emph{periodic orbit} for brevity and to distinguish it from fixed points. Obviously, the {$\omega$-limit set} of the trajectories $\{f^n(x^{(k)})\}$ $\forall k\in\{1,2,\cdots,m\}$ is the periodic orbit itself.

It is apposite to say that an $\omega(x_0)$ attracts a trajectory $\{f^n(x'_0)\}$ if $\omega(x'_0)\subseteq\omega(x_0)$ where $x'_0$ is some other initial phase point; the set of all such $x'_0$ constitutes the \emph{basin of attraction} of $\omega(x_0)$. One refers an $\omega$-limit set with a basin of attraction with non-zero measure as an \emph{attractor}. Obviously, an attractor is $f$-\emph{invariant} in the sense that $f(\omega(x_0))=\omega(x_0)$. This is due to the easily seen fact that if $x_0\in\omega(x_0)$, then $f^n(x_0)\in\omega(x_0)$ $\forall n\in\mathbb{N}$. A formal requirement---although it appears trivial in the light of the {$\omega$-limit sets} of periodic orbits---is that the attractor should contain a dense trajectory so that the attractor is \emph{minimal} closed set in the sense that none of its proper closed subsets is invariant and has a basin of attraction with non-zero measure.

The LE of any trajectory that is attracted towards a fixed point or a periodic orbit must be negative. The LE of a fixed point or a periodic orbit can be positive, zero, or negative respectively characterizing whether the set is \emph{unstable, neutrally stable, or stable}. Now coming to the trajectories that are bounded but are seemingly ever-wandering in the phase space in the sense that they are asymptotically attracted towards neither any fixed points nor any periodic orbits, in one dimensional maps, one can envisage two interesting possibilities: (i) The trajectory is \emph{quasiperiodic}, i.e., the trajectory is not asymptotically $m$-periodic ($m\in\mathbb{N}$) and has LE equal to zero; or (ii) the trajectory is \emph{chaotic}, i.e., the trajectory is not asymptotically $m$-periodic ($m\in\mathbb{N}$) and has positive LE. 

We are specifically interested in the chaotic trajectories as they lead to unpredictable behaviour leading to uncertainity. We want to know where do the chaotic trajectories settle if at all. Naturally, we are looking for an attractor associated with such trajectories. To this end, we define \emph{chaotic attractor} as follows: If $x_0\in\omega(x_0)$ for a chaotic trajectory $\{f^n(x_0)\}$, then $\omega(x_0)$ is called a \emph{chaotic set}; a chaotic set that is an attractor is called a chaotic attractor. By construction, a chaotic attractor has a chaotic orbit that being dense in the attractor, renders the attractor minimal as required. It should be kept in mind that not all systems showing chaos have chaotic attractors; however, here we shall be interested in the ones that do have them.

Obviously, the LE's corresponding to all the trajectories emanating in the basin of attraction of a chaotic attractor are all equal. Therefore, one could talk about a unique LE of an attractor. Such an LE is expected to be independent of the initial conditions; hence should be obtainable from the stationary pdf for the distribution of the phase points of the attractor; this pdf is called an \emph{invariant density} and is conventionally denoted as $\rho^*(x)$, independent of $x_0$. In other words, we are contemplating an ergodic chaotic attractor, $\mathbb{A}\subseteq\mathbb{P}$, with LE given by
\begin{eqnarray}
\lambda=\lim_{n\rightarrow\infty}\frac{1}{n}\sum_{k=0}^{n-1}\ln\left|\frac{df(x_k)}{dx}\right|=\int_\mathbb{A}\ln\left|\frac{df(x)}{dx}\right|\rho^*(x)dx.\,\,\qquad
\end{eqnarray}
Of course, if almost all the trajectories in the phase space $\mathbb{P}$---to be conveniently taken as some compact (closed and bounded)  interval of $\mathbb{R}$---go to the chaotic attractor then it makes sense to talk about a single LE for the entire map. %

Calculating the LE analytically is possible for only certain simple enough maps. Sometimes it helps to use the fact that two maps, $f:\mathbb{P}\rightarrow\mathbb{P}$ and $g:\mathbb{P}'\rightarrow\mathbb{P}'$---when \emph{topologically conjugate}---have same LE (considering typical orbits). Topological conjugacy means that $\exists$ a homeomorphism (i.e, one-to-one, onto, continuous function with continuous inverse) $h:\mathbb{P}\rightarrow\mathbb{P}'$, such that $f=h^{-1}\circ g\circ h$. In other words, a map $x_{n+1}=f(x_n)$ is  topologically conjugate to another map $y_{n+1}=g(y_n)$ through the existence of a homeomorphism $h$ such that $f(x)=h^{-1}(g(h(x)))$. Imposing fairly practical mild conditions on the derivative of $h$, it is easy to show that the LEs for the two orbits, $\{f^n(x_0)\}$ and $\{g^n(h(x_0))\}$, are same. However, how to find such homeomorphisms in general, even if they exist, is mostly a matter of trial-and-error. For example, consider the \emph{symmetric tent map},
\begin{equation}
x_{n+1}=T_r({x_n})\equiv
\left\{
\begin{array}{lll}
0.5{rx_n}&  \phantom{x} & 0\le x_n\le0.5\,,  \\
{0.5r(1- x_n)}&   \phantom{x} &0.5<x_n\le1,    
\end{array}
\right.
\end{equation}
and the \emph{logistic map}, 
\begin{equation}
y_{n+1}=L_r(y_n)\equiv ry_n(1-y_n)
\end{equation}
---both defined in the interval $[0,1]$. It is easy to see that $h(T_4(x))=L_4(h(x))$ $\forall x\in[0,1]$, for homeomorphism, $h(x)=\sin^2(\pi x/2)$. Thus, $T_4$ and $L_4$ are topologically conjugate and have same LEs, i.e., $\ln2$. It is interesting to note that the analytical form of chaotic orbits in $L_4$ is known exactly---$x_n=\sin^2(2^n\sin^{-1}\sqrt{x_0})$.

While the existence of chaotic attractor and hence positive LE for a finite measure of points is practically enough for a non-mathematician to call a system chaotic, there is no complete consensus on the definition of chaos. For the sake of completeness, we present one definition---\emph{Devaney's Chaos}---that has rather broad appeal among mathematicians: If $\mb{A}$ be an $f$-invariant subset of $\mb{P}$ that is endowed with a metric, then the continuous map $f:\mb{A}\rightarrow\mb{A}$ is called chaotic on satisfying following properties:
\begin{enumerate}
\item \emph{Topological transitivity}: For all open subsets $\mc{U}$ and $\mc{V}$ of $\mb{A}$, $\exists n\in\mb{N}$ such that $f^{n}(\mc{U})\cap\mc{V}\ne\phi$.
\item\emph{Dense periodic orbits}: The set of all its periodic orbits is dense in $\mb{A}$.
\item\emph{Weak sensitivity on initial conditions}: $\exists\varepsilon>0$ $\forall x_0$ and $\forall \delta>0$ such that $\exists n>0$ and $\exists x'_0$, $|x_0-x'_0|<\varepsilon$ and $|f^n(x_0)-f^n(x'_0)|>\delta$.
\end{enumerate}
Loosely speaking, the subsets of the chaotic attractor we discussed earlier are intuitively expected to follow the condition of topological transitivity as there is a dense orbit (although there exist examples when this implication may not be strictly true). Dense set of unstable periodic orbits are known to be embedded in a chaotic attractor; it has the effect that it pushes the trajectories in the attractor around and does not let them settle down. Positive LE is actually much a stronger requirement that implies the weak sensitivity on initial conditions. We point out the interesting fact that,  under not so restrictive conditions, the sensitivity---rather than being an independent requirement---follows from the topological transitivity and the existence of dense periodic orbits.
\subsection{Invariant Density}
The invariant density we are interested in essentially tells us about the probability of finding phase points of a trajectory in any closed subset $\mathcal{I}(x,\epsilon)\subset\mathbb{P}$ denoting an interval of length $\epsilon$ about the phase point $x\in\mathbb{P}$. The pdf of interest is, thus,
\begin{eqnarray}
\rho(x;x_0)=\lim_{n\rightarrow\infty}\frac{1}{n}\sum_{k=0}^{n-1}\delta(x-f^k(x_0))\label{eq:nd}
\end{eqnarray}
and the probability of finding a phase point of trajectory $\{f^n(x_0)\}$ in $\mathcal{I}(x,\epsilon)$ is
\begin{eqnarray}
\nu(\mathcal{I}(x,\epsilon);x_0)=\int_{\mathcal{I}(x,\epsilon)}d\nu=\int_{x-\frac{\epsilon}{2}}^{x+\frac{\epsilon}{2}}\rho(x;x_0)dx.\quad
\end{eqnarray}
Note we are not using the notation $P$ but $\nu$ because $P(\mathcal{I}(x,\epsilon))$ notationally may ambiguously mean the normalized Lebesgue measure, i.e., the length of $\mc{I}$ divided by the length of $\mathbb{P}$. The $\nu$ we are using, thus, is \emph{natural measure} generated by the map $f$:
\begin{eqnarray}
\nu(\mathcal{I}(x,\epsilon);x_0)=\lim_{n\rightarrow\infty}\frac{1}{n}{\rm card}\left[\{f^k(x_0)\}_{k=0}^{k=n} \cap \mathcal{I}(x,\epsilon)\right],\,\,\qquad
\end{eqnarray}
where `card' denotes the cardinality of the corresponding set. Note that when we use $d\nu=\rho dx$, we are implicitly assuming the measure is differentiable almost everywhere; otherwise, one should ideally use $\nu$ and omit any reference to $\rho$.

It is rather obvious that, by construction, the natural measure is invariant (under the action of $f$): $\nu(\mathcal{I}(x,\epsilon);x_0)=\nu(f^{-1}\mathcal{I}(x,\epsilon);x_0)$ $\forall x,\epsilon$. This is because all the iterates in the pre-image of the interval ends up being in the interval and no other points can enter the interval. It means that the corresponding density $\rho(x)$---suppressing $x_0$ for brevity---is invariant as well. Notationally, we therefore denote the invariant \emph{natural density} with an asterisk in the superscript: $\rho^*(x;x_0)$.

One can become vividly conscious about the density's dependence on $x_0$ on realizing that a chaotic attractor is known have countably infinite number of unstable periodic orbits embedded into it. Specifically, this fact means that if $x_0$ is a point of any such periodic orbit $\{x^{(k)}\}$, then the resulting density given by Eq.~(\ref{eq:nd}) is invariant. Thus, there are infinite invariant densities all dependent on $x_0$. The natural density we are interested in is independent of $x_0$ and describes the distribution of almost all points in the chaotic attractor. Now the question is starting from a generic distribution of phase points, is that natural density reached as time evolves?

As it turns out that just ergodicity is not always enough for an affirmative answer to this question. The existence of invariant density implies that certain amount of mixing should be present in the system---a set of initial conditions should be expanded and folded back into the phase space again and again over time just like kneading a flour dough. Since the definition of chaos itself not agreed upon, what kind of precise mixing property it has can only be dealt case by case; how convergence to the invariant density would occur due to that mixing is another question. Thus, we take a pragmatic route: We assume the systems we are going to consider have a rather \emph{strong mixing} property, i.e., mathematically,  $\forall\mc{U},\mc{V}\subset\mb{A}$, $\lim_{n\rightarrow\infty}\nu(f^{-n}(\mc{U})\cap\mc{V})=\nu(\mc{U})\nu(\mc{V})$. (Merely by taking $\mc{U}=\mc{V}=\mc{W}$, say an invariant set, strong mixing can be seen to imply ergodicity, i.e., $\nu(\mc{W})=0$~or~$1$.) Essentially, one is looking at all the points that arrived in the set $\mc{U}$ from infinite past and asking how much of them are filling any fixed set $\mc{V}$ in the same proportion $[\lim_{n\rightarrow\infty}\nu(f^{-n}(\mc{U})\cap\mc{V})]/[\nu(\mc{V})]$ as the set $\mc{U}$'s proportion is elsewhere, i.e., $\nu(\mc{U})$. 

In passing, we remark that there is a related concept of \emph{metric transitivity}---neither always implies nor is always implied by topological transitivity---that means for any two sets $\mc{U}$ and $\mc{V}$ with non-zero measures, $\nu(f^{-n}(\mc{U})\cap\mc{V})>0$ for some $n$. It is easy to show that it implies ergodicity (in fact, for measure preserving transformations, it is equivalent to ergodicity): Assume a metrically transitive system (also called \emph{metrically indecomposable} or \emph{metrically irreducible} system) that is not ergodic. So we have an invariant set $\mc{W}$ such that $\nu(\mc{W})\in(0,1)$. Then $\nu(\mc{W}^c)\in(0,1)$. Due to the assumed metric transitivity, for some $n$, $\nu(f^{-n}(\mc{W})\cap\mc{W}^c)>0$ which cannot be true because $f^{-n}(\mc{W})=\mc{W}$ and $\nu(\mc{W}\cap\mc{W}^c)=0$. Hence, by contradiction, metrically transitive systems are ergodic.

The evolution of any initial density under a map---$x_{n+1}=f(x_n)$---is given by the \emph{Frobenius--Perron equation}: 
\begin{eqnarray}
&&\rho_{n+1}(x)=L_{\rm FP}[\rho_n(x)]\equiv\int \delta(x-f(y))\rho_n(y)dy,\label{eq:FP1}\qquad\,\,\\
\implies&&\rho_{n+1}(x)=L_{\rm FP}[\rho_n(x)]=\sum_i\frac{\rho_n(y_i)}{\left|\frac{df(y_i)}{dx}\right|},\qquad
\end{eqnarray}
where $y_i$'s are the roots of the equation $x=f(y)$ [recall Eq.~(\ref{eq:conversionp})]. Note that this equation is just a compact way of writing how the probability density's functional form changes from $\rho_{n}(x_{n})$ to $\rho_{n+1}(x_{n+1})$ under the change of variable---$x_n\rightarrow x_{n+1}$; Eq.~(\ref{eq:FP1}) is basically the generalization of the obvious identity: $\delta(x-f(x_0))=\int \delta(x-f(y))\delta(y-x_0)dy$. Obviously, $\rho^*(x)$ is the eigenfunction of the \emph{Frobenius--Perron operator} $L_{\rm FP}$ with eigenvalue equal to one. For a mixing system, any sufficiently smooth initial density $\rho_0(x)$ approaches (typically exponentially fast) a limiting density $\rho_\infty(x)$ which is the invariant density $\rho^*(x)$ obtained using the evolution of a single trajectory ad infinitum with any generic initial condition. 

A useful fact is worth mentioning in this context. Since any trajectory of a map can be mapped to a corresponding equivalent trajectory of its topological conjugate map, this fact can be sometimes used to find invariant densities. For example, the invariant density $\rho_{T_4}^{*}(x)$ of the symmetric tent map, $T_4$, is known to be unity (proof is along the line done for the Bernoulli map later in this chapter); thus, the invariant density $\rho_{L_4}^{*}(y)$ of topologically conjugate logistic map, $L_4$, can be found by realizing that the two densities are essentially connected through a coordinate transformation: $y=h(x)$. Hence, $\rho_{L_4}^{*}(y)=[\pi\sqrt{y(1-y)}]^{-1}$.

Mixing leads to loss of memory in the system and this fact is nicely captured by the autocovariance $C(m)$ that vanishes exponentially fast with $m$ in such systems; here, $C(m)\equiv\lim_{n\rightarrow\infty}[n^{-1}\sum_{i=0}^{n-1}\Delta f^i(x_0)\Delta f^{i+m}(x_0)]$ with $\Delta f^k\equiv f^k(x_0)-\lim_{n\rightarrow\infty}[n^{-1}\sum_{i=0}^{n-1}f^i(x_0)]$. For ergodic maps, obviously, $C(m)$ can be written as integral using the natural invariant density. Arguably, such systems are somewhat similar to the Markov process---an analogy that can be made very precise for piecewise linear maps with $|df(x_k)/dx|>1$ (wherever defined) $\forall x_k\in\mb{P}=[0,1]$. In any such bounded, piecewise linear, expanding system, the invariant density should be a piecewise constant function. It is useful to consider a special type of partition---\emph{Markov partition}, $\mc{M}=\{m_i:i=1,2,\cdots,M\}$ having $M$ disjoint subintervals such that $m_i\cap f(m_j)\ne\phi$ if $m_i\subset f(m_j)$. The equivalent defining property of the Markov partition is that the end-points of the subintervals map onto other end-points and the interval between two end-points is mapped onto either a subinterval or a union of subintervals. A map can have more than one Markov partition.

We construct a probability row vector at time $n$: $\bs{\pi}(n)=[\pi_1(n)\,\pi_2(n)\,\cdots\,\pi_M(n)]$; $\pi_i(n)\equiv\int_{m_i}\rho_n(x)dx$. Then ${\bs{\pi}}(n)$ evolves under the map $f$ in accordance with the following relation:
\begin{eqnarray}
\pi_i(n+1)&=&\sum_{j=1}^M\pi_j(n){p}_{ji}\nonumber\\
&\equiv&\sum_{j=1}^M\pi_j(n)\left[\frac{\mu_L(m_j\cap f^{-1}(m_i))}{\mu_L(m_j)}\right],\quad
\end{eqnarray}
where $p_{ij}$ is the $(i,j)$th element of the row-stochastic $M\times M$ transition matrix ${\sf P}$ and where $\mu_L(m_i)$ is just the Lebesgue measure measuring the length of the subinterval $m_i$. While one can always (even with a partition that is not Markov) construct the coarse-grained picture of  a map through a Markov chain---$\bs{\pi}(n+1)=\bs{\pi}(n){\sf P}$, whether it is a true representation of the system under consideration not guaranteed. Is $\bs{\pi}^\infty$ of such a chain equal to $\rho_\infty=\rho^*$ of $\rho_{n+1}(x)=L_{\rm FP}[\rho_n(x)]$ written for the map under consideration? The magic of a Markov partition (if it can be found) is that the answer to the immediately preceding question is in affirmative even for finite $M$.

\subsection{Information and Fractal Dimensions}
The larger the value of the LE, the less predictable a system is in the following sense: If the state of a system is known with an accuracy of $\delta$ (say), then the state of the system can be predicted within the tolerance $\Delta$ (say) up to the time horizon of prediction, ${\lambda}^{-1}\ln(\Delta/\delta)$. Of course, we have assumed here that both $\delta$ and $\Delta$ are infinitesimal. The presence of the logarithm function means that the prediction time for strongly chaotic system is practically quite small and observing random unpredictable behaviour in such a system is a norm. 

Thus, the probabilistic description of chaos is inescapable because of the unpredictability introduced by the system's ultra-sensitive dependence on initial conditions. Naturally, the corresponding degree of surprise associated with a measurement performed on such a system could be expressed through the concept of the Shannon entropy that uses the underlying probabilistic description. The probabilistic distribution we are talking about is the one associated with the system's chaotic attractor because, after some transient period, any initial condition enters it and all the subsequent measurements are mathematically some function of the phase points in the attractor. Thus, it is the structure of the probabilistic structure of the chaotic attractor that we are interested in.

Interestingly, an attractor can be a \emph{strange attractor}, i.e., it is a \emph{fractal}; or in other words, its \emph{capacity or box dimension} (more accurately, \emph{Hausdroff dimension}), $D_B$, is greater than its \emph{topological dimension}, $D_T$. A chaotic attractor is often strange. A little discussion on this is useful for later purpose. $D_T$ of a set is the most common notion of dimension that we use everyday; its value for an object tells us about how many independent coordinates we require to specify a point uniquely on the object. It is a non-negative integer except for an empty set for which $D_T=-1$ by convention. In a rather different intuitive way---but still not the technically completely correct way---one may define $D_T$ of a set of points as one more than the $D_T$ of the smallest subset that when removed splits the set into two disjoint parts. One can invoke the idea of neighbourhoods to define $D_T$ properly but it is not needed for our purpose. Now, coming to $D_B$ of a set of points, we first consider it embedded in a Euclidean space of some topological dimension, say, $D$. Intuitive understanding of a dimension of an object may be informally summarized as: ${\rm bulk}\sim({\rm size})^{\rm dimension}$. For finding $D_B$ of a set, we find the minimum number $N(\epsilon)$ of $D$-dimensional hypercubes of infinitesimal side $\epsilon$ required to cover the set completely; the size of the set is proportional to $1/\epsilon$ in the unit of $\epsilon$ and its bulk has been captured by $N(\epsilon)$. Therefore, the dimension, $D_B\equiv\lim_{\epsilon\rightarrow0}[\log{N(\epsilon)}]/[\log(1/\epsilon)]$. $D_B$ can be any nonnegative real number.

We note that $D_B$ ignores the detailed nature of the distribution of points in the set of points and hence, it takes no account of the invariant density of an attractor. Therefore, $D_B$ cannot characterize a strange chaotic attractor completely for our purpose. Our purpose is to understand how much we can comprehend about a chaotic system by making physical measurements. Consider an ideal but conceptually non-trivial situation of an observer possessing detailed knowledge of the system to the extent that the invariant measure $\nu$ is known to her; moreover, she has a measuring instrument with uniform resolution $\epsilon$. This $\epsilon$ may be seen as the accuracy of the instrument and partially caused due to observational noise entering during the measurement process. This noise is assumed not to perturb the dynamical system. Thus, effectively, we have partitioned the phase space into similar cells of size $\epsilon$; this partition is called \emph{uncertainty partition} or \emph{U-partition}, $\mc{P}_U$. Let $N(\epsilon)$ number of such cells of the U-partition covers the attractor under question, then the average information gained by the aforementioned observer through a single measurement is $H(\mc{P}_U)=-\sum_{i=1}^{N(\epsilon)}\nu(c_i)\log\nu(c_i)$ where $c_i$ is the $i$th cell. The dimension defined through this is called the \emph{information dimension}, $D_I\equiv\lim_{\epsilon\rightarrow0}[H(\mc{P}_U)]/[\log(1/\epsilon)]$---a non-negative real number. Here, the size of the set is estimated by $1/\epsilon$ and the bulk by $\prod_{i=1}^{N(\epsilon)}[1/\nu_i^{\nu_i}]$ ($\nu_i\equiv\nu(c_i)$).

In case, the invariant measure is uniform, $D_B=D_I$ as expected. In general, however, $D_B\ge D_I$ (see below); if strict inequality holds, then the measure is called a \emph{fractal measure}. Looked from another angle, $D_I$ is a measure of information gain: the higher is the $D_I$ of an attractor, the more is the information gain on making an isolated measurement for a given infinitesimal resolution. Actually, the numbers $D_B$ and $D_I$ only partially characterize the corresponding attractor; in principle, an infinitely many such numbers are needed to characterize the attractor. 

To this end, it is been traditionally found helpful to use some auxiliary probability distributions---mimicking thermodynamic probability distributions---that in a way comb the natural invariant measure to extract its statistical characteristics. Specifically, such distributions are called \emph{escort distributions} given by
\begin{eqnarray}
p^{\rm ed}_i\equiv\frac{\nu_i^q}{\sum_i\nu_i^q};\quad \nu_i\ne0,\,q\in\mathbb{R}.
\end{eqnarray}
Observe that when $\nu(c_i)$ is written in the form $\exp(-b_i)$, $p^{\rm ed}_i$ looks exactly like the Gibbs' canonical distribution with $1/q$ and $b_i$ as respectively the analogues of temperature and energy; furthermore, $\sum_i\nu_i^q$ appears as the partition function, $Z(q)$. In this language, the R\'enyi entropy is nothing but $R_q=[\ln Z(q)]/(1-q).$
Here we allow $q$ to take negative values even though the corresponding negative-order R\'enyi entropies are further detached from the set of Shanon--Khinchin axioms.

A straightforward generalization comes to mind: Replace the Shannon entropy in the definition of $D_I$ to define a series of fractal dimensions, 
\begin{eqnarray}
D_q\equiv\lim_{\epsilon\rightarrow0}\frac{R_q}{\ln(\frac{1}{\epsilon})}=\frac{1}{q-1}\lim_{\epsilon\rightarrow0}\frac{\ln\sum_i\nu_i^q}{\ln\epsilon}\label{eq:dq2}.\quad
\end{eqnarray}
$D_q$ is called \emph{R\'enyi dimension} or \emph{generalized dimension} of order $q$. Obviously, $D_0=D_B$ and $D_1=D_I$; $D_2$ also has a special name---\emph{correlation dimension}, $D_c$. It is easy to see that $D_q\le D_{q'}$ $\forall q>q'$ by differentiating Eq.~(\ref{eq:dq2}) with respect to $q$ to find that the derivative, being equal to $-D(p^{\rm ed}_i||\nu_i)/(q-1)^2$, is always non-positive. Furthermore, $D_{+\infty}=\lim_{\epsilon\rightarrow0}{\ln(\max_i\nu_i)}/{\ln{\epsilon}}$ and $D_{-\infty}=\lim_{\epsilon\rightarrow0}{\ln(\min_i\nu_i)}/{\ln{\epsilon}}$, i.e., the former is a a measure of the most dense part of the fractal while the latter is a measure of the least dense part of the fractal. Actually, when a single dimension (i.e., $D_q$ with a particular $q$) is not enough to characterize a fractal, the fraction is essential nonuniform and is called a \emph{multifractal}. 

An alternate equivalent way of characterising a multifractal is to use the concept of \emph{pointwise dimension} and define \emph{singularity spectrum} as we do in what follows. The pointwise dimension is a local scaling exponent, $\alpha_i$, such that $\alpha_i=\lim_{\epsilon\rightarrow0}\ln\nu_i/\ln\epsilon$. Essentially, $\alpha_i$ (if it exists) captures the increase (via power law) in the number of points in the $i$th cell as its size increases. Suppose different cells have different pointwise dimensions whose collection is a continuum from, say,  $\alpha_{\rm min}$ to $\alpha_{\rm max}$. Assume that the number of cells with pointwise dimension between $\alpha$ and $\alpha+d\alpha$ is $dN(\alpha,\epsilon)\sim \epsilon^{-f(\alpha)}d\alpha$. $f(\alpha)$ is called the singularity spectrum; in this context, $\alpha$ is also called \emph{crowding exponent} or \emph{singularity exponent}. Note that $f(\alpha)$ is nothing but the box dimension of the set of cells with singularity exponent $\alpha$.

To see the connection between $D_q$ and $f(\alpha)$, we first note that, since $\nu_i\sim\epsilon^\alpha$, 
\begin{equation}
\sum_i\nu_i^q=\int_{\alpha_{\rm min}}^{\alpha_{\rm max}}g(\alpha')[\epsilon^{\alpha'}]^q\epsilon^{-f(\alpha')}d\alpha',
\end{equation}
where $g(\alpha')$ is some function independent of $q$. As $\epsilon\rightarrow 0$, $\min_{\alpha'}[q\alpha'-f(\alpha')]$ contributes the most to the integral. Taking derivatives with respect of $\alpha'$ at the assumed solution $\alpha(q)$ yielding the minimum gives: $f'(\alpha(q))=q$ and $f''(\alpha(q))<0$. This tells us that using this $\alpha$ in the singularity spectrum makes the latter a concave function with a maximum at $q=0$. Next, plugging in the solution of the integral in the definition of $D_q$, we find
\begin{equation}
\tau_q\equiv (q-1)D_q=q\alpha(q)-f(\alpha(q)). \label{eq:tauq}
\end{equation}
$\tau_q$ is sometimes called \emph{mass exponent}. Thus, $\tau_q$ and $f(\alpha)$ form a Legendre dual. Furthermore, it is clear from Eq.~(\ref{eq:tauq}) that $\alpha(\pm\infty)=D_{\pm\infty}$. 

The inverse Legendre transform is performed by noticing that $d\tau_q/dq=\alpha$:
\begin{equation}
f(\alpha)=q(\alpha)\frac{d\tau_{q(\alpha)}}{dq}-\tau_{q(\alpha)}.
\end{equation}
From this it is immediately clear that $f(\alpha)$ at $q=0$ is $D_0$, i.e., the maximum of the singularity spectrum is the box dimension of the multifractal. We conclude by bringing to the readers' attention another point: Since $f(\alpha)$ is concave, and $df(\alpha)/d\alpha=+\infty$ and $df(\alpha)/d\alpha=-\infty$ at $q=+\infty$ and  $q=-\infty$, respectively, it is clear that $\alpha_{\rm min}$ and $\alpha_{\rm max}$ must be $\alpha({+\infty})$ and $\alpha({-\infty})$, respectively. Thus, a generic singularity spectrum curve looks like a concave curve that has a maximum at $\alpha(0)$ and terminates at $\alpha_{\rm min}$ and $\alpha_{\rm max}$ vertically.
\subsection{Measurement and Dynamical Entropies}
When following a trajectory, the information loss (denoted by $-\Delta I$) on a subsequent measurement has a revealing connection with the LE of the trajectory. Consider the phase space to be U-partitioned into $N(\epsilon)$ number of cells. If the initial condition $x_0$ is equiprobable in any of the cells, then $H(\mc{P}_U)=\log N(\epsilon)$. As the system evolves, under a map $f$, the interval $\epsilon$ about $x_0$ changes by a factor $|df(x_0)/dx|$ leading to a change in the effective resolution such that the total number of new cells where the next iterate can be is $N(\epsilon)/|df(x_0)/dx|$. Let the resulting new partition be denoted by $\mc{P}_{fU}$. We have $H(\mc{P}_{fU})=\log [N(\epsilon)/|df(x_0)/dx|]$. Therefore, the information loss realized on making the second measurement is $-\Delta I\equiv -[H(\mc{P}_{fU})-H(\mc{P}_{U})]=\log|df(x_0)/dx|$. Obviously, the information loss averaged over the trajectory---i.e., replacing $x_0$ with its different iterates and averaging $-\Delta I$ over them---is nothing but, by definition, the LE of the trajectory. Given the positivity of the LE for the chaotic trajectories, chaos leads to information loss in the above sense. This conclusion based on two consecutive measurements in a chaotic system basically hints at the fact that here, in general, the U-partition becomes coarser: $ \mc{P}_{fU}\succ\mc{P}_{U}$.

Alternatively, chaos may be seen as a source that generates information when one conditions information gainable at a certain time step on all the measurements made in the earlier time steps. The idea is that in a series of sequential measurements, the new incremental information about the chaotic system is constantly acquired; this is not so in non-chaotic dynamics because after a while no new information about the system can be acquired through measurements. Mathematically, as the U-partition $\mc{P}_U$ evolves to a new partition $\mc{P}_{fU}$ under the map $f$, the information gained post the two measurements is given by the joint entropy, $H(\mc{P}_U\vee\mc{P}_{fU})$. Since $\mc{P}_U\vee\mc{P}_{fU}\prec \mc{P}_U\,{\rm or}\,\mc{P}_{fU}$, $H(\mc{P}_U\vee\mc{P}_{fU})\ge H(\mc{P}_U)\,{\rm or}\,H(\mc{P}_{fU})$. As the mixing of phase points happens in chaotic dynamics, we expect the partition products to become finer and finer over time.

We now formalize this for an infinite set of sequential measurements along an entire trajectory. Since the map may be non-invertible, we should use $f^{-1}$ rather than $f$ to characterize the aforementioned idea. One defines the \emph{topological entropy} as
\begin{eqnarray}
h_{\rm T}(f)\equiv \sup_{\mc{P}}\lim_{n\rightarrow\infty}\frac{1}{n}H\left(\bigvee_{k=0}^{n-1}f^{-k}\mc{P}\right).\quad\label{eq:te1}
\end{eqnarray}
The joint Shannon entropy here considers the elements of the partition to be equally probable. Obviously, there is no effect of (natural) measure in this definition and hence, the adjective---topological. Consequently, one is looking at how the number of elements of the partition product $\bigvee_{k=0}^{n-1}f^{-k}\mc{P}$ is increasing with $n$, i.e., how one is locating a phase point more precisely in the light of many measurements taken together. The division by $n$ and the limit makes the definition interpretable as an asymptotic rate. Furthermore, supremum takes out any dependence on the partition---i.e., the effect of the resolution of measurement apparatus---from the consideration as one wants $h_{\rm T}(f)$ to be an inherent property of the dynamical system. Unless the number of elements of the partition product $\bigvee_{k=0}^{n-1}f^{-k}\mc{P}$ increases exponentially with $n$, $h_{\rm T}(f)$ vanishes. By definition, $h_{\rm T}(f)\ge0$ for any map. Since calculating supremum is often not practical, the following alternative definition---proven to be equivalent---is useful:
\begin{eqnarray}
h_{\rm T}(f)\equiv \lim_{m\rightarrow\infty}\lim_{n\rightarrow\infty}\frac{1}{n}H\left(\bigvee_{k=0}^{n-1}f^{-k}\mc{P}_m\right).\quad\label{eq:te2}
\end{eqnarray}
Here $\{\mc{P}_m\}_{m=1}^{m=\infty}$ is a \emph{refining sequence} defined as the one such that $\mc{P}_{m+1}\prec\mc{P}_{m}$ $\forall m$ and for any arbitrary partition $\mc{P}$, $\exists$ some $\mc{P}_m\prec \mc{P}$. Just like for the LE, for calculating  $h_{\rm T}(f)$, sometimes it helps to use the fact that topologically conjugate maps have same topological entropies.

It is a natural next step to generalize the topological entropy so as to include the invariant measure of the chaotic attractor so that the probabilistic structure of the attractor is taken into consideration. To this end, one defines the \emph{Kolmogorov--Sinai} (KS) entropy:
\begin{eqnarray}
&&h_{\rm KS}(\nu,f)\equiv \sup_{\mc{P}}\lim_{n\rightarrow\infty}\frac{1}{n}H_\nu\left(\bigvee_{k=0}^{n-1}f^{-k}\mc{P}\right),\quad\label{eq:ks}\\
\rm{or,}\,
&&\,h_{\rm KS}(\nu,f)\equiv\lim_{m\rightarrow\infty}\lim_{n\rightarrow\infty}\frac{1}{n}H_\nu\left(\bigvee_{k=0}^{n-1}f^{-k}\mc{P}_m\right),\qquad\label{eq:ks2}
\end{eqnarray}
where $H_\nu(\mc{P})\equiv-\sum_i\nu(c_i)\log\nu(c_i)$, $c_i$'s being the elements of $\mc{P}$. If the measure is uniform, then it reduces to the definition of $H$ used in Eq.~(\ref{eq:te1}), as expected. In fact, $h_{\rm KS}(\nu,f)$---which is always a non-negative real number---can never be greater than $h_{\rm T}(f)$. Also, for a chaotic attractor the KS entropy can never exceed the LE. 

The KS entropy's analogy with the source entropy used in the information theory is rather apparent and given the use of invariant measure, it may be related to a sequence of conditional entropies just as the source entropy is alternatively defined through conditional entropy for stationary sources. The KS entropy is zero for non-chaotic deterministic system. If the KS entropy is not zero but finite, the corresponding system is defined to be chaotic. It is an alternative useful definition of chaos. For a stochastic system (with continuous valued random variable), the KS entropy can be shown to be infinite. In passing, we mention that the KS entropy is also called \emph{measure-theoretic entropy} or even \emph{metric entropy} although only measure (and not metric) has been used to define the entropy. 

The KS entropy measures the average extra amount of information gained about an initial condition with each subsequent measurement. This is due to the use of the \emph{dynamical refinement}---$\bigvee_{k=0}^{n-1}f^{-k}\mc{P}$---of an initial partition $\mc{P}$ in its definition. This should be apparent on recasting Eq.~(\ref{eq:ks}) as follows:
\begin{eqnarray}
&&h_{\rm KS}(\nu,f)= \sup_{\mc{P}}\lim_{n\rightarrow\infty}\frac{1}{n}\sum_{j=1}^{j=n-1}\left[H_\nu\left(\bigvee_{k=0}^{j}f^{-k}\mc{P}\right)\right.\nonumber\\
&&\phantom{xxxxxxxxxxxxxxxxxxxxxxxx}-\left.H_\nu\left(\bigvee_{k=0}^{j-1}f^{-k}\mc{P}\right)\right].\,\,\qquad
\end{eqnarray}
In order to understand this better, let us assume that the partition has $N$ elements denoted by $c_i$'s ($i=1,2,\cdots,N$). We pick an initial condition $x_0$ in one of the elements, say, $c_{i_0}$ ($i_0\in\{1,2,\cdots,N\}$). Then the trajectory $\{f^k(x_0)\}_{k=0}^{k=\infty}$ at a time $n=1$ is in another element, say, $c_{i_1}$. Naturally, $f^{-1}(c_{i_1})$ ($i_1\in\{1,2,\cdots,N\}$) must contain $x_0$; i.e., $x_0$ is in $c_{i_0}\cap f^{-1}(c_{i_1})$ that is element of the first refinement of $\mc{P}$, i.e., $\mc{P}\vee f^{-1}(\mc{P})$. Thus, $n$ subsequent measurements, in principle, locate $x_0$ in some element of $\bigvee_{k=0}^{n}f^{-k}\mc{P}$. In case $\bigvee_{k=0}^{\infty}f^{-k}\mc{P}$ has singleton sets as its elements, then any initial condition can be located precisely after infinite measurements. Any such special partition $\mc{P}$ for that this happens is called a \emph{generating partition} $\mc{G}$; since any dynamical refinement of $\mc{G}$ is also a generating partition, it makes sense to use the one with minimum number of elements. A generating partition, if fortunately found, can be used to study the corresponding system using \emph{symbolic dynamics}.

We choose any arbitrary partition, $\mc{P}=\{c_1,\,c_2,\cdots,\,c_N\}$, of the phase space: the size and shape of the elements, $c_i$'s, of partition can be arbitrary; $N$ is also chosen to suit the purpose. A finite symbol sequence $\varsigma_0,\varsigma_1,\varsigma_2,\varsigma_3,\cdots,\varsigma_{n-1}$ of symbols ($1$ to $N$) denotes all the time evolved sequences of length $n$, starting at some $x_0\in c_{\varsigma_0}$ such that $x_i\in c_{\varsigma_i}$. Although it is convenient to choose---as we do here---the symbols as numbers $1$ to $N$, it is not necessary in general: They can be a set of any distinct symbols that identifies the  corresponding elements of the partition uniquely. The set of all initial $x_0$'s that generate the same symbol sequence (of length $n$) is called an \emph{$n$-cylinder} and hence the probability of occurrence of a symbol sequence is determinable from the probability of the initial distribution of $x_0$. Thus, for ergodic map, it makes sense to work with the natural invariant density because then the probability of the symbol sequences would be stationary.

Given a symbol sequence $\varsigma_0,\varsigma_1,\varsigma_2,\varsigma_3,\cdots,\varsigma_{n-1}$, it is obvious that $\cap_{k=0}^{n-1}f^{-k}c_{\varsigma_k}$ locates initial condition, $x_0$, much more precisely than what $c_{\varsigma_0}$ alone does. One says that $\varsigma_0,\varsigma_1,\varsigma_2,\varsigma_3,\cdots,\varsigma_{n-1}$ is an \emph{$n$-word}, $W_n(\mathcal{P})$, induced by the partition $\mathcal{P}$. Therefore,
\begin{eqnarray}
H_\nu\left(\bigvee_{k=0}^{n-1}f^{-k}\mc{P}\right)=-\sum_{\{W_n(\mathcal{P})\}}P(W_n(\mathcal{P}))\ln P(W_n(\mathcal{P})),\nonumber\\\label{eq:Hnufin}
\end{eqnarray}
where $P(W_n(\mathcal{P}))$ denoted the probability of a particular $n$-word (which can denote more than one dynamically allowed sequences) and the sum is over all such $n$-words. When one explicitly realizes that
\begin{eqnarray}
P(W_n(\mathcal{P}))=\int_{n\textrm{-cylinder}}d\nu,
\end{eqnarray}
Eq.~(\ref{eq:Hnufin}) becomes more transparent.

A generalization in the line of generalized dimensions is straightforward: One defines \emph{generalized R\'enyi entropies} or \emph{dynamical R\'enyi entropies} or \emph{generalized Kolmogorov entropies} of order $q$ as
\begin{eqnarray}
K_q\equiv -\sup_{\mc{P}}\lim_{n\rightarrow\infty}\frac{1}{n}\frac{1}{q-1}\ln\left[\sum_{\{W_n(\mathcal{P})\}}P(W_n(\mathcal{P}))^q\right].\nonumber\\
\end{eqnarray}
Here, $q\in\mathbb{R}$. $K_q$ is a non-increasing function of $q$. Obviously, $K_0$ is the topological entropy and $K_1\equiv\lim_{q\rightarrow1}K_q$ is the KS entropy. In case the partition is generating, the supremum is automatically achieved; in that case, taking supremum is not needed. One can also define a spectrum of fluctuations around the KS entropy using the Legendre transformation of $(q-1)K_q$, just like the correspondence between the singularity spectrum, $f(\alpha)$ and the generalized dimension, $D_q$.
\subsection{Case Study of Bernoulli Map}
Let us now revise the concepts discussed till now through the case study of the \emph{Bernoulli map}, $B:[0,1)\rightarrow[0,1)$, given below:
\begin{equation}
x_{n+1} = B(x_{n})= \left\{ \begin{array}{lcl}
2x_{n} & \mbox{for} & 0\leq x_{n}<0.5,
\\ 2x_{n}-1 & \mbox{for} & 0.5\leq x_{n}<1.
\end{array}\right.
\end{equation}
We note that it is piecewise linear, expanding map with $dB/dx=2$ (except at $x=0$ and $x=0.5$ where $B(x)$ has one-sided derivatives equal to $2$) leading to $\lambda=\lim_{n\rightarrow\infty}{n^{-1}}\sum_{k=0}^{n-1}\ln|{dB(x_k)}/{dx}|=\log 2>0$ for any trajectory that is aperiodic. Therefore, there are chaotic orbits in the Bernoulli map (except, of course, when $x_0$ is a fixed point or a periodic point or the point of discontinuity $0.5$). Since the LE is independent of $x_0$, we expect that this map is ergodic and there exists a chaotic attractor with invariant density $\rho^*(x)$. 

In fact, the Bernoulli map is chaotic in Devaney's sense. The positivity of the LE already ensures that the defining condition of the sensitive dependence on initial conditions is satisfied by the map. Next we check the other two requirements for the existence of Devaney's chaos in the map. The map is topologically transitive as can be argued as follows: Consider any open subinterval of $[0,1)$. The action of the map is to double its size at each time step until the length exceeds unity, when only the extra length beyond one is folded back inside the phase space. It is, thus, obvious that the open subinterval under question must overlap at some time step with any other fixed open subinterval of $[0,1)$ and hence the map is topologically transitive. Lastly, the map has a dense set of periodic orbits. This fact may be comprehended by noticing that the periodic points can be graphically found by locating the intersections between $B^n(x)=x$ line and the lines corresponding to $B^n(x)$ in $B^n(x)$ versus $x$  plots ($\forall n\in\mb{N}$) because any $m$-periodic point is a solution of $x=B^m(x)$. It is easy to see that there are $2^{m}$ such equispaced intersection points for every $m$ and the distance between any two neighbouring ones becomes progressively smaller as $m$ increases. Therefore, for any subinterval of $[1,0)$---howsoever small---one can find large enough $m$ such that an $m$-period point lies in that subinterval. In other words, the set of periodic orbits are dense in $[1,0)$. The periodic orbits are all unstable.

Coming back to the issue of the invariant density, the Frobenius--Perron equation for the Bernoulli map is
\begin{equation}
\rho_{n+1}(x) = \frac{1}{2}\left[\rho_{n}\left(\frac{x}{2}\right)+\rho_{n}\left( \frac{1+x}{2}\right)\right].\label{eq:bmfp}
\end{equation}
Therefore, $\rho^*(x)$ must satisfy
\begin{equation}
\rho^*(x) = \frac{1}{2}\left[\rho^*\left(\frac{x}{2}\right)+\rho^*\left( \frac{1+x}{2}\right)\right],
\end{equation}
yielding $\rho^*(x)=1$ as the solution for the (normalized) invariant density. We can use this to find the LE alternatively by using $\lambda\equiv\int_\mathbb{A}\ln\left|\frac{dB(x)}{dx}\right|\rho^*(x)dx$ and realising that $\mathbb{A}=[0,1]$ for the Bernoulli map: The LE is found to be $\ln 2$ as should be the case.

However, whether any sufficiently smooth (normalized) $\rho_0(x)$ evolves to $\rho_\infty=\rho^*$ is a question that requires us to look back at Eq.~(\ref{eq:bmfp}) to find
\begin{equation}
\rho_{n}(x) = {2^{-n}}\sum_{k=1}^{2^{n}}\rho_{0}\left(\frac{k-1}{2^{n}}+\frac{x}{2^{n}}\right)\label{eq:bmrhon}
\end{equation}
by mathematical induction. Arguably, as $n\rightarrow\infty$, $\rho_\infty=\int_0^1\rho_0(x)dx=1$ (independent of initial condition) by changing the sum to integral. Hence, $\rho^*=\rho_\infty$, implying that this system is mixing. We further note that this convergence is exponentially fast due to presence of $2^{-n}=e^{-(\ln 2)n}$ factor. We can see another signature of it in the loss of memory of any chaotic trajectory.

To this end, let us find the autocovariance $C(m)$ for a chaotic trajectory $\{B^n(x_0)\}$. Defining $\left<B(x_0)\right>\equiv\lim_{n\rightarrow\infty}[n^{-1}\sum_{i=0}^{n-1}B^i(x_0)]$, $C(m)$ can be cast as follows:
\begin{eqnarray}
C(m) &&= \lim_{n\to \infty}\frac{1}{n}\sum_{i=0}^{n-1}\left[B^{i}(x_{0})B^{i+m}(x_{0})- \left<B(x_0)\right>^{2}\right]\nonumber\\
&&=  \lim_{n\to \infty}\frac{1}{n}\sum_{i=0}^{n-1}\left[B^{i}(x_{0})B^{m}(B^{i}(x_{0}))- \left<B(x_0)\right>^{2}\right].\nonumber\\
\end{eqnarray}
The Bernoulli map is mixing and hence ergodic; so, by making use of the invariant density that is independent of $x_0$, we have
\begin{eqnarray}
&&C(m) = \int_0^1\rho^*(x)xB^m(x)dx-\left[\int_0^1\rho^*(x)xdx\right]^2\qquad\\
\implies &&C(m)=\left(\frac{1}{12}\right)2^{-m}.\label{eq:cm2-m}\qquad
\end{eqnarray}
We have found the final expression (Eq.~(\ref{eq:cm2-m})) by mathematical induction. We note that the loss of correlation between two iterates is exponential in their time separation and the decay rate is same as that found for the convergence of $\rho_0$ to $\rho^*$.

The Bernoulli map is, thus, analogous to a Markov process and it so happens that it possesses a Markov partition $\mc{M}$ given by $\mc{M}=\{m_1,m_2\}\equiv\{[0,0.5),[0.5,1)\}$. We note $B(m_1)=B(m_2)=[0,1)$; hence, $m_i\cap B(m_j)\ne\phi$ iff $m_i\subset B(m_j)$ $\forall i,j\in\{1,2\}$ is trivially satisfied as required by a Markov partition. Also, it is obvious that $p_{ij}\equiv \mu_L(m_i\cap B^{-1}(m_j))/\mu_L(m_i)=0.5$ $\forall i,j\in\{1,2\}$. In other words, the transition matrix ${\sf P}$ is a $2\times2$ matrix with its every element equal to $0.5$. If $\bs{\pi}(n)$ is the two-component probability row vector, then we have:
\begin{equation}
\left[
\begin{array}{cc}
 \pi_1(n+1) & \pi_2(n+1)
\end{array}
\right]
=
\left[
\begin{array}{cc}
 \pi_1(n) & \pi_2 (n)
\end{array}
\right]
\left[
\begin{array}{cc}
0.5&0.5 \\
0.5& 0.5   
\end{array}
\right].\qquad
\end{equation}
 The Markov chain under consideration is irreducible and as it is finite, it is positive recurrent as well. Therefore, there exists a unique stationary distribution. Obviously, the stationary state of this Markov chain is $\bs{\pi}^*=[0.5\,0.5]$. (The result that the uniform distribution happens to be the stationary distribution is the consequence of the general fact that the necessary and sufficient condition for a stationary distribution of a Markov chain to be uniform is that the transition matrix be \emph{doubly stochastic}, i.e, $\sum_ip_{ij}=1$.) Furthermore, this Markov chain is aperiodic and so the stationary distribution is limiting distribution as well. Indeed, $\bs{\pi}(n)=\bs{\pi}(0){\sf P}^n=\bs{\pi}(0){\sf P}=[0.5(\pi_1(0)+\pi_2(0))\,0.5(\pi_1(0)+\pi_2(0))]=[0.5\,0.5]$ and hence $\bs{\pi}^\infty=[0.5\,0.5]=\bs{\pi}^*$ which is compatible with a coarse-grained version of $\rho^*=1$. In passing, we also note that since the detailed balance is trivially satisfied, the stationary distribution is an equilibrium distribution as well. This $\rho^*$ is however not a fractal measure as $D_B=D_I=1$ for the chaotic attractor of the Bernoulli map. 
 
 What is most interesting for the case of the Bernoulli map is that the Markov partition found above is also the generating partition $\mc{G}$ that induces a symbolic dynamics. Let $\mc{G}=\{g_0,g_1\}\equiv\{[0,0.5),[0.5,1)\}$.  Let us introduce a binary symbol $\varsigma_{0}\in\{0,1\}$ such that $\varsigma_{0}=0$ and $\varsigma_{0}=1$ respectively indicate that $x_0\in g_{0}$ and $x_0\in g_{1}$. First dynamical refinement of $\mc{G}$ is $\mc{G}\vee B^{-1}(\mc{G})=\mc{G}^{(1)}$ (say): explicitly, $\mc{G}^{(1)}=\{g_{00},g_{01},g_{10},g_{11}\}\equiv\{[0,1/4),[1/4,1/2),[1/2,3/4),[3/4,1)\}$. ($\mc{G}^{(0)}$ is defined to be $\mc{G}$.) So we can use a binary sequence---$\varsigma_{0},\varsigma_{1}$---to express that a trajectory $\{x_n\}_{n=0}^{n=1}$ locates $x_0$ in $g_{\varsigma_0\varsigma_1}$ where $\varsigma_{1}\in\{0,1\}$: The value of  $\varsigma_{1}$ tells us in which element of $\mc{G}^{(0)}$ (not $\mc{G}^{(1)}$) $x_1$ resides. Consequently, a trajectory $\{x_n\}_{n=0}^{n=\infty}$ through the infinite dynamical refinement $\mc{G}^{(\infty)}$ with elements $g_{\varsigma_0\varsigma_1\cdots}$ locates $x_0$ precisely as a unique number represented by a semi-infinite binary sequence $\{\varsigma_{n}\}_{n=0}^{n=\infty}$ in $[0,1)$. 
 
 At this point we note that we can use a binary representation, $0.\varsigma_0\varsigma_1\varsigma_2\varsigma_3\cdots$ (whose decimal equivalent number is $\sum_{j=0}^\infty \varsigma_{j}/2^{j+1}$, where $\varsigma_{j}\in\{0,1\}$), truncated up to the $n$th digit to locates $x_0$ is some element of $\mc{G}^{(n-1)}$. Furthermore, we observe that the Bernoulli map can be equivalently seen as shift operator $T$ acting on binary numbers $0.\varsigma_0\varsigma_1\varsigma_2\varsigma_3\cdots$ (where each $\varsigma_k\in\{0,1\}$ and $k\in\mathbb{Z}_{\ge0}$) at each time step to shift the decimal point towards right by one place, i.e., $T(0.\varsigma_0\varsigma_1\varsigma_2\varsigma_3\cdots)=0.\varsigma_1\varsigma_2\varsigma_3\cdots$. Thus, the sequence $\varsigma_0,\varsigma_1,\varsigma_2,\varsigma_3,\cdots$ represents a trajectory of the Bernoulli map such that when $\varsigma_{j}$ is zero, it means that $x_j=B^j(x_0)$ is in $g_0$ and when $\varsigma_{j}$ is one, it means that $x_j=B^j(x_0)$ is in $g_1$. In rather technical words, one says that the set $\{0,1\}$ of symbols is \emph{alphabet} induced by the partition $\mc{G}$. This alphabet is used by the partition to induce different $n$-words, e.g., a 3-word 0,0,1 and a 5-word 1,0,0,1,1.
 
It is intriguing to realize that the map can be seen equivalent to a fair coin toss such that the binary numbers with $0$ as the first digit after decimal point correspond to head and the binary numbers with $1$ as the first digit after decimal point correspond to tail. Hence, the shift operator generates a sequence of heads and tails; in other words, if any random semi-infinite sequence of heads and tails is fed into the Bernoulli map as a sequence of zeros and ones---zero corresponding to head and one to tail---then the map exactly generates the sequence in the course of the time evolution of $x_0$ taken as the sequence (with a decimal point put in front of it). This is an illuminating connection between chaos and randomness.

Now in order to find the topological and the metric entropies, we note that the sequence of partitions $\{\mc{G}^{(n)}\}$ is a refining sequence. Therefore, we can use Eq.~(\ref{eq:te2}) and Eq.~(\ref{eq:ks2}) to find the entropies by replacing $\mc{P}_m$ by $\mc{G}^{(m-1)}$. We have $n^{-1}[H(\vee_{k=0}^{n-1}B^{-k}\mc{G}^{(m-1)})]=n^{-1}[ \log(2^{n-1}2^m)]=\log 2$ for $\forall m$ as $n\rightarrow\infty$. Thus, $h_{\rm T}=\log 2$. In the case of the Bernoulli map, $H_\nu=H$ since the invariant density is uniform and consequently, $h_{\rm KS}(\nu,B)=\log 2$. We  emphasize the observation that the entropies being independent of $m$ implies that the supremum over partitions or the limit over refining sequences are not required to define the entropies in such cases. This is a defining property of generation partition, i.e., any partition for which this observation is true may be defined as a generating partition of the corresponding map. 

It is worth point out that $h_{T}(B)=D_B\log N$ and $h_{KS}(\nu,B)=D_I\log N$  ($N$=2 for the Bernoulli map) are not a mere coincidences but are rather general facts in any general symbolic dynamics: While $\log N$ is the average information gained on observing one of the $N$ equiprobable symbols, $D_I$ quantifies the non-uniformity of the probability of the sequences over the chaotic attractor. Since $D_I$ of a set is dependent on the probability distribution within the set and also on the spatial distribution of the set's elements (or in other words, also on the metric properties of the set), the name `metric entropy' for the KS entropy is justified.
\section{Uncertainty in Statistical Mechanics}
\hskip 2.5cm``\emph{Information is physical.}''---Landauer
\\
\\
\indent{\LARGE\textgoth{T}}he word `entropy' existed almost a century ago before the Shannon entropy came into existence. The Shannon entropy actually extended the use of the deep concept, entropy, to the areas way beyond the realm of thermodynamics where entropy had been first coined by Clausius. Over the time, many closely related definitions of entropy were formulated; in fact, there exists a formalism---Jaynes' formalism---which uses the Shannon entropy to mathematically connect with thermodynamics via statistical mechanics.
\subsection{Thermodynamics: Clausius Entropy}
Thermodynamics is mainly the study of energy, work done, and heat exchanges in macroscopic systems where the microscopic details of the systems are ignored while adopting a coarse-grained pragmatic description for them. A thermodynamic system can be \emph{isolated}, \emph{closed}, or \emph{open}, respectively, meaning that the walls bounding the system are \emph{adiabatic} (allow transfer of neither heat or matter), \emph{diathermal} (allow transfer of only heat), or neither (i.e., allow transfer of both heat and matter). The state of such a system is characterized by thermodynamic variables/coordinates: \emph{extensive} variables/coordinates (e.g., energy, volume, polarization, magnetization, and number of particles that are proportional to the size of the system) and \emph{intensive} variables/coordinates (e.g., pressure, electric field, magnetic field and chemical potential that are independent of the size of the system). Let us consider only \emph{homogeneous} system such that value of any of the thermodynamic variables is unchanged across the system. The \emph{first law of thermodynamics}, which essentially is statement of the conservation of energy, states: $dE=\textit{\dj}Q+\textit{\dj}W$, where $dE$ is the internal energy, $\textit{\dj}Q$ is the heat supplied to the system, and $\textit{\dj}W$ is the (generalized) work done on the system.The symbol `$\textit{\dj}$\,' is used to denote inexact differential: the total heat transferred and the total work done depend on the details/path of the process. 

There is a special thermodynamical state, called \emph{equilibrium state}, such that its thermodynamic coordinates do not appear change over time scales of observation period that is much greater than microscopic time scales but naturally not infinitely long. When in \emph{thermodynamic equilibrium}, a system is simultaneously in \emph{thermal equilibrium} (uniform constant temperature), \emph{mechanical equilibrium} (balanced mechanical forces), \emph{chemical equilibrium} (constant chemical composition), and \emph{phase equilibrium} (no phase changes like melting and evaporation). It may appear a bit of a circular argument that the concept of thermal equilibrium (defined using the concept of temperature) in the light of the \emph{zeroth law of thermodynamics}, expounds the existence of temperature $T$ as a state variable. The zeroth law states that the thermal equilibrium among thermodynamic systems is a transitive property; thus, two systems in thermal equilibrium have identical temperatures. In theoretical thermodynamics, the temperature is expressed in \emph{thermodynamic scale} (in the unit of Kelvin) by choosing the triple point of water-ice-steam system to be $273.16$ Kelvin. The thermodynamic temperature must always non-negative otherwise it can be shown to lead to the violation the \emph{second law of thermodynamics} which we state below.

Suppose a \emph{quasistatic} process on a system is performed. It means that the process is so slow that each step of the process the system is given enough time to settle down practically into an equilibrium state and consequently, the specific values of the thermodynamic variables can define it unambiguously. In such a process, the infinitesimal work done, $\textit{\dj}W$, due to change in the value of an extensive variable $\mc{D}_i$ under the influence of an intensive variable $\mc{F}_i$ (not constructed by dividing an extensive variable by system's volume or mass) can be written as $\textit{\dj}W=\mc{F}_id\mc{D}_i$; thus, $\mc{D}_i$ may be seen as a \emph{generalized displacement} and $\mc{F}_i$ as its \emph{conjugate generalized force}. However, what about the conjugate variable of temperature $T$, an intensive variable? We consider a further refinement of quasistatic process: \emph{reversible process}. A reversible process is a quasistatic process that when run backward in time from the final state leads to the unchanged initial state. It was established by Clausius that for such a process $\textit{\dj}Q/T$ is an exact differential, $dS$, where $S$ is an extensive variable and called it entropy; we specifically call it \emph{Clausius entropy} in order to differentiate it from the other definitions of entropy. Therefore, the first law can be written as: $dE=TdS+\sum_i\mc{F}_id\mc{D}_i$. For any arbitrary process (not necessarily reversible) leading to heat input in the system, $dS\ge \textit{\dj Q}/T$; here, $dS$ is the change in entropy assuming a virtual reversible process. It should be noted that the Clausius entropy is defined only for the equilibrium states.

The futile endeavours of inventing perfect engine (which transforms heat completely into work) and perfect refrigerator  (which transfers heat from cold body to hot body without any energy consumption) led to an empirical law---\emph{the second law of thermodynamics}---that is often expressed using the Clausius entropy: Entropy of an isolated thermodynamic system can not decrease in any (spontaneous) process, i.e., $dS\ge0$. The entropy of a system with its isolated subsystems at respective distinct equilibria, goes to an equilibrium state of higher entropy, following irreversible exchanges of heat inside it once all the adiabatic walls are removed between the subsystems. The intermediate non-equilibrium states are captured by the time-varying values of $\mc{D}_i$'s. Thus, stated another way, in the final equilibrium state, the entropy of an isolated system at the fixed total energy is maximized by the set of $\mc{D}_i$'s that take their equilibrium values in the final state. This is the \emph{maximum entropy principle}.

In passing, we observe that the Clausius entropy is defined only up to an additive constant. In this context, we mention that there is one more law---the \emph{third law of thermodynamics}---that states that the entropy of any system is a universal constant (conveniently taken as zero) at $T=0$. It can be shown that the third law implies unattainability (in a finite number of steps) of the absolute zero temperature, and also implies vanishing of thermal expansivities and heat capacities of the system at $T=0$. However, since at very low temperature the classical mechanics is not strictly valid, the third law explicitly requires quantum mechanics for its justification.

The Clausius entropy for a system in equilibrium may be envisaged to be determinable from $E$ and $\mc{D}_i$'s, i.e., $S=S(E,\{\mc{D}_i\})$---called \emph{(entropic) fundamental relation}---whose specific form would depend on the system in hand. All the intensive variables can, thus, be obtained as first order partial derivatives of $S$ with respect to $\mc{D}_i$'s. The relations relating the intensive with the extensive variables are called the \emph{equation of states}. Furthermore, inverting the preceding equation to write the \emph{(energetic)  fundamental relation}, $E=E(S,\{\mc{D}_i\})$, we can mathematically write the assumed extensivity of energy as $E(\lambda S,\{\lambda\mc{D}_i\})=\lambda E(S,\{\mc{D}_i\})$ ($\lambda>0$) which is valid in case there is short-range interactions between constituent particles of the system because then the energy is additive and extensive (in general, additivity and extensivity need not imply each other). Along with the first law, it implies the \emph{Euler relation}, $E=TS+\sum_i\mc{F}_i\mc{D}_i$, that in turn leads to the \emph{Gibbs--Duhem relation}: $SdT+\sum_i\mc{D}_id\mc{F}_i$=0. It should be kept in mind that the fundamental relations are not derived from thermodynamic considerations, but are rather found phenomenologically using insights from experiments performed on the thermodynamic system under question. A basic assumption of thermodynamics is that only a few coordinates are enough to specify thermodynamic equilibrium.

At this juncture, it is of use to comment that the maximum entropy principle is equivalent to the \emph{minimum energy principle}: In the final equilibrium state, the energy of a closed system at the fixed total entropy is minimized by the set of $\mc{D}_i$'s that take their equilibrium values in the final state. In practice, the tuneable thermodynamic variables may not be (some of) the extensive variables and so $E$ or $S$ may not be the right quantities for finding the equilibrium state variables. Therefore, with the Euler relation in the back of ones mind, one uses the Legendre transformation to define \emph{enthalpy} $\mc{H}\equiv E-\sum_i\mc{F}_i\mc{D}_i$, \emph{Helmholtz free energy} ${F}\equiv E-TS$, \emph{Gibbs free energy} $G\equiv E-TS-\sum_i\mc{F}_i\mc{D}_i$, and \emph{grand potential} $\mc{G}\equiv E-TS-\sum_i\mu_iN_i$. Note here we are not considering the pairs $(\mu_i,N_i)$---the chemical potential of a species and the corresponding number of particles---in the set of the pairs $(\mc{F}_i,\mc{D}_i)$; in other words, we are separating chemical work from mechanical work. ${\mc H}$, $F$, $G$ and ${\mc G}$ are respectively minimized during adiabatic transformation with mechanical work at constant generalized forces, isothermal transformation in absence of mechanical work, isothermal transformation with mechanical work at constant generalized forces, and isothermal transformation with chemical work at constant chemical potentials. In view of the afore-discussed minimization schemes, $E$, $\mc{H}$, $F$, $G$, and $\mc{G}$ are appositely called \emph{thermodynamic potentials} in analogy with the role of the potential energy in the Newtonian dynamics.

If we can find the fundamental relation in terms of the Clausius entropy (or, equivalently, in terms of the thermodynamic potentials), then all the macroscopic thermodynamic properties of the system can be evaluated.
\subsection{The Second Law and Markov Chain}
The constituent particles (molecules) in a thermodynamic system are too large (practically infinity) in number to be tracked individually and the interactions among them make the problem even harder.  If the total particle number is $N$, then the \emph{microstate} (the complete microscopic specification) of the system at any instant can be denoted by a phase point $(\bs{q},\bs{p})\equiv(\{\bs{q}_i\}_{i=1}^{i=N},\{\bs{p}_i\}_{i=1}^{i=N})$ in a $6N$ dimensional phase space, $\Omega$. Here, $\bs{q}_i$ and $\bs{p}_i$ are, respectively, the three dimensional generalized coordinate and conjugate generalized momentum of $i$th particle. 

Now we consider an isolated system so that the total energy of the system is conserved. Depending on the resolution of the measurement device at hand, a particular microstate corresponding to any given \emph{macrostate} (macroscopic thermodynamic state) with constant $E$ and $N$ can only be located in a $(6N-1)$-dimensional hypercube (of the constant energy manifold, $\Omega_E$) whose size is set by the resolving power of the measurement device. We assume nonexistence of any other conserved quantity for simplicity. Therefore, the \emph{accessible} phase space---basically the constant energy manifold in this case---is U-partitioned into finite number (say, $M$) of hypercubes. The underlying classical dynamics decides how the microstate moves from one element of the partition to another. In reality, the macrostate may also change over time; if it does not over a significant amount of time, then the macrostate corresponds to a thermodynamic equilibrium.

Let us ignore any specificity of the dynamics, and model the evolution of the probability distribution, $\bs{\pi}=[\pi_i]_{1\times M}$, of the occupancy of the elements by the microstate as a time-homogenous Markov chain $X_n$ ($n\in\mb{Z}_{\ge0}$) with stationary transition matrix $\textsf{P}=[p_{ij}(n)]_{M\times M}$: $\bs{\pi}(n+1)=\bs{\pi}(n){\sf P}$. The state space of the chain is then the set $\{1,2,\cdots,M\}$ any element of which may be denoted by $x_n$ at time step $n$. We furthermore assume the existence of a stationary distribution $\bs{\pi}^*$ that is limiting as well $\bs{\pi}^\infty$.  Since we now have a well-defined probability distribution, we can define the information-theoretic entropies---Shannon, relative, conditional, etc.---using it and see their temporal behaviours. 

First, we consider the relative entropy, $D(\bs{\pi}(n)||\bs{\pi}'(n))$, between two distributions $\bs{\pi}$ and $\bs{\pi}'$ at time $n$. We have the chain rule,
\begin{eqnarray}
&&D(p(x_n,x_{n+1})||p'(x_n,x_{n+1}))=\nonumber\qquad\\
&&D(p(x_n)||p'(x_n))+D(p(x_{n+1}|x_n)||p'(x_{n+1}|x_n)).\label{eq:Dm}\qquad\quad
\end{eqnarray}
Here, $p$ and $p'$ denotes the joint probability mass functions; in this notation, $p(x_n)$ denotes the mass distribution $\bs{\pi}$ and $p'(x_n)$ denotes the mass distribution $\bs{\pi}'$. Also,  $p(x_{n+1}|x_n)=p'(x_{n+1}|x_n)=p_{x_nx_{n+1}}$ (element of the transition matrix ${\sf P}$) by construction of the Markov chain and hence, $D(p(x_{n+1}|x_n)||p'(x_{n+1}|x_n))=0$. Keeping this in mind and the fact $D(p(x_n,x_{n+1})||p'(x_n,x_{n+1}))$ has an alternative expression (just swap positions of $x_n$ and $x_{n+1}$) in line with the chain rule, we get from Eq.~(\ref{eq:Dm}) ,%
\begin{eqnarray}
&&D(p(x_n)||p'(x_n))=D(p(x_{n+1})||p'(x_{n+1}))\nonumber\qquad\qquad\qquad\\
&&\qquad\qquad\qquad\qquad+D(p(x_{n}|x_{n+1})||p'(x_{n}|x_{n+1})).\label{eq:Dm'}
\end{eqnarray}
Due to the non-negativity of the second term in the R.H.S. of Eq.~(\ref{eq:Dm'}), it is evident that $D(p(x_{n+1})||p'(x_{n+1}))\le D(p(x_n)||p'(x_n))$, i.e., the relative entropy $D(\bs{\pi}(n)||\bs{\pi}'(n))$ is always non-increasing. In passing, we point out that we have not used the time-homogeneity of the Markov chain while reaching the above conclusion.

Note that in the special case of $\bs{\pi}'(n)$ being $\bs{\pi}^*$, we may conclude that any arbitrary initial distribution can get closer to the stationary distribution over time. Furthermore, if the stationary distribution is uniform (equivalently meaning that $\textsf{P}$ is doubly stochastic), then $D(\bs{\pi}(n)||\bs{\pi}^*)=\log M-H(X_n)$ implying that the Shannon entropy $H(X_n)$ is non-decreasing quite like what the second law of thermodynamics states for the Clausius entropy. In analogy with what happens in the thermodynamic system at equilibrium, if we also demand that the detailed balance ($\pi_i^*p_{ij}=\pi_j^*p_{ji}$ $\forall i,j\in\{1,2,\cdots,M\}$) is satisfied, rendering the stationary distribution equilibrium distribution as well, then the doubly stochastic transition matrix must be symmetric.

Once the stationary distribution is achieved $H(X_n)$ no longer changes but intriguingly the conditional entropies $H(X_0|X_n)$ and $H(X_n|X_1)$ can be shown to be capable of increasing as time passes. Let us prove for the former conditional entropy first. Employing the chain rule to $I(X_0,X_{n-1};X_n)$, it is easy to verify that $I(X_0;X_{n-1})+I(X_0;X_{n}|X_{n-1})=I(X_0;X_{n})+I(X_0;X_{n-1}|X_n)$. Also, we have because of the Markovity, $p(x_0,x_n|x_{n-1})=p(x_0|x_{n-1})p(x_n|x_{n-1})$ meaning that $X_0$ and $X_n$ are conditionally independent given $X_{n-1}$; and consequently, $I(X_0;X_{n}|X_{n-1})\equiv H(X_0|X_{n-1})-H(X_0|X_{n},X_{n-1})=0$. Therefore, in the light of non-negativity of $I(X_0;X_{n-1}|X_n)$, we have $I(X_0;X_{n-1})\ge I(X_0;X_{n})$ (also called \emph{data processing inequality}) which implies $H(X_0|X_{n-1})\le H(X_0|X_n)$, the relation we were after. Proof of non-decreasing $H(X_n|X_1)$ can be done similarly but there is an alternative simpler way: $H(X_n|X_1)\ge H(X_n|X_1,X_2)=H(X_n|X_2)=H(X_{n-1}|X_1)$, hence proven. The first equality is due to Markovity and the second one due to stationarity.

In summary, we discover the emergence of the second law of thermodynamics in the analogous set-up of a system with underlying Markov process at a microscopic level where the entropy has been defined in an information-theoretic way. Recall how some chaotic systems can be approximated by Markov processes. Also, it is a well-known fact that the real $N$ particle deterministic Newtonian dynamics generically has nonlinear interactions and is more than capable of showing chaotic dynamics. So, could one use the insight, just gained, to define entropy for the thermodynamical systems as a phase space concept and relate to the Clausius entropy? In other words, how to define entropy in a microscopic theory (i.e., statistical mechanics) and how to relate it to the macroscopic theory (thermodynamics) should be our next natural question.

 \subsection{Statistical Mechanics: Boltzmann Entropy}
Within the paradigm of classical mechanics, we may consider the particle dynamics to be given by some Newtonian dynamics where the interaction forces are derivable from some time-independent potential and so the particle dynamics is governed by a Hamiltonian $H(\bs{q},\bs{p})$ (not to be confused with the Shannon entropy's symbol); the exact form of the potential, however, is not known precisely. Since there can be many {microstates} corresponding to a given {macrostate}, a set of many phase points---constrained by the values of the thermodynamic variables at a given point of time---would correspond to a single macrostate (defined by the very values of the thermodynamics variables) at that time instant; this set is called an \emph{ensemble} of microstates, each of which is a copy of the macrostate.  One can define the phase space density $\rho(\bs{q},\bs{p},t)$ of the ensemble such that $\rho(\bs{q},\bs{p},t)\prod_{i=1}^Nd\bs{q}_id\bs{p}_i$ is the fraction of all possible microstates present in the elementary phase space volume $\prod_{i=1}^Nd\bs{q}_id\bs{p}_i$. As per the \emph{Liouville theorem} for the Hamiltonian systems, the time evolution in the phase space $\Omega$ is such that $d\rho/dt=0$. The microscopic dynamics, being Newtonian, is \emph{reversible}. In other words, since the invariance of the equation of motion under time reversal---i.e., $(\bs{q},\bs{p},t)\rightarrow(\bs{q},-\bs{p},-t)$---is manifest, $\rho(\bs{q},\bs{p},t)=\rho(\bs{q},-\bs{p},-t)$. For the system under consideration, another feature of the phase space dynamics is the \emph{Poincar\'e recurrence}:  After some finite (but maybe very long) time, almost every microstate on the constant energy manifold $\Omega_E$ returns arbitrarily near to its initial position.

Many paradoxes related to the microscopic statistical justification of the second law of thermodynamics have been put forward over the years and they have actually helped in developing deeper insights into the subject matter. They mainly hover around the topic of the intriguing emergence of macroscopic or thermodynamic irreversibility, sometimes fancily known as the concept of thermodynamic \emph{arrow of time}. Two of the earliest ones are \emph{Loschmidt's paradox} (due to time-reversibility of the Hamiltonian dynamics, increase in entropy with time cannot occur for all the phase points) and \emph{Zermelo's paradox} (due to the Poincar\'e recurrence, entropy of a system cannot be non-decreasing at all times). However, these are not quite paradoxes; in fact, these are compatible with the second law of thermodynamics as the law does not claim to be  applicable either for all phase points or for infinite time. 

Before we proceed further, we must first clearly understand the meaning of thermodynamic equilibrium from the microscopic picture. First, we realize that $\Omega$ can be partitioned into a number of \emph{macrosets}, $\Gamma_\mu$'s ($\mu=1,2,\cdots$): $\Omega=\bigcup_\mu\Gamma_\mu$, each macroset corresponds to one distinct macrostate. Each macroset, $\Gamma_\mu$, is a collection of microstates, $\omega_{i_\mu}$'s ($\omega_{i_\mu}\in\Omega$; $i_\mu=1,2,\cdots$). On every $\Omega_E$, suppose there exists a $\Gamma_\mu=\Gamma_{\rm eq}$ (say) such that it is almost as big as the $\Omega_E$. Then this $\Gamma_{\rm eq}$ corresponds to the thermodynamic equilibrium because microstates from other macrosets (of relatively very small sizes) are expected to arrive into $\Gamma_{\rm eq}$ quickly and the almost all microstates in $\Gamma_{\rm eq}$ are expected to stay inside for a long time. Of course, owing to the reversibility and the Poincar\'e recurrence, sometimes some microstates in $\Gamma_{\rm eq}$ may leave $\Gamma_{\rm eq}$, only to come back to it at a later time. Summarizing, in a little different jargon, we could say that the equilibrium macrostates are \emph{typical} and the nonequilibrium ones are \emph{atypical}.

It should be emphasized that the objective of the statistical mechanics is to study a single experimental system. The thermodynamics variables may fluctuate for any realistic finite system; usually (but not always) the fluctuations dies off  in the \emph{thermodynamic limit} defined as $N\to\infty$ and $N/V={\rm constant}$ ($V$ denotes physical volume).  In this limit, one expects the mathematical machinery of the statistical mechanics to reproduce the corresponding thermodynamic properties observed in experiments. When we measure any macroscopic property experimentally, we essentially report its time-averaged version---a frequentist viewpoint. The ensemble is a mathematical way of looking at the underlying probabilities from the point of view of a classical/logical probabilist. There is no a priori reason to believe that the ensemble averaging would give same results as the time averaging. If the averages are in fact equal, then equilibrium thermodynamic variables may be postulated to be the ensemble averages  found using some time-independent phase space density, $\rho(\bs{q},\bs{p})$. Given the very limited knowledge about the precise microscopic dynamical evolution of $N$ particles, it is impossible to rigorously prove the aforementioned equality of averages for every arbitrary thermodynamical system in the universe. Hence, this equality is postulated and is known as the \emph{ergodic hypothesis}. It should be noted that an additional implicit assertion that the ergodic hypothesis makes is that all measurable thermodynamical quantities can actually be seen as their respective ensemble averages. However, the choice of $\rho(\bs{q},\bs{p})$ remains to be ascertained. 

For isolated thermodynamical system with constant energy ($E$), particle number ($N$), and volume ($V$), Boltzmann associated a time independent phase space density $\rho(\bs{q},\bs{p})=\rho_{\rm mc}\equiv \frac{1}{\Omega(E)}\delta(H(\bs{q},\bs{p})-E)$ with $\Gamma_{\rm eq}$ in the spirit of the principle of insufficient reasons. Here, the \emph{statistical weight}, $\Omega(E)\equiv \int\int\delta(H(\bs{q},\bs{p})-E)d\bs{q}d\bs{p}$, depends on $E$, $V$, and $N$ but does not depend on the phase space coordinates at constant energy. This postulate by Boltzmann is called \emph{equal a priori probability} hypothesis and the ensemble is called \emph{microcanonical ensemble}. This invariant microcanonical ensemble density is automatically compatible with the Liouville theorem and the microscopic time-reversibility. The intuitive justification behind this postulate is as follows: We are tracking a single thermodynamic system's time evolution as a microstate's evolution in $\Omega_E$, and given the complexity of the system, it is most possibly a highly chaotic system such that almost all parts of the accessible phase space is visited. In the light of the fact that there is no perceivable bias towards any particular microstate, one could assume that the trajectory spends equal time in equisized hypercubes. Then one could furthermore define a measure such that any sized hypercube's measure is fraction of time spent in the hypercube. We note that this means that the measure density is constant. Also, the immediate consequence of this picture is that the (infinite) time average of some thermodynamic quantity is equal to the microcanonical ensemble average of the quantity---something we dearly need for relating the statistical mechanics to the thermodynamics. 

Which dynamical systems are ergodic and which dynamics actually evolve an arbitrary phase space density in such a fashion that an invariant density is reached are extremely difficult questions. The justification of the ergodic hypothesis and the equal a priori probability is best attributed to the tremendous success of the scheme of statistical mechanics in explaining thermodynamics properties of varied systems. In passing, we remark that the typicality of macrostates corresponding to $\Gamma_{\rm eq}$ manifests itself, in the thermodynamic limit, as the fact that the mean and the most probable values of a thermodynamic quantity converge towards each other; or in other words, the mean square fluctuations of thermodynamic quantities are negligibly small in the thermodynamic limit. In fact, the macroscopic arrow of time---in the backdrop of microscopic reversibility---arises because we practically track the most probable behaviour of the thermodynamic system: The increase of entropy as encapsulated in the second law of thermodynamics is a typical behaviour.

The last profound piece needed to connect with thermodynamics is to get the (entropic) fundamental relation of thermodynamics. But the path to that has to start with defining entropy in the statistical mechanical setting. This calls for an epoch-making hypothesis. Defining the \emph{Boltzmann entropy} in the units of the Boltzmann constant as $S_B(\omega_{i_\mu})\equiv S_B(\Gamma_\mu)\equiv\log({\rm vol}(\Gamma_\mu))$ where `$\textrm{vol}$' is the appropriate hypervolume of $\Gamma_\mu$, the hypothesis can be cast as: $S=S_B(\Gamma_{\rm eq})$. We note that through this definition of entropy (which is additive and extensive), we can define $S_B$ for every phase point in the phase space $\Omega$; i.e., every distinct physical thermodynamic system can be associated with an $S_B$. The microcanonical phase space density and the definition of Boltzmann entropy can be together used to argue that $\partial S_B/\partial E$ ($N$ and $V$ held constant) remains same between two systems in thermal equilibrium with each other, and hence the notion of temperature $T$ is introduced as $1/T=\partial S_B/\partial E$ in compatible with thermodynamics.

We cast $S_B$ in a slightly different form so as its insightful analogy with information-theoretic Shanon entropy comes out crystal clearly. Consider a single particle $6$-dimensional phase space, $\mho_1$, where each point is specified by the generalized coordinates and momenta of  the particle. We create a partition of $\mho_1$ into $r$ cells of volume $\Delta$: $\mho_1=\bigcup_{j=1}^{j=r}c_j$. Any $N$-particle microstate $\omega_{i_\mu}$ corresponds to a particular \emph{arrangement} of the particles in a distribution, $(n_1,n_2,\cdots,n_r)$ with $\sum_{j=1}^rn_j=N$, that states that $n_j$ particles are in $c_j$ cell; the {arrangement} specifies which particle is in which cell. Thus, two distinct microstates can have the same distribution but they cannot have the same arrangement. Obviously many arrangements may correspond to a single macroset, $\Gamma_\mu$; each distribution corresponds to a specific macrostate. Considering the particles to be \emph{distinguishable}, the number of arrangements for the macroset $\Gamma_\mu$ is $W(\Gamma_\mu)=N!/\prod_{j=1}^{j=r}n_j!$. Naturally, the volume of the macroset $\Gamma_\mu$ is $W(\Gamma_\mu)\Delta^N$ and hence $S_B(\Gamma_\mu)=\log W(\Gamma_\mu)$ within an additive constant. Next, in the limit of very large $n_j$'s, we can use \emph{Stirling's approximation}, i.e, $\log (n_j!)\approx n_j\log n_j-n_j$, to easily show that $S_B(\Gamma_\mu)=-N\sum_{j=1}^{j=r}\frac{n_j}{N}\log\left(\frac{n_j}{N}\right)$ within an additive constant. Obviously, $n_j/N$ is the probability of finding a single particle in $j$th cell. Naturally, we can view the single particle microstate as a random variable with alphabet $\{c_j\}_{j=1}^{j=r}$ with probability mass function $p_i\equiv p(c_j)\equiv n_j/N$. Therefore, $S_B/N$ is the Shannon entropy that quantifies average information gained on measuring a single particle's phase coordinates in $\mho_1$.

\subsection{Gibbs Entropy and Jaynes' Formalism}
Although we have introduced the concept of microcanonical ensemble while discussing the Boltzmann entropy, we should emphasize that the viewpoint in focus in that discussion has been an \emph{individualist} (or \emph{Boltzmannian}) one---we are tracking a single system over time. The Boltzmann entropy is property of a system that is mathematically specified in $\Omega$ or $\Omega_E$. We note that if we define the entropy as $-\int_\Omega\rho_{\rm mc}\log\rho_{\rm mc}d\bs{q}d\bs{p}$, then this entropy would be equal to $\log({\rm vol}(\Omega_E))$ which is actually $S_B(\Gamma_{\rm eq})$ in the thermodynamic limit because $\Gamma_{\rm eq}$ is the lion's share of $\Omega_E$.

However, there can be many different kinds of ensembles, e.g., \emph{canonical ensemble} that describes the possible states of some closed system in thermal equilibrium with a heat bath at a constant temperature, and \emph{grand canonical ensemble} that describes the possible states of some open system in thermal and chemical equilibria with a heat bath. It motivates to propose another, seemingly more general, definition of entropy---\emph{Gibbs entropy}: $S_G\equiv-\int_\Omega\rho_{g}(\bs{q},\bs{p})\log\rho_{g}(\bs{q},\bs{p})d\bs{q}d\bs{p}$ (a functional of $\rho_{g}(\bs{q},\bs{p})$). We have explicitly put the subscript `$g$' to emphasize that it is an \emph{ensemblist} (or \emph{Gibbsian}) viewpoint. Arguably $\rho_{g}$ is not a system property; it is not completely clear what $\rho_{g}$ is appropriate and what even its interpretation is. $\rho_{g}(\bs{q},\bs{p})$ may have a very subjective meaning in the sense that it may characterize an observer's belief that the system is at $(\bs{q},\bs{p})$. It could also relate it to the system preparation procedure that yields some random phase point and $\rho_g$ happens to be its distribution. 

Moreover, when a system is in equilibrium state, i.e., its thermodynamical variables are time-independent, the definition of $S_G$ with appropriate $\rho_{g}$ may be made mathematically compatible with the underlying Hamiltonian dynamics; but when one wants to talk about time evolution of this entropy, it possess a problem because in the light of the Liouville theorem, $dS_G/dt$ is always zero. Thus, it does appear paradoxical that the second law of thermodynamics stated with the Gibbs entropy in mind should be compatible with the reasonable assumption that the underlying dynamics is Hamiltonian (recall $dS_G/dt=0$ due to the Liouville theorem). A way around this is to realize that a blob of initial conditions in the phase space, while keeping its (hyper)volume preserved in line with the Liouville theorem, can get stretched into fine  long threads reaching distant parts of the accessible phase space. One can define a \emph{coarse-grained entropy} using a coarse-grained phase space density---the fine-grained phase space density averaged out in each element of U-partition. By construction, this coarse-grained entropy is expected to increase in a system that is dynamically mixing (a requirement even stronger than ergodicity).

In order to see the connection between $S_B$ and $S_G$, we recall two points: Firstly, the macroset $\Gamma_{\rm eq}$ may be the largest but it may not be the only one present and secondly, the thermodynamic variables---including (Clausius) entropy---are mean (and the most probable) values of the corresponding statistical quantities. Therefore, let's consider all the macrosets $\Gamma_\mu$ partitioning $\Omega_E$ and assume that each macroset has $W(\Gamma_\mu)$ microstates. The average Boltzmann entropy, $\langle S_B\rangle$, of the system is $\langle S_B\rangle=\sum_\mu P(\Gamma_\mu)\log W(\Gamma_\mu)$; here $P(\Gamma_\mu)\equiv W(\Gamma_\mu)p_{i_\mu}=W(\Gamma_\mu)p_{\mu}$ is the probability of the macroset where we have assumed that $p_{i_\mu}$---the probability of the microstate $\omega_{i_\mu}$---is some constant $p_{\mu}$  $\forall i_\mu=1,2,\cdots,W(\Gamma_\mu)$. Note that for this system, by the definition of the Gibbs entropy, $S_G=-\sum_\mu\sum_{i_\mu}^{W(\Gamma_\mu)}p_{i_\mu}\log p_{i_\mu}$. Thus, we have after some straightforward mathematical steps:
\begin{equation}
S_G=\langle S_B\rangle-\sum_\mu P(\Gamma_\mu)\log P(\Gamma_\mu).
\end{equation}
Evidently, $S_G\ge \langle S_B\rangle$ with the equality holding when the Shannon entropy corresponding to the probability distribution of the macrostates vanishes, i.e., when $-\sum_\mu P(\Gamma_\mu)\log P(\Gamma_\mu)=0$. Actually, $-\sum_\mu P(\Gamma_\mu)\log P(\Gamma_\mu)$ is nothing but a quantity characterizing the fluctuations in $S_B$ about its mean value and may be termed the \emph{entropy of fluctuations}.

At this point, it is correct to say that for an arbitrary thermodynamic system, the choice of $\rho_g$ is a matter of trial and error, and educated guess; sometimes, one can get to one $\rho_g$ from another one through mathematical and logically plausible manipulations, e.g., microcanonical ensemble to canonical ensemble. It is presently an unsolved quest how to find a $\rho_g$ appropriate for making connection with the measured thermodynamical quantities starting from the underlying Hamiltonian dynamics. In fact, even if somehow the underlying dynamics is proven to be ergodic with a limiting invariant density that may be taken as a $\rho_g$, ergodicity obtained through chaos is neither (practically) sufficient nor necessary for ensemble averaged quantities to be equal to the thermodynamical variables measured in equilibrium state. In slightly different words, ergodicity may neither be cause nor effect of thermodynamic behaviour although both can stem from the same root: nonlinear interactions between particles. 

Let us elaborate a bit what we want to convey in the preceding terse paragraph. Experts have argued that approach towards and sustenance of equilibrium may be governed by the unavoidable external perturbations---and not ergodicity due to chaos; after all the time taken by the deterministic Hamiltonian trajectory to cover almost all of the phase space may be orders of magnitude higher that the time scale over which the time average seems to practically agree with the (mathematical) ensemble average. So effectively, even if a system is mathematically ergodic, it may not be sufficient for realization of equivalence between the time and the ensemble averages. Next, as far as the necessity of ergodicity is concerned, practically, one only ever requires the time and the ensemble averages to be  same for some relevant thermodynamical variables, not all mathematically possible generic functions (as implied by `ergodicity'). In fact, it is known that for a special set of observables (which includes pressure and kinetic energy) in certain non-interacting and very weakly interacting many particle systems, the time and the ensemble averages may differ only in phase space's subsets of almost zero measure as number of particles tend to infinity. This should be a practically acceptable solution for the issue under consideration because it means that if for some negligible small set of initial conditions, ergodicity is not realized, even then correspondence between thermodynamics and ensemble picture of statistical mechanics is practically almost always realized in certain cases. Thus, it is not surprising that some experts maintain that the single most important reason behind the success of microscopic models in describing thermodynamics is the consideration of almost infinite number of degrees of freedom. If we consider a thermodynamics system as consisting of many independent thermodynamic subsystems, then an additive thermodynamic variable's per-particle or per-volume average value becomes sharply peaked in line with the central limit theorem.

Remarkably, one can sideline all these formidable conundrums and dilemmas about the justification of the statistical mechanics by turning to a subjective view as advocated by Jaynes. We notice the formal similarity between the expressions of the Gibbs entropy and the Shannon entropy, and then recast the entire problem of connection between the thermodynamics and the statistical mechanics as follows: Let us say for a system in equilibrium, we measure certain thermodynamic variables that we assume to be equal to some ensemble averaged quantity and hence denote them is as $\langle f_k\rangle$'s. We assume that each thermodynamic variable is a function of a (random) variable $X$ with finite alphabet $\{x_i\}$; each $x_i$ corresponds an element of the set of countably finite set of all possible distinctly resolvable microstates, $\omega_i$'s. Each $\omega_i$ is realizable---from an ensemblist viewpoint---with probability $p_i$. Hence, mathematically, we have the constraints: $\sum_i p_i f_k(x_i)=\langle f_k\rangle$ $\forall k$. There may be many probability mass functions compatible with this much input. However, if we are interested in the most unbiased inference given this much input, then we should adopt the maximum entropy principle, i.e., maximize $S_G=-\sum_ip_i\log p_i$. Therefore, we arrive at $p_i=\exp(-\sum_k\lambda_kf_k(x_i))/Z$ where $\lambda_k$'s are Langrange multipliers and the \emph{partition function}, $Z$, is the normalization constant. Since $Z=\sum_i\exp(-\sum_k\lambda_kf_k(x_i))$, $\lambda_k$'s can be determined by making use of the following relation: $\langle f_k\rangle=-\partial \log Z/\partial \lambda_k$, $\forall k$. If $f_k$'s additionally depend on other experimentally tuneable thermodynamical parameters (like volume, externally applied fields, etc.), say $\alpha_l$'s for the system under consideration, then it can be shown that $\langle \partial f_k/\partial\alpha_l\rangle=(-1/\lambda_k)(\partial \log Z/\partial \alpha_l)$. For completeness, we mention that the maximized entropy is given by $S\equiv\log Z+\sum_k\lambda_k\langle f_k\rangle$.

For example, in microcanonical ensemble there is no constraint (other than the normalization condition, $\sum_ip_i=1$), and it leads to uniform probability mass function. For the canonical ensemble, the only constraint (in addition to the normalization condition) that one has is $\sum_i p_i E(x_i)=\langle E\rangle$. For the grand canonical ensemble, in addition to the energy-constraint, one as $\sum_i p_i N(x_i)=\langle N\rangle$ where $N$ is the number of particles (all of them assumed to be of same species; this can be easily generalized for many species of particles). The entire connection with thermodynamic follows on making following identifications: Firstly, the Lagrange multipliers associated with the energy and the particle number respectively are $1/T$ ($=\partial S/\partial \langle E\rangle$) and $-\mu/T$ ($=\partial S/\partial \langle N\rangle$) where $T$ is the temperature of the heat bath and $\mu$ is the chemical potential; and secondly, $-T\log Z$ can be identified with the Helmholtz free energy in the canonical ensemble and the grand potential in the grand canonical ensemble. In this context, it is interesting to note that in a canonical ensemble where $T$, $V$, and $N$ are held fixed, and the equilibrium distribution is $\exp[-E(x_i)/T]/Z$, one can expand $D(p_i||\exp[-E(x_i)/T]/Z)$ (where $p_i$ is arbitrary distribution) to see that $TD(p_i||\exp[-E(x_i)/T]/Z)-T\log Z=\langle E\rangle-TS$ (where $S\equiv-\sum_ip_i\log p_i$): The decrease in the Helmholtz free energy---now recognized to be a KL-divergence---as equilibrium is approached, is equivalent to the principle of decrease of the relative entropy. Of course, at equilibrium, the relative entropy takes the minimum value possible, i.e., zero, and the equilibrium free energy is $-T\log Z$.

We conclude with a few comments. Under Jaynes' formalism, in a way, the equilibrium statistical mechanics can be seen as a special case of statistical inference: If we know $\langle f_k\rangle$'s, then we want to infer about some arbitrary function $g(x)$---say, $\langle g(x)\rangle$---for which we somehow have to find an appropriate probability distribution. This approach describes our knowledge about a system but not the system itself. Furthermore, the maximum entropy principle can always be applied but it has predictive power for the experimental results only when the maximum entropy probability distribution is found have a sharp peak with negligible variance. Having found the distributions for various ensembles using the maximum entropy principle, we may make the philosophical observation that at the macroscopic level the measurements provide as much information as the equations of motion for the constituent particles could have. Finally, note that there is no mention about the nuances of ergodicity and mixing dynamics. In fact, even if the systems were not ergodic, in the maximum entropy method there is no rationale for ignoring the contribution from any region of the phase space because we have no explicit evidence to do so.

\subsection{The Second Law and Landauer's Principle}
Historically, the detailed experience with real heat engines led to various statements of the second law of thermodynamics. For example, 
\emph{Kelvin's statement}: Complete conversion of heat to work can never be the sole effect of any process, and
\emph{Clausius's statement}: Transfer of heat from a colder to a hotter body can never be the sole effect of any process. These two statements are logically equivalent. Moreover, no engine working between two heat reservoirs can be more efficient than a Carnot engine that works between the same two reserviors---a fact known as  \emph{Carnot's theorem} which can be proven using Clausius's statement. A \emph{Carnot engine} is any cyclic reversible engine with all of its heat exchanges taking place at the temperatures of the two heat reservoirs. The Carnot engine and Kelvin's statement can be used to prove \emph{Clausius's theorem} that states that in any (reversible or irreversible) cyclic process, if $\textit{\dj Q}$ is the incremental heat supplied to the system at temperature $T$, then $\oint \textit{\dj Q}/T\le 0$; for reversible cycle, the equality holds. Finally, by making use of Clausius' theorem, the existence of Clausius entropy---a state variable $S$---is realized, and the well-known statement of the second law in terms of the entropy arises: Entropy of an isolated thermodynamic system can not decrease in any process, i.e., $dS\ge0$.

A version of the second law in the setting of a closed system in thermal equilibrium with a heat reservoir at temperature $T$ is also quite useful. Suppose the Helmholtz free energy, $F$, of the system changes by $\Delta F$ when the system absorbs heat $Q$ from the reservoir. Therefore, $\Delta F=\Delta E-T\Delta S\le (W+Q)-T(Q/T)=W$ where $W$ is the work done on the system. So, for reversible process when the equality holds, if $\Delta F$ is negative, then its magnitude is the work done by the system. Hence, $F$ is the measure of energy `free' to be converted to work. Also, if no work is done on or by the system, then $\Delta F\le 0$ so that $F$ decreases to reach a minimum value at equilibrium. Anyway, the second law is then $W\ge\Delta F$ or equivalently, the \emph{total entropy production}, $\Delta S_{\rm s+r}\equiv\Delta S-Q/T\ge0$ (system plus reservoir is isolated and $S_{\rm s+r}$ denotes its entropy). Here, $-Q/T$ can be regarded as change in the reservoir's entropy. Furthermore, we can define \emph{thermodynamically reversible} process as the one for which $\Delta S_{\rm s+r}=0$; otherwise the process is \emph{thermodynamically irreversible}.

Keeping the above discussion in the back of our minds, let us now perform a thought experiment: Consider a horizontal cylindrical box of volume $2V$ containing an ideal gas consisting of a single particle and in thermal equilibrium with a heat reservoir at fixed temperature $T$. Let there be a vertical impenetrable movable diaphragm creating a symmetric partition of the box---two cylinders of volume $V$ each. The particle's location---whether in left (say, state $0$) or in right side (say, state $1$) of the diaphragm---serves as a memory with one bit of information. We plan to erase this information and observe its physical consequence. In other words, we plan to reset the system to a specific state, say $0$, irrespective of what the initial state was before the resetting. The physical processes involved are stepwise as follows: 
\begin{enumerate}
\item[(i)] We pull the diaphragm vertically out and let the gas come to equilibrium---same one whether we start with state 0 or state 1. This leads to gas's expansion that is adiabatic (no heat exchange) and free (no work done). Naturally, there is no change in internal energy either; and because the gas is considered ideal, there is no change in temperature either. (For ideal gas, internal energy depends only on the temperature.) The adiabatic free expansion is an irreversible process, so to find the change in its entropy (which is a state variable), we can find the corresponding entropy change in a reversible isothermal expansion. Consequently, the entropy change of the gas is $T^{-1}\int_V^{2V}PdV=\log 2$. Note that this is one bit of average Shannon information.
\item[(ii)] Subsequently, to erase the one bit of information from memory, we plan to confine the particle in one specific side of the box irrespective of what the initial state was. To this end, we do reversible isothermal compression of the gas by frictionlessly moving the right (or left) wall of the box. This means work $W=-\int_{2V}^{V}PdV=T\log 2$ is done on the system. Again, since temperature is unchanged in isothermal process, so is the internal energy of ideal gas. Therefore, the entropy changes by $\Delta S=-\log2$; and in the light of the relation, $\Delta S=Q/T$, it means that $T\log2$ amount of heat is dumped into the heat reservoir. In conclusion, the information erasure requires work and produces heat.
\end{enumerate}

Information is physical: Information processing is in effect a physical process and hence there is no reason that it should not obey the laws of thermodynamics. Thus, it is not surprising that the afore-discussed observation leads to a principle analogous to the conventional statements of the second law of thermodynamics: A weaker form of the \emph{Landauer principle} states that the erasure of information can never be the sole result of any process. The Landauer principle, in a stronger form, states that $T\log 2$ (known as the \emph{Landauer limit/bound}) is the minimum amount of energy (work) required for erasing a single bit of information and same amount of heat is released into a heat bath. In order to understand the essence of the principle, let's first digress a bit.

A \emph{memory} may be considered to have a set of logic states in which information is stored. A \emph{computational process} involves an algorithm that changes an input logical state to an output logical state and can be seen as a map (or a \emph{gate}), $C:\mb{M}_i\to\mb{M}_o$, where $\mb{M}_i$ and $\mb{M}_o$ are respectively the sets of input and output states. A computational process is defined to be \emph{logically reversible} iff it is an injection. In other words, logical reversibility implies that we can know the input state unambiguously from the knowledge of the output state. For example, a one bit input set $\{0,1\}$ and a NOT gate ($0\mapsto1$ and $1\mapsto0$) constitute a logically reversible process; so is a two bits input set $\{00,01,10,11\}$ and a CNOT gate ($00\mapsto00$, $01\mapsto01$, $10\mapsto11$, and $11\mapsto10$). However, a one bit input set $\{0,1\}$ and ERASE gate ($0\mapsto0$ and $1\mapsto0$) is clearly logically irreversible. In all these three cases, $\mb{M}_i=\mb{M}_o$. As an example where $\mb{M}_i\ne\mb{M}_o$, we consider $\mb{M}_i=\{00,01,10,11\}$ and $\mb{M}_o=\{0,1\}$, and a XOR gate ($00\mapsto0$, $01\mapsto1$, $10\mapsto1$, and $11\mapsto0$). This process is obviously logically irreversible. 

Interestingly, any irreversible computational process can be extended to a reversible one. It is known that given any $C:\mb{M}_i\to\mb{M}_o$, $\exists$ at least one finite set $\mb{M}_e$ (\emph{ancilla} or \emph{environment}) and a corresponding logically reversible map, $C_e:\mb{M}_i\to\mb{M}_o\times\mb{M}_e$, such that the map's restriction on $\mb{M}_i\to\mb{M}_o$ is equivalent to $C$. For example, we  can extend irreversible ERASE using $\mb{M}_e=\{0,1\}$ and reversible $C_e$ ($0\mapsto00$, and $1\mapsto01$); furthermore, we note that the restriction of $C_e$ on $\mb{M}_i\to\mb{M}_o$ is equivalent to ERASE. This particular extension has a very apparent interpretation: The erased bit is kept in ancilla. In order to connect with the concept of entropy, it is natural to define \emph{logical entropy}, $H(\mb{M})$, of a set $\mb{M}$ as the Shannon entropy calculated using the probability distribution over the elements (logical states) of $\mb{M}$. A computational process may change the logical entropy. From the general properties of the Shannon entropy, it is crystal clear that a logically reversible process does not change the logical entropy but a logically irreversible one decreases the entropy. Thus, for ERASE the logical entropy decreases; however, if one uses the reversible extension of the ERASE, the logical entropy remains unchanged because the decrease in entropy of the input set is balanced by the increase in entropy of the ancilla.

The fundamental contribution of the Landauer principle is actually the assertion that heat is released in a logically irreversible information processing and that the {logical entropy} should be given the same status as the thermodynamic entropy; the two entropies must be considered together in any relevant system while determining its total entropy change. In passing, we remark that the logical irreversibility need not imply thermodynamic irreversibility. In this context, we note that the total entropy production in the quasistatic information erasure in the preceding thought experiment is actually zero, meaning that the process is thermodynamically reversible although not so logically.

The most celebrated application of the Landauer principle is in exorcism of \emph{Maxwell's demon}. Maxwell wrote: ``One of the best established facts in thermodynamics is that it is impossible in a system enclosed in an envelope which permits neither change of volume nor passage of heat, and in which both the temperature and the pressure are everywhere the same, to produce any inequality of temperature or of pressure without the expenditure of work. This is the second law of thermodynamics, and it is undoubtedly true as long as we can deal with bodies only in mass, and have no power of perceiving or handling the separate molecules of which they are made up. But if we conceive a being whose faculties are so sharpened that he can follow every molecule in its course, such a being, whose attributes are still as essentially finite as our own, would be able to do what is at present impossible to us. For we have seen that the molecules in a vessel full of air at uniform temperature are moving with velocities by no means uniform, though the mean velocity of any great number of them, arbitrarily selected, is almost exactly uniform. Now let us suppose that such a vessel is divided into two portions, A and B, by a division in which there is a small hole, and that a being, who can see the individual molecules, opens and closes this hole, so as to allow only the swifter molecules to pass from A to B, and only the slower ones to pass from B to A. He will thus, without expenditure of work, raise the temperature of B and lower that of A, in contradiction to the second law of thermodynamics.'' (Later Kelvin christened the being---Maxwell's intelligent demon.) 

Of course, there is a simple argument that says the demon cannot keep doing what it is supposed to do: The demon is part of the system and on observing a molecule of gas, it get hit and heated up. Soon the demon will be so erratic about its own position and velocity that it can't make out which molecule is slower and which one is faster. Anyway, there are many versions of the demon in the literature, all with same purpose---apparent violation of the second law of thermodynamics. In order to comprehend how the Landauer principle formally comes to rescue the second law, it suffices to consider and understand the `thermodemonics' of a simple version of the demonic activity: \emph{Szilard's engine}.

Szilard's engine contains an ideal gas consisting of a single particle in a horizontal cylindrical box (of volume $2V$) that is in a thermal equilibrium with a heat reservoir at temperature $T$. There is an air-tight piston at each flat end of the box; the pistons can slide frictionlessly. Now four steps follow:
\begin{enumerate}
\item[(i)]A demon inserts a vertical impenetrable movable diaphragm in the middle of the box to create a symmetric partition of the box. \item[(ii)]Subsequently, it measures (without any error, say) the particle's position. 
\item[(iii)]Suppose the particle is in the left side then the right piston is moved in at no energy cost up to the diaphragm that is then pulled out. The gas expands reversibly and does work ($T\log2$) in pushing the piston out to its initial position. 
\item[(iv)]Finally, the gas is allowed to thermalize so as to come back to its initial state. 
\end{enumerate}
So, the thermodynamic process we just witnessed is cyclic and hence the change in the Helmholtz free energy, a state variable, is zero. However, work has been extracted in direct contradiction with the second law of thermodynamics. The catch is that we have not considered the demon's one-bit memory wherein information is stored; the true cycle is accomplished only if in a final step the data in the demon's memory is erased. This erasure requires a minimum $T\log2$ of work to be done in accordance with the Landauer principle.

We must however remark that the discussion in this section has involved a symmetric partition wherein the states of memory and (error-free) measurement outcomes coincide. The system with the symmetric partition can be seen as a configuration of a particle in a symmetric double-well potential with very high potential barrier. It can be shown that for an asymmetric double-well potential, the Landauer principle may not hold; however, in such cases, the minimum total work required for measurement and erasure still is $T\log2$. As stated, the Landauer principle holds when the information measurement doesn't involve any energy.
 \section{Geometrizing Uncertainty}
``\emph{Geometry is not true, it is advantageous.}''---Poincar\'e
\\
\\
\indent{\LARGE\textgoth{A}}s we have discussed earlier, Jaynes' formalism of statistical mechanics presents it as a problem of statistical inference which is a very well developed topic in the field of \emph{mathematical statistics}. What we aim to highlight now is that the probability distributions can be visualized geometrically and one can develop this idea further to build a beautiful geometric construction of information, sometimes called \emph{information geometry}; and apply it to give a geometrical understanding of the phase transitions, whose theory is a remarkable success story of the equilibrium statistical mechanics. For this purpose, we first review some basic ideas in statistical inference and witness in it the utility of the information theory through the concept of \emph{Fisher Information} which is connected with the rather general concept of entropy. It is the Fisher information that brings forth the information geometry.
\subsection{Estimating Uncertain Parameters}
Let us first discuss the bare basics of inferences based on observations involving repeated trials. It essentially brings the abstract formalist view of probability into application and relates to the modelling of uncertain observations in the real world. Naturally, both the frequentist and the Bayesian views depend heavily on statistics as the former is all about deriving probabilities based to trials and the latter is all about updating beliefs based on series of evidences.

As the simplest case, we are going to consider an experimental setup that churns out iid random variables ($X_i$'s all equal to some fixed $X$) for some observable: e.g., repeated toss of a coin or measurement-error affected repeated measurements of a property (mass, volume, etc.) of an object. The main goal of statistics is either (i) estimating the value of the random variable $X$ at a future instant when the underlying probability distribution is completely known or (ii) estimating some unknown parameter $\theta$ of the known probability distribution supposedly modelling the underlying phenomenon leading to the observed sequence $\{X_i\}_{i=1}^{i=N}$ of random variables. This is why sometimes this line of study is also called \emph{estimation theory}. 

The first kind of estimation is also called \emph{prediction}---which can be either \emph{point prediction} or \emph{interval prediction}---because we are trying to predict the value of the random variable $X$ corresponding to a future outcome. In the point prediction, one aims to determine a constant $c$ such that \emph{error} $(X-c)$ is minimal in some sense; whereas in the interval prediction, one intends to find two constants $c_1$ and $c_2$ such that $P(c_1<X<c_2)=\gamma$. $\gamma$ is called \emph{confidence coefficient} and $1-\gamma$ is called \emph{confidence level}; the open interval $(c_1,c_2)$ is called \emph{$\gamma$ confidence interval of $X$}. Note that the task of predicting an outcome effectively means estimating its corresponding random variable's value using some optimally chosen constants, viz., $c$, $c_1$, and $c_2$.

Actually the point prediction is practically useless in predicting outcome of any trial; it makes much more sense to try to predict whether the outcome would lie between some bounds with almost certainty. The interval prediction has the dichotomy that a good estimation would mean high confidence coefficient but smaller confidence interval. In order to strike a balance between the two, for a chosen confidence coefficient (usually taken to be close to unity), the minimal confidence interval---the so-called optimal confidence interval---is sought. While in general this involves some trial and error on top of formal mathematical analysis, for the special case of a unimodal density function symmetric about its mean, the optimal confidence interval for a given confidence coefficient (or equivalently for a given confidence level $\delta$) is simply the open interval $(x_{\delta/2},x_{1-\delta/2})$, where recall the notation that $x_u$ denotes the $u$ percentile of $X$. However, for our purpose, we need to concern ourselves mostly with the point prediction and estimation.

In the point prediction, if error is minimized in the mean square sense, i.e., ${er}\equiv E[(X-c)^2]$ is minimized then $c$ turns out to be $E(X)$. This can be easily seen by setting $d(er)/dc=0$. Sometimes there may be another random variable $X'$ defined on the same sample space. In such cases, the knowledge of $X'$ can reduce the uncertainly about $X$. For example, suppose Neeraj Chopra has a set of javelins of varying weights and lengths. Let this set be the sample space. The length and the weight can be considered as two random variables $X$ and $X'$ (with known probability distributions) respectively. One can try to estimate the length of an arbitrarily picked javelin. Again, if error in the estimation is to be minimized in the mean square sense, then the length must be the expectation value of $X$. However, if suppose the weight of the picked javelin is known, say, $X'=x'$, then obviously the length can be estimated better because there should be a relationship between length and weight. More precisely, now $c$ is a function of $x'$ and the error, given by $E[(X-c(x'))^2]$, must be defined using joint pdf $p_{XX'}(x,x')=p_{X|X'}(x|x')p_{X'}(x')$ of $X$ and $X'$. The estimate, when the error is minimal in mean square sense, may then be found as $c(x')=E[X|x']$---the \emph{conditional mean} of the lengths of many javelins with same weight $x'$. Of course, if there exists a functional relationship, $X=f(X')$, then $c(x')$ should trivially be $f(x')$.

The jargon, prediction, involves mathematical analysis where the underlying probability distribution is completely known right from the beginning. What if only the form of the probability distribution is known and some parameter $\theta$ required to specify the distribution precisely is not known? (For our present discussion, we shall be concerned only with a single parameter of a single variable model.) Then definitely prediction makes no sense until we have found out $\theta$. There are two main ways of going about the task of finding $\theta$ as best as one can: 
\begin{itemize}
\item \emph{Classical estimation}: $\theta$ is treated as a completely unknown constant. Being part of the model probability distribution, it is supposed to leave its mark on observed random samples: The idea is that the random samples could be used to estimate $\theta$.
\item \emph{Bayesian estimation}: $\theta$ is treated as a random variable. It is distribution not assumed to be completely unknown; some prior belief about $\theta$ is assumed. Subsequently, the belief is updated in the light of the observed random samples. Basically, in this method, one is involved in the prediction of the random variable $\theta$.
\end{itemize}
Both the methods of estimation are expected to give same result if the sample size tends to infinity; also, the result in that limit is expected to be independent of the choice of the prior. As is the case with prediction, each of the above estimations can be classified into two classes: \emph{point estimation} and \emph{interval estimation}.

Few definitions are in order for understanding the classical point estimation formally. Any real valued function of the random sample $\{X_i\}_{i=1}^{i=N}$ is called a \emph{statistic}. $\hat{\theta}_N\equiv T_N(x_1,x_2,\cdots,x_N)$, where $(x_1,x_2,\cdots,x_N)$ contains observed values, is called a \emph{point estimate} which is a specific value of a statistic, the \emph{point estimator}---the function $T_N$ of the random variables $(X_1,X_2,\cdots,X_N)$. Estimators are particular statistics which are used to estimate the unknown parameter. The \emph{bias} of an estimator is defined as $b\equiv E[T_N]-\theta$. An \emph{unbiased estimator} has $b=0$. The estimator is said to be \emph{consistent} if $\lim_{N\rightarrow\infty}P(|\hat{\theta}_N-\theta|>\epsilon)\rightarrow0\,\forall\epsilon>0$.  A particular set of two sufficient conditions for a point estimator to be consistent are $\lim_{N\rightarrow\infty}E[T_N]=\theta$ and  $\lim_{N\rightarrow\infty}E[(T_N-\theta)^2]=0$. Since realistically, one almost always deals with finite observations, a diluted but more convenient concept than consistent estimator is that of \emph{best estimator} (in some optimal sense) defined, e.g. in mean square sense, as the one with minimal error, $E[(T_N-\theta)^2]$. If such a best estimator in the mean square sense is also unbiased, then it is called a \emph{(most) efficient} estimator. Since the main idea is to find some statistic based on observation that estimates $\theta$ in the most optimal way, a natural estimate of  interest is the \emph{maximum likelihood estimate}, $\hat{\theta}^{\rm(ML)}_N\equiv \arg\sup_\theta p_{\{X_i\}}(\{x_i\};\theta)$ (practically the parameter value that maximizes of the \emph{likelihood function}---the joint pdf---conventionally denoted as $\mathcal{L}(\theta|\{x_i\})$ in this context), i.e., the value for the parameter that most likely yielded the observed sequence. Another important concept is that of the \emph{sufficient statistic}. A statistic $T_N(\{X_i\})$ is called sufficient statistic for a parameter $\theta$ if the conditional pmf or pdf $p_{\{X_i\}|T}(\{x_i\};\theta|T_N=t)$ is independent of $\theta$. In simpler words, a sufficient statistic contains all information about the unknown parameter that the observed data contains.

Bayesian point estimation takes a fundamentally different route. In this approach, one assumes that the value of $\theta$ is a particular realization of a random variable $\Theta$; and in the light of some background information about $\theta$, one associates a \emph{prior} probability distribution $p_{\Theta}(\theta)$ with it. Now the observed data is another multivariate random variable $\{X_i\}$. Under the implicit assumption that a particular realization of  $\{X_i\}$ (say, $\{x_i\}$) narrows down estimation of $\theta$ because the former contains information about the latter, the point estimation minimizing the mean square error should be the conditional mean of $\theta$, $E[\Theta|\{X_i=x_i\}]$: This result should be obvious when one recalls the mathematically similar question asked in classical (prediction) estimation of a random variable $X$ when one knows the value of a related random variable $X'$. Here, suppressing unnecessary subscripts, $p(\theta|\{x_i\})=p(\{x_i\}|\theta)p(\theta)/p(\{x_i\})$ (i.e., \emph{posterior}=\emph{likelihood}$\times$\emph{prior}/\emph{evidence}) should be used for the evaluation of the conditional mean. $E[\Theta|\{X_i=x_i\}]$ is called \emph{Bayes' estimate} and $E[\Theta|\{X_i\}]$ is called \emph{Bayes' estimator}. An interesting point to note is that, despite its deceptive appearance, the likelihood is not a probability density. This is simply because, owing to the Bayesian view, the data is held fixed and $\theta$ is varying; and thus, the integral of the likelihood over the range of $\theta$ need not be equal unity as required for a probabilistic interpretation.
\subsection{Fisher Information}
In the context of statistical inferencing, there is a paramount concept of information, viz., the \emph{Fisher information}. This term was present in the literature before Shannon's information theory and the term \emph{information} is used synonymously for it in mathematical statistics. The derivative of the \emph{log-likelihood function}, $\ln p(x;{\theta})$, with respect to the parameter is known as \emph{score} which is a random variable. The Fisher information, $J(\theta)$,  is defined as the variance of the score, i.e.,
\begin{equation}
J(\theta)\equiv E\left[\left(\frac{\partial\ln p(x;{\theta})}{\partial\theta}\right)^2\right].
\end{equation}
In this context, one should note that expected value of the score is zero, whose simple proof requires the condition that the summation (or integration) and the differentiation can be commutated. In case the random variable $X$ with a probability distribution involves $M$ parameters, say, set $\{\theta_1,\theta_2,\cdots,\theta_M\}$ denoted by $\boldsymbol{\theta}$, then
\begin{equation}
J_{ij}(\boldsymbol{\theta})\equiv E\left[\frac{\partial\ln p(x_i;\boldsymbol{\theta})}{\partial\theta_i}\frac{\partial\ln p(x_i;\boldsymbol{\theta})}{\partial\theta_j}\right]\label{gij}
\end{equation}
 is defined as the $(i,j)$th component of the \emph{Fisher information matrix}, ${\sf J}(\boldsymbol{\theta})$. 

The reason behind calling the Fisher information, `information', is best understood in the context of statistical inferencing: It is a measure of the quantity of information that an observed value of a random variable possess about an unknown parameter of the probability distribution modelling the random variable. To comprehend this better, consider a one-parameter family of probability density functions, $p(x;\theta)$ (below, sometimes, written only as $p$ for brevity); both $x$ and $\theta$ are allowed to take continuous values. Let $T_N(\{X_i\})$ be an unbiased point estimator so that $E[T_N]=\theta$ $\forall\theta$. The corresponding point estimates may be denoted as $\hat{\theta}_N$. Let us simplify the situation further by considering the case of $N=1$. Now, it is fairly straightforward to see that
\begin{subequations}
\begin{eqnarray}
&&\int \hat{\theta}_1p(x;\theta)dx=\theta,\\
\implies&&\frac{\partial}{\partial\theta}\int (\hat{\theta}_1-\theta)p(x;\theta)dx=0,\\
\implies&& \int \left[(\hat{\theta}_1-\theta)\sqrt{p}\right]\left[\sqrt{p}\frac{\partial}{\partial\theta}\ln p\right]dx=1.\label{eq:CSI}\quad\label{eq:CSIP}
\end{eqnarray}
\end{subequations}
As one may note that we have assumed that two conditions hold for $p(x;\theta)$: Firstly, derivative of $\frac{\partial}{\partial\theta}\int\hat{\theta}_N\prod_i^{N}p(x_i;\theta) \prod_i^{N}dx_i=\int\hat{\theta}_N\prod_i^{N}\frac{\partial}{\partial\theta}p(x_i;\theta) \prod_i^{N}dx_i$; and secondly, $E[(T_N-\theta)^2]$ is finite. 
Squaring both the sides of Eq.~(\ref{eq:CSI}) and invoking the Cauchy--Schwartz inequality, we arrive at
\begin{eqnarray}
&&\int \left[(\hat{\theta}_1-\theta)\sqrt{p}\right]^2dx\int\left[{\sqrt p}\left(\frac{\partial\ln p}{\partial\theta}\right)\right]^2dx\ge1,\qquad\,\,\label{eq:cse}\\
\implies && E[(T_1-\theta)^2]\ge\frac{1}{J(\theta)},
\end{eqnarray}
which is an example of the \emph{Cram\'er--Rao inequality} (also called \emph{information inequality}). 

Let us now try to gain a bit of physical insight of what it is trying to say. First, let us define the \emph{efficiency} of an unbiased estimator as  
\begin{eqnarray}
e(T_1)\equiv\frac{J(\theta)^{-1}}{E[(T_1-\theta)^2]}
\end{eqnarray}
that measures how close the variance of estimator comes to the \emph{Cram\'er--Rao lower bound} set by ${J(\theta)^{-1}}$. Evidently, $e(T_1)\in[0,1]$; and the lower the variance of an estimator, the \emph{efficient} it is. Now the important fact is that there is no guarantee that the most efficient estimator (i.e., $e(T_1)=1$) would always exist; but when it does, it happens to be the maximum-likelihood (ML) estimator. (All ML-estimators need not be the most efficient; in fact, they need not even be unbiased.) This is not very hard to see. To this end,  we go back to Eq.~(\ref{eq:cse}) to note that equality holds---i.e., $e(T_1)=1$---iff 
\begin{equation}
\frac{\partial \ln p}{\partial \theta}=(\hat{\theta}_1-\theta)f(\theta),\label{eq:er}
\end{equation}
where $f(\theta)$ is some, at least once differentiable, function.  Since we are working with unbiased estimators, by taking one more derivative of this relation with respect to $\theta$ and subsequently taking the expectation values, one sees that $f(\theta)$ is actually $J(\theta)$. Ergo, $f(\theta)>0$. Hence, finally, Eq.~(\ref{eq:er}) at $\theta=\hat{\theta}_1^{\rm(ML)}$ yields $T_1=T_1^{\rm(ML)}$ as the only solution. This is because, by definition, the ML-estimator, $T_1^{\rm(ML)}$, is the one such that the score, $\partial \ln p/\partial \theta$, calculated at $\theta=\hat{\theta}_1^{\rm(ML)}$, vanishes. Furthermore, it can be seen as a consequence of Eq.~(\ref{eq:er}) and the fact that $f(\theta)=J(\theta)$, that the form of pdf of the most efficient estimator must be $p(x;\theta)=q(x)\exp[\int J(\theta)(T_1(x)-\theta)d\theta]$---an exponential family that is known to be very relevant in statistical mechanics.

Now, consider the Cram\'er--Rao inequality for the most efficient ML-estimator so that the equality holds in the relation. Since logarithm is a monotonic function, consider either the likelihood or the log-likelihood function such that it has a unimodal shape---in a plot versus $\theta$---with peak at the maximum-likelihood estimate, $\hat{\theta}_1^{(\rm ML)}$. The width of the curve is indicated by the L.H.S.---the variance term---of the Cram\'er--Rao (in)equality. Of course, more the width, the less should be the curvature around the peak. In this context,  note that  $E[\frac{\partial^2\ln p}{\partial\theta^2}]$ is equal to $-E[(\frac{\partial\ln p}{\partial\theta})^2]=-J(\theta)$. In other words, $J(\theta)$ is measure of the curvature. Therefore, less curvature means more $J(\theta)^{-1}$ which is in line with the Cram\'er--Rao (in)equality. If the estimator is not the most efficient, then the estimates are worse: The spread of the estimates, as quantified by the variance, can not beat the limit $J(\theta)^{-1}$ which is only reached (and not beaten) even by the most efficient estimators. Thus, the Cram\'er--Rao inequality puts a fundamental bound, given by the Fisher Information, on how much `information' about the parameter under question can be extracted using statistical inferencing.

Some well-known forms of the Cram\'er--Rao inequality are as follows: If the $N$ samples come from iid distributions, then the Fisher information is easily shown to be $NJ(\theta)$; and the corresponding Cram\'er--Rao inequality comes out to be
\begin{eqnarray}
 E[(T_N-\theta)^2]\ge\frac{1}{NJ(\theta)}.
\end{eqnarray}
For any estimator---unbiased or biased---one can find
\begin{eqnarray}
 E[(T_1-\theta)^2]\ge\frac{\left\{1+\frac{\partial}{\partial\theta}(E[T_1]-\theta)\right\}^2}{J(\theta)}+(E[T_1]-\theta)^2.\nonumber\\\label{eq:bcri}
\end{eqnarray}
Moreover, in the multidimensional parameter case, the Cram\'er--Rao inequality takes the following form:
\begin{eqnarray}
 {\sf C}(\boldsymbol{T}_1)\ge{{\sf J}(\boldsymbol{\theta})}^{-1},\label{eq:mcri}
\end{eqnarray}
where ${\sf C}(\boldsymbol{T}_1)$ is the $M\times M$ covariance matrix with $E[\{(\boldsymbol{T}_1)_i-\theta_i\}\{(\boldsymbol{T}_1)_j-\theta_j\}]$ as its $(i,j)$-th element and the L.H.S. is the inverse of the Fisher information matrix. The extensions of Eq.~(\ref{eq:bcri}) and Eq.~(\ref{eq:mcri}) for the case of $N$ iid samples are also possible. It should not be forgotten that not all pdf's result in the Cram\'er--Rao lower bound: E.g., as in the case of uniform distribution, $U(0,\theta)$ (i.e., $p(x;\theta)=1/\theta$ $\forall x\in[0,\theta]$), the aforementioned first condition concerned with the swapping of integral and differentiation does not always hold.

In passing, we mention an important usage of the Fisher information in the Bayesian estimation. This concerns what to choose as the prior, $p(\theta)$. Usually, the priors that influence the posterior the least are preferred. In the absence of any bias, such a \emph{non-informative} prior is a \emph{flat prior}, i.e., a uniform distribution assigning constant probability to each allowed value of $\theta$.  The non-informative prior may appear to be a rather misnomer term as some information is required to arrive at it. Probably, it is better seen as a not-very-informative prior or as an objective prior in the sense that everyone chooses the same prior when presented with the corresponding statistical inference problem. In any case, the flat prior has two serious flaws. Firstly, it is not invariant: A transformation $\phi=h(\theta)$ may lead to inequivalent inferences. Secondly, in higher dimensions ($M>1$), the more of the mass of distribution accumulates near the end of the ranges. In other words, the flat prior becomes biased in high dimensions. The issue of invariance is bypassed on choosing \emph{Jeffrey's prior}---$p_J(\boldsymbol{\theta})\propto\sqrt{{\rm det}[{\sf J}(\boldsymbol{\theta})]}$---written for $M$-dimensional case. For the one-dimensional case, it is obvious that since $p(\phi)=p(\theta)|d\theta/d\phi|$ is required for the invariance of the Bayes' rule, the fact that ${J(\phi)}={J(\theta)}|d\theta/d\phi|^2$ means that the corresponding Jeffrey's prior is invariant. For higher dimensions, sometimes Jeffrey's prior suffers from the same flaw as the flat prior in that it becomes biased. Moreover, many a time, this prior is \emph{improper}---the integral/sum of the prior values diverges. While certainly not very desirable, it is sometimes justified because if the evidence in the Bayes' rule converges, the posterior probabilities may still add up to unity even if the prior probabilities do not. Obviously, thus, all one needs is a specification of the prior in the correct proportion. The issue of the choice of appropriate prior has a huge literature and we dare not delve into it here any further.


\subsection{Fisher--Rao Geometry}
%
Note that while the Shannon entropy is defined for all kinds of probability distributions, the Fisher information is defined only for the parametrized probability distrbutions. Fisher information matrix can be extracted from the notion of relative entropy. To this end, let us consider random variable $X$ with a probability distribution consisting of $M$ parameters, say, set $\{\theta_1,\theta_2,\cdots,\theta_M\}$ denoted by $\boldsymbol{\theta}$. We assume that the corresponding pmf is a continuous and sufficiently differentiable function of $\boldsymbol{\theta}$; e.g., we can take the example of pmf of the normal distribution which is a smooth function of the parameters---mean and variance. Suppose we ask that given there is a `true' distribution $p(x;\boldsymbol{\theta})$, how a nearby `guess' distribution---$p(x;\boldsymbol{\theta}+\delta\boldsymbol{\theta})$---fares. The appropriate quantity to look for then is $D(p(x_i;\boldsymbol{\theta})||p(x;\boldsymbol{\theta}+\delta\boldsymbol{\theta}))$ (denoted as $D_{\rm KL}$ below for brevity, along with using $p$ for $p(x;\boldsymbol{\theta})$). Here, the logarithm in its expression is taken to be the natural logarithm as is customary while studying the Fisher information. In the limit $\delta\boldsymbol{\theta}\rightarrow 0$ and up to the first non-vanishing order in $\delta\boldsymbol{\theta}$, this relative entropy can be expanded and simplified to arrive at the following:
\begin{subequations}
\begin{eqnarray}
&&D_{\rm KL}
=-\frac{1}{2}\sum_{x\in\mathcal{A}}\sum_{i=1}^M\sum_{j=1}^M
p\delta\theta_i\delta\theta_j \frac{\partial^2\ln p}{\partial\theta_i\partial\theta_j},\quad\\
\implies &&D_{\rm KL}=-\frac{1}{2}\sum_{i=1}^M\sum_{j=1}^M\delta\theta_i\delta\theta_j E\left[\frac{\partial^2\ln p}{\partial\theta_i\partial\theta_j}\right],\label{eq:fiii}\\
\implies &&D_{\rm KL}=\frac{1}{2}\sum_{i=1}^M\sum_{j=1}^M\delta\theta_i\delta\theta_j E\left[\frac{\partial\ln p}{\partial\theta_i}\frac{\partial\ln p}{\partial\theta_j}\right],\qquad\label{eq:fiiii}\\
\implies &&D_{\rm KL}=\frac{1}{2}\sum_{i=1}^M\sum_{j=1}^MJ_{ij}(\boldsymbol{\theta})\delta\theta_i\delta\theta_j.
\end{eqnarray}
\end{subequations}
Here, we have assumed some mathematical regularity conditions including that the summation and the differentiations can be commutated, and used the definition of the Fisher information matrix, $J_{ij}$.  Thus, the Fisher information matrix can be seen as the Hessian of relative entropy---specifically, 
\begin{equation}
\left.\frac{\partial^2 D(p(x;\boldsymbol{\theta^*})||p(x;\boldsymbol{\theta}))}{\partial \theta_i\partial \theta_j}\right|_{\boldsymbol{\theta}=\boldsymbol{\theta^*}}=J_{ij}(\boldsymbol{\theta^*}),
\end{equation}
as can be verified by direct calculation.

In more compact matrix notations, Eq.~(\ref{gij}) can be written as ${\sf J}=E[\boldsymbol{ \vartheta}\boldsymbol{\vartheta}^T]$, where $\boldsymbol{ \vartheta}$ is an $M$-dimensional column vector with ${\partial\ln p}/{\partial\theta_i}$'s as components. ${\sf J}$ has the property that for any non-zero $M$-dimensional column vector ${\bf  u}$, ${\bf u}^T{\sf J}{\bf u}={\bf u}^TE[\boldsymbol{ \vartheta}\boldsymbol{\vartheta}^T]{\bf u}=E[{\bf u}^T\boldsymbol{ \vartheta}\boldsymbol{\vartheta}^T{\bf u}]=E[(\boldsymbol{ \vartheta}^T{\bf u})^T(\boldsymbol{ \vartheta}^T{\bf u})]=E[(\boldsymbol{ \vartheta}^T{\bf u})^2]\ge0$ on noting that $\boldsymbol{ \vartheta}^T{\bf u}$ is just a real number. In other words, ${\sf J}$ is in general a positive semidefinite matrix. In case it is positive definite, then it can induce an \emph{inner product}  $\left< \boldsymbol{\theta},\boldsymbol{\theta}'\right>\equiv \boldsymbol{\theta}^T{\sf J}\boldsymbol{\theta}'$ between any two parameter vectors---$\boldsymbol{\theta}$ and $\boldsymbol{\theta}'$---in the parameter space where the \emph{norm} is then automatically given by $||\boldsymbol{\theta}||\equiv\sqrt{\left< \boldsymbol{\theta},\boldsymbol{\theta}\right>}$ that in turn induces a metric $g(\boldsymbol{\theta},\boldsymbol{\theta}')\equiv||\boldsymbol{\theta}-\boldsymbol{\theta}'||$ in the resultant \emph{Riemannian manifold} so that $g_{ij}=J_{ij}$ is effectively a \emph{Riemannian metric tensor}. We remind ourselves that an $n$-dimensional \emph{manifold} is a topological space such that for every open set about a point in the space, there exists a continuous map which maps the open subset onto an open subset of $\mathbb{R}^n$. In the statistical scenario under consideration, this Riemannian manifold is called \emph{statistical manifold} where each point is a probability distribution parametrized by $\boldsymbol{\theta}$. Thus, one could say that the distance---within an inconsequential multiplicative factor of $1/2$---between two infinitesimally nearby points on the statistical manifold is nothing but the KL-divergence between them. Thus, there is some justification for the common practice of terming the relative entropy as \emph{KL-distance} even though it is not a true metric itself as it is not symmetric and does not satisfy the triangle inequality.

The above relations between the relative entropy and the Fisher information matrix may straightforwardly be generalized for multivariate distributions; moreover, they easily carry over to the case of continuous-state systems---the summations just get replaced by integrals. In that case, $p(x;\boldsymbol{\theta})$ is a pdf. Without any loss for generality, let us henceforth consider pdfs for concreteness. Rather than working with a space of distributions as discussed above, it is more useful to work in a Hilbert space of square integrable functions. Recall that a Hilbert space is an inner product space that is a complete metric space with respect to the norm defined by the inner product; a complete metric space is a metric space where every Cauchy sequence converges within the space. It is known that $L^2(\mathbb{R})$---a space of square integrable functions on real line---is an infinite dimensional Hilbert space. Thus, we note that if we consider \emph{square-root map}: $p(x)\mapsto\xi(x)=\sqrt{p(x)}$, then the set of all $\xi(x)$ belongs to $L^2(\mathbb{R})$ because $\int [\xi(x)]^2dx=1$ due to the normalization condition of pdf. This normalization condition also implies that the set of $\xi(x)$'s lie on a unit (hyper)sphere, $\mathbb{S}\subset{L}^2(\mathbb{R})$. The inner product between two elements, $\xi_1$ and $\xi_2$ (say), of $\mathbb{S}$ can simply be defined as 
\begin{equation}
\left< \xi_1(x),\xi_2(x)\right>\equiv\int \xi_1(x)\xi_2(x)dx.
\end{equation}
As an aside, we remark that one defines the \emph{Bhattacharyya spherical distance} between the two elements as $\cos^{-1}\left< \xi_1(x),\xi_2(x)\right>$. 

Please observe that in the immediately preceding few mathematical relations we have not tried suppress the parameters just for notational convenience, rather we have emphasized that the results are true for any pdf irrespective of how they are parametrized (if at all). Now if the pdfs depend on parameters, we need to further restrict ourselves to a subspace $\boldsymbol{\Theta}\subset\mathbb{S}$, in which the elements are now denoted aptly by $\xi_{\boldsymbol{\theta}}(x)=\sqrt{p(x;\boldsymbol{\theta})}$. These are assumed to be at least twice differentiable with respect to the parameters. Now, it is easy to see that the distance squared, $ds^2$, between two functions $\xi_{\boldsymbol{\theta}}(x)$ and $\xi_{\boldsymbol{\theta}+d\boldsymbol{\theta}}(x)$ ($d\boldsymbol{\theta}\to\boldsymbol{0}$) is given by 
\begin{subequations}
\begin{eqnarray}
&&ds^2=\left<\xi_{\boldsymbol{\theta}}(x)-\xi_{\boldsymbol{\theta}+d\boldsymbol{\theta}}(x),\xi_{\boldsymbol{\theta}}(x)-\xi_{\boldsymbol{\theta}+d\boldsymbol{\theta}}(x)\right>,\\
\implies&&ds^2=\left[\int \partial_i\xi_{\boldsymbol{\theta}}(x)\partial_j\xi_{\boldsymbol{\theta}}(x)dx\right]d\theta^id\theta^j,\qquad\qquad\qquad
\end{eqnarray}
\end{subequations}
where $\partial_i\equiv\partial/\partial\theta_i$ and the \emph{Einstein summation convention} has been assumed. One notes that
\begin{equation}
ds^2=\frac{1}{4}J_{ij}d\theta^id\theta^j,\label{eq:frm}
\end{equation}
where $J_{ij}/4$ is called the \emph{Fisher--Rao metric}. However, in practice, the factor $1/4$ is usually absorbed in $J_{ij}$. Together with this metric, the family of pdfs spanned by the parameters $\theta_i$'s---now recognized as coordinates $\theta^i$'s on the corresponding statistical manifold---generates the \emph{information geometry}, in particular, the \emph{Fisher--Rao geometry}. Note that technically $J_{ij}$ is rank-2 covariant tensor and the corresponding rank-2 contravariant tensor $J^{ij}$ is related through the relation $J_{ik}J^{kj}=\delta_i^j$ where $\delta_i^j$ is mixed \emph{Kronecker tensor}.

How to define distance between any two distributions on statistical manifold? For this purpose, Fisher--Rao geometry facilitates defining  \emph{Fisher--Rao distance} (or \emph{Rao's distance}) between two probability distributions on the manifold as that of curve joining them with minimized length (according to the Fisher--Rao metric), i.e., the \emph{geodesic} distance  between them. This distance is non-negative and symmetric, and satisfies triangle inequality; it is a proper distance unlike KL-divergence. Now-a-days, Rao's distance has garnered attention in the field of machine learning. Now to find the geodesic between two points, say,  $A,B\in\boldsymbol{\Theta}$, we need to minimize the distance of the paths between them, i.e., $\delta\int_A^Bds=0$. Simple mathematical manipulations using Eq.~(\ref{eq:frm}) yield the standard result that the geodesics must follow the following \emph{geodesic equation}:
\begin{equation}
\frac{d^2\theta^i}{ds^2}+\Gamma^{i}_{jk}\frac{d\theta^j}{ds}\frac{d\theta^k}{ds}=0,\label{eq:geoeqn}
\end{equation}
where, 
\begin{equation}
\Gamma^{i}_{jk}\equiv\frac{1}{2}J^{il}\left(\partial_j J_{lk}+\partial_k J_{lj}-\partial_l J_{jk}\right)
\end{equation} 
is known as \emph{Christoffel symbol} of the second kind (which is not a tensor). Note that $\Gamma^{i}_{jk}=\Gamma^{i}_{kj}$. Technically speaking, we are only extremizing to reach to the geodesic equation; in order to ensure minimization, the second variation needs to be checked. However, for Riemannian manifold, there can be no maximum path between two points as one can always construct a relatively longer path by doing some detour on the manifold. 

An important property of a manifold is its curvature at every point. A single number at a point does not suffice to describe curvature of a Riemannian manifold of dimension more than two. Consequently, a rank-4 tensor, {Riemann curvature tensor}, $R^{i}_{jkl}$, is used for this purpose. A Riemannian manifold has zero curvature ($R^{i}_{jkl}=0$) at a point iff there exists a distance preserving (local) bijective map between an open set about the point of manifold and the Euclidian space; a manifold with zero curvature everywhere is said to be \emph{flat}. $R^{i}_{jkl}$ is defined through \emph{covariant derivative}, $\nabla_i$, 
\begin{equation}
\nabla_iV_j\equiv \partial_iV_j-\Gamma_{ij}^kV_k,
\end{equation}
where $V_i$ is any covariant vector. One defines
\begin{equation}
R^{i}_{jkl}V_i\equiv [\nabla_k\nabla_l-\nabla_l\nabla_k]V_j,
\end{equation}
which can be more explicitly expressed as
\begin{equation}
R^{i}_{jkl}=\partial_k\Gamma^i_{jl}-\partial_l\Gamma^i_{jk}+\Gamma^i_{km}\Gamma^m_{jl}-\Gamma^i_{lm}\Gamma^m_{jk}.
\end{equation}
Two related quantities are \emph{Ricci curvature tensor}, $R_{ij}$, and \emph{Ricci scalar curvature}, $R$, defined as follows:
\begin{equation}
R_{ij}\equiv R^{k}_{ikj}~{\rm and}~R\equiv J^{ij}R_{ij}.
\end{equation}
If the manifold is flat, then all three---$R^{i}_{jkl}$, $R_{ij}$, and $R$---are zero:  $R^{i}_{jkl}=0\implies R_{ij}=0\implies R=0$. The converse is dependent on the dimension of the Riemannian manifold. One dimensional manifold is always flat and all three are always zero. A two dimensional manifold is flat if $R=0$; vanishing of $R_{ij}$ and $R^i_{jkl}$ follows automatically. In three dimensional manifold, however, vanishing of the Ricci scalar need not imply flatness but $R_{ij}=0$ implies that the manifold is flat. In a manifold of dimension more than three, neither the Ricci tensor nor the Ricci scalar can determine sufficient condition for finding whether the manifold is flat or not.
\subsection{Statistical Manifold of Normal Distributions}
Let us now study a canonical example of Fisher--Rao geometry on statistical manifold. Consider the univariate Normal or Gaussian distribution, $\mathcal{N}(\mu,\sigma^2)$, as given in Eq.~(\ref{eq:gdnd}). Considering $\theta^1\equiv \mu$ and $\theta^2\equiv\sigma$ as the coordinates of the manifold,  we can use all the relations discussed in the last section to find the curvature and the geodesics. 

First and foremost, the Fisher--Rao metric should be found. It is easy to check that
$\partial_1\ln{\mathcal{N}}=(x-\mu)/\sigma^2$ and $\partial_2\ln{\mathcal{N}}=(x-\mu)^2/\sigma^3-1/\sigma$. Therefore, 
\begin{subequations}
\begin{eqnarray}
&&J_{11}=\int_{-\infty}^{\infty}\frac{e^{-\frac{(x-\mu)^2}{2\sigma^2}}}{\sqrt{2\pi\sigma^2}}\left(\frac{x-\mu}{\sigma^2}\right)\left(\frac{x-\mu}{\sigma^2}\right)dx,\label{eq:gdnd}\nonumber\\
\\
&&J_{12}=\int_{-\infty}^{+\infty}\frac{e^{-\frac{(x-\mu)^2}{2\sigma^2}}}{\sqrt{2\pi\sigma^2}}\left(\frac{x-\mu}{\sigma^2}\right)\left(\frac{(x-\mu)^2}{\sigma^3}-\frac{1}{\sigma}\right)dx,\nonumber\\
\\
&&J_{22}=\int_{-\infty}^{+\infty}\frac{e^{-\frac{(x-\mu)^2}{2\sigma^2}}}{\sqrt{2\pi\sigma^2}}\left(\frac{(x-\mu)^2}{\sigma^3}-\frac{1}{\sigma}\right)^2dx.\nonumber\\
\end{eqnarray}
\end{subequations}
On evaluating these simple integrals and using the fact that $J_{12}=J_{21}$, we arrive at the metric,
\begin{eqnarray}
\sf{J}=\left(
\begin{array}{cc}
   J_{11} & J_{12}  \\
J_{21}   & J_{22}
\end{array}
\right)
=
\left(
\begin{array}{cc}
   \frac{1}{\sigma^2} & 0  \\
0    & \frac{2}{\sigma^2}
\end{array}
\right),
\end{eqnarray} 
when expressed in matrix form. Now calculating the Christoffel symbols and thence Ricci scalar curvature is straightforward. Specifically, one finds that in this case, $R=-1$, implying that the manifold is a homogeneous one with constant negative curvature, i.e., it is a \emph{hyperbolic space}.

The implications of being hyperbolic is manifested as non-straight line geodesics. To see this, we write down the geodesic equations [Eq.~(\ref{eq:geoeqn})] explicitly:
\begin{subequations}
\begin{eqnarray}
&&\frac{d^2\mu}{ds^2}-\frac{2}{\sigma}\frac{d\mu}{ds}\frac{d\sigma}{ds}=0,\label{eq:g1}\\
&&\frac{d^2\sigma}{ds^2}+\frac{1}{2\sigma}\left(\frac{d\mu}{ds}\right)^2-\frac{1}{\sigma}\left(\frac{d\sigma}{ds}\right)^2=0.\label{eq:g2}
\end{eqnarray}
\end{subequations}
Now, for notational convenience, allowing prime symbol to denote derivative with respect to $s$, we arrive at
\begin{subequations}
\begin{eqnarray}
&&\left(\frac{\mu'}{\sigma}\right)'-\frac{\mu'}{\sigma}\frac{\sigma'}{\sigma}=0,\label{eq:g1'}\\
&&\left(\frac{\sigma'}{\sigma}\right)'+\frac{1}{2}\frac{\mu'}{\sigma}\frac{\mu'}{\sigma}=0,\label{eq:g2'}
\end{eqnarray}
\end{subequations}
on dividing geodesic equations by $\sigma$. Multiplying Eq.~(\ref{eq:g1'}) by $\mu'/\sigma$ and Eq.~(\ref{eq:g1'}) by $2\sigma'/\sigma$, and adding the results, gets us
\begin{eqnarray}
\left[\left(\frac{\mu'}{\sigma}\right)^2+2\left(\frac{\sigma'}{\sigma}\right)^2\right]'=0.
\end{eqnarray}
It implies that 
\begin{eqnarray}
\left(\frac{\mu'}{\sigma}\right)^2+2\left(\frac{\sigma'}{\sigma}\right)^2=2a^2, \label{eq:222}
\end{eqnarray}
where $2a^2$ is integration constant. Now on dividing Eq.~(\ref{eq:g1'}) by $\sigma$, we find that the equation boils down to $(\mu'/\sigma^2)'=0$, leading to $\mu'={\sqrt{2}}b\sigma^2$, where ${\sqrt{2}}b$ is an integration constant. Consequently, Eq.~(\ref{eq:222}) can be cast as
\begin{eqnarray}
b^2\sigma^2+\left(\frac{\sigma'}{\sigma}\right)^2=a^2. \label{eq:ab}
\end{eqnarray}
Two cases emerge.

The first case is when $b=0$, which implies $\mu$ is constant and $\sigma\propto \exp(as)$ [see Eq.~(\ref{eq:ab})] with $a\ne0$. Therefore, the geodesics are straight lines parallel to $\sigma$-axis in the $\mu$-$\sigma$ plane that is $\mathbb{R}\times(0,\infty)$.  The second case is when $b\ne 0$. In this case, first one notes that, owing to Eq.~(\ref{eq:ab}), $b\sigma\le a$. This means that one can represent $\sigma(s)$ as $\frac{a}{b}\sin\gamma(s)$ where $\gamma$ is some function; because of the property of $\sin$, $\sigma$ cannot exceed $a/b$. This form of $\sigma$ helps reducing Eq.~(\ref{eq:ab}) to $\gamma'^2=a^2\sin^2\gamma$. Obviously, $\gamma'\ne0$ rendering $\gamma$ invertible. Furthermore, assuming without any loss of generality, $\gamma'>0$, we get $\gamma'=a\sin\gamma$. This, in turn, implies $\sigma'/\sigma=a\cos\gamma$, and thence, using Eq.~(\ref{eq:222}), $\mu'=\sqrt{2}a\sigma\sin\gamma$ which can be integrated to arrive at $\mu/\sqrt{2}=-\frac{a}{b}\cos\gamma+c$ ($c$ is integration constant). Hence, the conclusion is that the geodesics can be represented by the equations:
\begin{subequations}
\begin{eqnarray}
&&\frac{\mu(s)}{\sqrt{2}}=-\frac{a}{b}\cos\gamma(s)+c,\\
&&\sigma(s)=\frac{a}{b}\sin\gamma(s).
\end{eqnarray}
\end{subequations}
Consequently, using the notation, $\mu^*\equiv\mu/\sqrt{2}$, one can see the geodesics as semi-circles of radius $a/b$ centered at $\mu^*$-axis  on the $\mu^*$-$\sigma$ plane.

Now finding Rao's distance, $D_{\rm FR}$, between two Gaussian distributions, $\mathcal{N}(\mu_1,\sigma_1^2)$ and $\mathcal{N}(\mu_2,\sigma_2^2)$, is essentially finding length of the semi-circle's arc joining the two corresponding points (say, $p_1$ and $p_2$) on the statistical manifold. The result is
\begin{equation}
D_{\rm FR}(p_1,p_2)={2}{\sqrt{2}}{\rm arctanh}\left(\sqrt\frac{(\mu_1-\mu_2)^2+2(\sigma_1-\sigma_2)^2}{(\mu_1-\mu_2)^2+2(\sigma_1+\sigma_2)^2}\right).
\end{equation}
In case $\mu_1=\mu_2$, then geodesic becomes a straight line ($\mu={\rm constant}$) and $D_{\rm FR}(p_1,p_2)$ simplifies to ${\sqrt{2}}|\ln(\sigma_1)/\sigma_2|$.

The $\boldsymbol{\theta}=(\theta^1,\theta^2)=(\mu,\sigma)$ coordinates are not special---another choice sometimes can be more useful or tractable. An interesting form, in which the pdf of Gaussian distribution can be expressed, is
\begin{equation}
p(x;\boldsymbol{\theta})=\frac{\exp\left(-\sum_{i=1}^2\theta^iE_i(x)\right)}{Z(\boldsymbol{\theta})},
\end{equation}
where $E_1\equiv x^2$, $E_2\equiv x$ $(\theta^1,\theta^2)\equiv(1/2\sigma,-\mu/\sigma^2)$ and 
\begin{equation}
\ln Z\equiv \frac{1}{8}\left(\frac{\theta^2}{\theta^1}\right)^2-\ln\left(\frac{\sqrt{2\pi}}{2\theta^1}\right).
\end{equation}
This form of pdf reminds one of the canonical distribution in equilibrium statistical mechanics and hence, the specific coordinate system may be said to have been expressed using \emph{canonical parametrization}---$(\theta^1,\theta^2)$---in this case. Such coordinate systems facilitate easier manipulations of the tensorial relations as we now mention.

In general, let us consider equilibrium distributions of the canonical form---
\begin{equation}
p(x;\boldsymbol{\theta})=\frac{q(x)\exp\left(-\sum_{i=1}^M\theta^iE_i(x)\right)}{Z(\boldsymbol{\theta})},\label{eq:pc}
\end{equation}
where $x$ is phase space point, $q(x)$ is the equilibrium state at $\boldsymbol{\theta}=\boldsymbol{0}$; $\theta^i$'s are parameters like (inverse) temperature, pressure, chemical potential, etc.; and $E_i$'s are extensive variables like energy, volume, particle number, etc. Of course, due to normalization condition of pdf, $Z$ is nothing but the integral of the numerator of the right hand side of Eq.~(\ref{eq:pc}) with respect to  $x$ over the entire phase space. Under this canonical parametrization, $\boldsymbol{\theta}$, on $\boldsymbol{\Theta}$, one finds by direct calculations that the corresponding Fisher--Rao metric, $J_{ij}$ is $\partial_i\partial_j\ln Z$; the Christoffel symbol (of the first kind), $\Gamma_{ijk}\equiv J_{il}\Gamma^l_{jk}$ is $\frac{1}{2}\partial_i\partial_j\partial_k\ln Z$; and ergo, the Ricci scalar curvature, in case the manifold is two dimensional, is
\begin{equation}
R=
-\frac{1}{2|{\sf J}|^2}\left|
\begin{array}{lll}
 \partial_1\partial_1\ln Z &   \partial_1\partial_2\ln Z  &  \partial_2\partial_2\ln Z   \\
  \partial_1\partial_1\partial_1\ln Z  &  \partial_1\partial _1\partial _2\ln Z  &  \partial_1\partial _2\partial _2\ln Z  \\
  \partial_1\partial _1\partial _2\ln Z  &  \partial_1\partial _2\partial _2\ln Z  &  \partial_2\partial _2\partial _2\ln Z 
\end{array}
\right|,\qquad
\end{equation}
where $|{\sf J}|$ is determinant of the metric. 

We note an interesting point: Since under canonical parametrization, $\ln Z$ may be seen as related to free energy, any discontinuity in the free energy or its lower order derivatives manifest as singularity in the metric and hence, in the Riemann curvature tensor. It means that thermodynamic phase transitions should manifest as singularities of the statistical manifold because at phase transitions, the free energy is singular.

\acknowledgments
 The author is indebted to Saikat Ghosh for the motivation, enthusiasm and criticism that initiated and shaped this write-up. Furthermore, the author is thankful to all those students who audited or credited the course---partially or fully---that was closely based on this write-up. 
\section*{References}
Unfortunately, the author did not keep track of all the references used to prepare this write-up. Hence, the set of references listed below is supposed to serve only as the bare minimum authoritative introductory material on the subject:

\begin{itemize}
\item C. E. Shannon and W. Weaver, {\it The Mathematical Theory of Communication}, University of Illinois Press (1962). 
\item A. I. Khinchin, {\it Mathematical Foundations of Information Theory}, Dover Publications (1957). 
\item T. M. Cover and J. A. Thomas, {\it Elements of Information Theory}, Wiley-Interscience (2006).
\item A. Papoulis and S. U. Pillai, {\it Probability, Random Variables and Stochastic Processes}, McGraw Hill Education (2002). 
\item M. Cencini, F. Cecconi, and A. Vulpiani, {\it Chaos: From Simple Models to Complex systems}, World Scientific (2009).
\item C. Beck and F. Schl\"ogl, {\it Thermodynamics of Chaotic Systems: An Introduction}, Cambridge University Press (1995).
\item H. G. Schuster, {\it Deterministic Chaos: An Introduction}, Wiley VCH (1995).
\item R.D. Rosenkrantz (Ed.), {\it E. T. Jaynes: Papers on Probability, Statistics and Statistical Physics}, Springer Netherlands (1989). 
\item E. T. Jaynes, {\it Probability Theory: The Logic of Science}, Cambridge University Press (2003).
\item M. J. Titelbaum, {\it Fundamentals of Bayesian Epistemology 1}, Oxford University Press (2022).
\item P. Adriaan and J. van Benthem (Eds.), {\it Philosophy of Information (Handbook of the Philosophy of Science)}, North Holland (2008). 
\item A. H\'ajek, {\it Interpretations of Probability}, Stanford Encyclopaedia of Philosophy (2002).
\item D. C. Brody and D. W, Hook, {\it Information Geometry in Vapour-Liquid Equilibrium}, Journal of Physics A: Mathematical and Theoretical 42, 023001 (2009).
\item S. R. Eddy, {\it What is Bayesian statistics?}, Nature Biotechnology 22, 1117 (2004).
\end{itemize}
\begin{center}
	$\symking\symqueen\symrook\symknight\symbishop\sympawn\epsdice{1}\,\epsdice{2}\,\epsdice{3}\,\epsdice{4}\,\epsdice{5}\,\epsdice{6}\,\spadesuit\heartsuit\clubsuit\diamondsuit$
\end{center}
 \end{document}